\documentclass[onecolumn]{svjour3}

\journalname{Journal of Computational Physics}

\usepackage{amssymb,amsfonts}
\usepackage{amsmath,mathtools}
\usepackage{upgreek} 
\usepackage{graphicx}
\usepackage{color}
\usepackage{bm}
\usepackage{float}
\usepackage{subfiles}
\usepackage{soul}
\usepackage{subcaption}
\usepackage{fancybox}
\usepackage[linesnumbered]{algorithm2e}

\RequirePackage{fix-cm}
\smartqed

\newcommand{\heading}[1]{\smallskip\noindent\textbf{#1}}

\newcommand{\ddt}[1]{\frac{{\rm d}{#1}}{{\rm d}t}}

\newcommand{\ba}{\boldsymbol{\alpha}}
\newcommand{\mbm}{{\textbf{m}}}
\newcommand{\mbk}{{\textbf{k}}}

\usepackage[
backend=biber,
style=numeric-comp,
sorting=none
]{biblatex}
\begin{document}
\title{Minimizing information loss reduces spiking neuronal networks to differential equations} 
\author{Jie Chang$^{1}$ \and Zhuoran Li$^{1}$ \and Zhongyi Wang$^{2}$   \and Louis Tao$^{1,3,\dagger}$ \and Zhuo-Cheng Xiao$^{4,5,\dagger}$}

\institute{  
           $^1$ Center for Bioinformatics, National Laboratory of Protein Engineering and Plant Genetic Engineering, School of Life Sciences, Peking University, Beijing, 100871, China; \\
           $^2$ Courant Institute of Mathematical Sciences, New York University, NY, USA 10003; \\
           $^3$ Center for Quantitative Biology, Academy for Advanced Interdisciplinary Studies, Peking University, Beijing, 100871, China; \\
           $^4$ NYU-ECNU Institute of Mathematical Sciences, New York University Shanghai, Shanghai, 200124, China; \\
           $^5$ NYU-ECNU Institute of Brain and Cognitive Science, New York University Shanghai, Shanghai, 200124, China; \\
           $^\dagger$ Corresponding authors: 
              Z.-C.~Xiao \email{xiao.zc@nyu.edu};
              L.~Tao \email{taolt@mail.cbi.pku.edu.cn}.           
}

\date{Received: date / Accepted: date / Edited: date}

\maketitle

\begin{abstract}
Spiking neuronal networks (SNNs) are widely used in computational neuroscience, from biologically realistic modeling of local cortical networks to phenomenological modeling of the whole brain. Despite their prevalence, a systematic mathematical theory for finite-sized SNNs remains elusive, even for idealized homogeneous networks. The primary challenges are twofold: 1) the rich, parameter-sensitive SNN dynamics, and 2) the singularity and irreversibility of spikes. These challenges pose significant difficulties when relating SNNs to systems of differential equations, leading previous studies to impose additional assumptions or to focus on individual dynamic regimes. In this study, we introduce a Markov approximation of homogeneous SNN dynamics to minimize information loss when translating SNNs into ordinary differential equations. Our only assumption for the Markov approximation is the fast self-decorrelation of synaptic conductances. The system of ordinary differential equations derived from the Markov model effectively captures high-frequency partial synchrony and the metastability of finite-neuron networks produced by interacting excitatory and inhibitory populations. Besides accurately predicting dynamical statistics, such as firing rates, our theory also quantitatively captures the geometry of attractors and bifurcation structures of SNNs. Thus, our work provides a comprehensive mathematical framework that can systematically map parameters of single-neuron physiology, network coupling, and external stimuli to homogeneous SNN dynamics.
\end{abstract}


\section*{Highlights}
\begin{itemize}
    \item With minimal assumptions, a Markov framework counting the number of neurons in each state captures population dynamics of finite-sized, homogeneous spiking neural networks. 
    \item The expectation of the random path is derived as a set of ordinary differential equations (ODE) for discretized neuronal states. 
    \item Our ODE system well-approximates the dynamical statistics and attractors of the corresponding spiking neuronal networks, including near bifurcation points.
\end{itemize}

\section{Introduction}
\label{Sect1-Intro}
Spiking neuronal networks (SNNs) are widely used in computational neuroscience to model and simulate neuronal networks ranging from local patches in cortex or the hippocampus to larger-scale neural pathways and processes involved in sensory integration and decision making \cite{wang2002probabilistic,izhikevich2008large,eliasmith2012large,potjans2014cell,zenke2015diverse,markram2015reconstruction,bezaire2016interneuronal,chariker2016orientation,ramirez2018dissecting,schmidt2018multi,billeh2020systematic,siegle2021survey,zenke2021remarkable,chariker2022computational}. In contrast to rate-based network models, where the population firing rates is the carrier of information, SNN captures the fine-grained dynamics where neurons communicate with discrete electrical impulses, i.e., action potentials or spikes. In the brain, spikes may carry information not only through mean frequencies but also via precise timing \cite{bialek1989reading,mainen1995reliability,decharms1996primary,BialekWarlandSteveninck1997,singer1999neuronal,tiesinga2008regulation}. For instance, temporal precision is critical in learning mechanisms such as spike-timing dependent plasticity (STDP), where the relative timing of spikes between neurons instructs the strengthening or weakening of their synaptic connections \cite{bi1998synaptic,song2000competitive,dan2004spike,brzosko2019neuromodulation,payeur2021burst}.

Furthermore, biological neurons are sensitive to spike timing when they receive inputs \cite{bialek1989reading,KonigEtAl1996,singer1999neuronal,brette2012computing}. This sensitivity can induce partial and transient synchrony within neural circuits, leading to temporally coordinated activities \cite{singer1999neuronal, wang2010neurophysiological}. The precision of neuronal spikes is essential for understanding how neural circuits encode, transmit, and process information \cite{johansson2004first,fries2015rhythms,panzeri2017cracking,lundqvist2016gamma}. Neural oscillations arising from synchronized neuronal activity play a pivotal role in coordinating communication between different brain regions and facilitating cognitive processes \cite{buzsaki2004neuronal,siegle2021survey}. Similarly, previous studies on SNNs also show that small variability in spike timing can lead to vastly different overall network dynamics, including significant changes in synchrony and the emergence or suppression of neural oscillations \cite{izhikevich2004model,schmidt2018multi}. Thus, a unified mathematical theory for SNNs is highly sought after. Ideally, such a theory would provide a mapping that directly relates external inputs, network architecture, and single-neuronal physiology to the dynamics and function of the SNN, allowing for accurate predictions of network synchrony and neural oscillatory behavior.
However, developing a comprehensive mathematical theory faces significant challenges. 

First, SNNs can exhibit a wide range of dynamical behaviors, from stable, fixed points to oscillatory, metastable, and chaotic dynamics \cite{van1996chaos,izhikevich2004model,gerstner2014neuronal,nobukawa2017chaotic}. Previous studies show that SNNs are highly sensitive to parameter choices \cite{brunel2000dynamics,izhikevich2008large,mejias2010irregular,landau2016impact,xiao2021data}. For example, Xiao et al. (2021) report that a mere 1\% change in synaptic coupling weights can disrupt biologically realistic firing rates. Second, the singularity and irreversibility of spikes lead to great difficulty for classical analytical tools based on differential equations \cite{ostojic2011spiking,schwalger2017towards,MontbrioEtAl2015,buice2013dynamic}. In SNNs, spikes are often modeled as singular signals between neurons.  
The physiological process of a neuron generating a spike—surpassing a threshold and then resetting—is irreversible, unlike mutual interactions in Newtonian systems \cite{hodgkin1952quantitative,izhikevich2004model}. This irreversibility and singularity, combined with the asymmetry of spike projections, means that the precise timing of even a single spike may significantly influence the overall SNN dynamics \cite{DiesmannEtAl1999,zillmer2006desynchronization,london2010sensitivity}. Additionally, real brain networks are neither too small nor infinitely large. This makes SNNs high-dimensional systems, but assuming $N \to \infty$ neglects fluctuations and variations inherent in local neural circuits with finite numbers of neurons. This is especially the case in fundamental functional units containing only $O(10)-O(10^3)$ neurons, such as minicolumns and hypercolumns \cite{mountcastle1997columnar,rockel1980basic,buxhoeveden2002minicolumn,feldmeyer2013barrel}.

Many approaches have been proposed to address these challenges. These include neural field models, Fokker-Planck equations, kinetic theories, refractory density methods, Kuramoto-type methods, and so on \cite{wilson1972excitatory,wilson1973mathematical,cai2006kinetic,vinci2023self,schwalger2017towards,MontbrioEtAl2015,buice2013beyond}. Each of these methods typically requires additional assumptions that facilitate the derivation of differential equations for the population dynamics. For example, neural field equations assume predefined forms of interactions between average rates of spiking events, and therefore, does not apply to spiking dynamics \cite{buice2013dynamic,ostojic2011spiking}; Fokker-Planck and kinetic theory models assume infinite neuron numbers, weak coupling, and low degrees of transient synchrony; moreover, refractory density methods assume identical inputs to all neurons. These assumptions come with a trade-off: the further they deviate from biological realism, the more the resulting theory is confined to specific dynamic regimes, making the model less generally applicable to the complex dynamics observed in real brain networks.

In this study, we propose a Markovian framework that reduces SNNs comprising of multiple homogeneous populations to a set of differential equations. We minimize the information loss of SNN dynamics by \textit{only} assuming the fast decorrelation of synaptic conductances from their own history. This assumption, necessary for coarse-graining, makes neurons in the same state interchangeable. It is also a natural assumption for neuronal networks with homogeneous architecture. To approximate SNN dynamics, our theory employs a Markov model with discrete states, and the irreversible spike-generating process and singularity of spiking impact become merely state transitions within the Markov model.
Its Markovian nature also provides unique analytical benefits, such as invariant probability measures, transitions between multiple attractors, further model reductions, and more (see \textbf{Discussion}). We then derive a set of ordinary differential equations (termed discrete-state ODEs or dsODEs) from the coarse-grained Markov model by counting the number of neurons in each state. It provides both quantitative predictions for SNN dynamics and convenience for mathematical analysis. The whole approach is demonstrated on leaky, integrate-and-fire neuronal networks that are widely used in computational neuroscience and brain modeling, and can also be extended to other types of SNNs with more biological details.

Compared to previous mathematical theories of SNNs, our framework directly addresses challenges of biologically realistic modeling setups, such as partially synchronous dynamics, fluctuation-driven dynamics, variations from a finite number of neurons, and strong recurrent couplings. These advantages are gained by discretizing and fixing neuronal states, so one can directly count the neuronal flux between states for arbitrarily small timestep $\delta t$. In other words, we sacrifice the ``spatial accuracy" of neuronal state to gain the flexibility to account for transient synchrony and noise, represented by the neuronal flux between states (see more in Sect.~\ref{Sect5-Comparison}).  

This paper is structured as follows. Section \ref{Sect2-Markov} presents the Markov approximation of SNNs on leaky integrate-and-fire (LIF) networks and introduces its coarse-grained version. In Section \ref{Sect3-SDE}, we derive the dsODEs from the coarse-grained Markov model. Readers not interested in mathematical details can directly proceed to Section \ref{Sect4-Simulation}, where we systematically compare dsODEs with LIF network simulations across parameter space. Section \ref{Sect5-Comparison} presents the comparison between dsODE and previous population methods, including Wilson-Cowan neural field equations, Fokker-Planck equations, and refractory density methods. Section \ref{Sect6-Diss} concludes our results and discusses future directions.

\section{A Markovian approximation to SNNs}
\label{Sect2-Markov}
We present a Markovian framework on SNNs consisting of leaky integrate-and-fire (LIF) neurons. Potential extensions to other types of spiking neurons are discussed in Sect \ref{Sect6-Diss}. 

In this section, we first recall the general formulation of an LIF network, before introducing our Markov framework. Then, for an SNN consisting of homogeneous neural populations, we derive a coarse-grained (CG) version of the Markov model focusing on the number of neurons in each state. The only assumption for the CG model is the fast self-decorrelation of recurrent synaptic drives to each neuron, leading to the interchangeability of neurons within each state. 

\subsection{Leaky-Integrate-Fire networks}
\label{Sect2.1-LIF}
The formulation of LIF networks is standard and can be found in \cite{gerstner2014neuronal}.
Throughout this paper, we focus on LIF networks consisting of homogeneous neuronal populations (that is, within each sub-population, all neurons have identical physiological properties). Here we introduce an example of a network with two populations, 1 excitatory (E) and 1 inhibitory (I). Similar setups can be directly extended to networks with fewer or more populations.

Consider an E population with $N^E$ neurons and an I population with $N^I$ neurons. The dynamics of neuron $i$ of type-$Q$ ($Q\in\{E,I\}$) are described by its membrane voltage $V_i$, externally injected current $I_i^{\mathrm{ext}}$, and synaptic conductances $(g_i^E, g_i^I)$:
\begin{subequations}
\label{Sect2.1-Eqn1-LIF}
    \begin{eqnarray}
    \label{Sect2.1-Eqn1-LIFv}
            \ddt{V_i} &=& \underbrace{g^{Q\mathrm{leak}}(\varepsilon^{\mathrm{rest}}-V_i)}_{\mathrm{leaky}} + \underbrace{I_i^{\rm{ext}}+g_i^E\cdot\left(\varepsilon^{E}-V_i\right)}_{\mathrm{excitatory}}+
    \underbrace{g_i^I\left(\varepsilon^{I}-V_i\right)}_{\mathrm{inhibitory}}, \\
    \label{Sect2.1-Eqn1-LIFext}
    I_i^{\rm{ext}}(t) &=& S_i^{\rm{ext}}\sum_{\mu_i^{\rm{ext}}}G^{\rm{ext}}(t-t_{\mu_i^{\rm{ext}}}), \\
    \label{Sect2.1-Eqn1-LIFE}
    g_i^E(t) &=& \sum_{\substack{j\in E\\j\neq i}}S_{ij}^{E}\sum_{\mu_j^{E}}G^E(t-t_{\mu_{j}^{E}}), \\
    \label{Sect2.1-Eqn1-LIFI}
    g_i^I(t) &=& \sum_{\substack{j\in I\\j\neq i}}S_{ij}^{I}\sum_{\mu_j^{I}}G^I(t-t_{\mu_{j}^{I}}).
    \end{eqnarray}
\end{subequations}
$V_i$ is driven towards the rest potential $\varepsilon^{\mathrm{rest}}$ by the leak  current, towards the excitatory reversal potential $\varepsilon^{E}$ by the excitatory current, and towards the inhibitory reversal potential $\varepsilon^{I}$ by the inhibitory current. When $V_i$ arrives at the threshold potential $V^{th}$, a type-$Q$ spike ($Q\in\{E,I\}$) is instantly released from neuron $i$ and projected to its postsynaptic neurons. After that, $V_i$ stays in a refractory period for $\tau^{\rm{ref}}$ ms before reset to $\varepsilon^{\mathrm{rest}}$. Neurons in the refractory period do not respond to any stimuli. A reduced-dimensional choice of $(\varepsilon^{\mathrm{rest}}, V^{th}, \varepsilon^{E}, \varepsilon^{I})$ is $(0, 1, 14/3, -2/3)\times100$ to best fit electrophysiological experiments \cite{koch1999biophysics,mclaughlin2000neuronal}.

While the leak conductances $g^{Q\mathrm{leak}}$ are constants, the two synaptic conductances $(g_i^E, g_i^I)$ and external stimulation current $I_i^{\mathrm{ext}}$ are time dependent and reflect spiking signals received by neuron $i$. They are weighted sums of Green's functions of corresponding spike trains coming from external sources and other E/I neurons ($\{t_{\mu_i^{\rm{ext}}}\}$, $\{t_{\mu_j^E}\}$, and $\{t_{\mu_j^I}\}$, respectively). In practice, the Green's functions, which represent the dynamics of the post-synaptic receptors, are usually chosen as 
\begin{align}
\label{Sect2.1-Eqn2-Green}
    G^R(t) = \frac{1}{\tau^R} e^{-t/\tau^R} \cdot H(t),
\end{align}
where $\tau^R$ is the timescale of the type-$R$ synaptic dynamics ($R\in\{\mathrm{ext},E,I\}$), and $H(t)$ is the Heaviside function.  

Finally, $S_i^{\rm{ext}}$ stands for the strength of input synaptic coupling weight to neuron $i$.  The network architecture is encoded by all recurrent synaptic coupling weights ($S_{ij}^{E}$ and $S_{ij}^{I}$). According to the Dale's law, a neuron can only be either excitatory or inhibitory \cite{dale1935pharmacology,kandel2000principles}. Hence, at least one of $S_{ij}^{E}$ and $S_{ij}^{I}$ should be 0, depending on the cell-type of neuron $j$.

\subsection{A Markovian approximation of LIF networks}
\label{Sect2.2-MarkovIF}
This subsection presents our Markov framework as a surrogate of LIF networks described in Sect.~\ref{Sect2.1-LIF}. The primary idea is to discretize the voltage of each neuron ($V_i$) into a finite state space. The drives $(I_i^{\mathrm{ext}},g_i^E, g_i^I)$ are discretized accordingly, and their consequence on voltages induce transitions between states. For refractory periods, we apply a special treatment to render them memoryless in implementation. Finally, we introduce the homogeneous network setup.

When the number of states approaches infinity, we expect the Markov model to converge to the corresponding LIF model, at least in the weak sense. This is supported by our numerical experiments with both single neurons and networks (data not shown). 
However, rigorous proof of convergence and an analytical estimation of convergence speed are well beyond the scope of the current study. The main purpose of this paper is to present the concept of the Markovian framework for SNNs and demonstrate its feasibility in reducing SNNs to ODEs.

\subsubsection{Markovian Dynamics from Single Neurons to Networks}

\heading{Discretized Membrane Potentials.} We discretize the domain of membrane potential $(\varepsilon^{I}, V^{th}]$ into a state space $\Gamma$. For a type-Q neuron $i$, its discretized membrane potential ($v_i$) takes value in:
$$
v_i \in \Gamma := \{-m^I, -m^I+1, \ldots, -1, 0, 1, \ldots, m^{th}\} \cup \{\mathcal{R}\}.
$$
Here, $-m^I$, $m^{th}$, and $\mathcal{R}$ correspond to $\varepsilon^{I}$, $V^{th}$, and the refractory state, respectively. Other states $m \in \Gamma$ are chosen to cover $V_i \in [ma, (m+1)a)$ for the corresponding LIF neuron. Here, $a = (V^{th} - \varepsilon^{I})/M$ is the ``bin size" of each voltage state, and $M = m^{th} + m^I$. Upon reaching $m^{th}$, $v_i$ enters the refractory state $\mathcal{R}$, emitting a type-Q spike to its postsynaptic neurons.

\heading{Approximating Conductances.} One of the most important features of the Markov model is that the continuous drive variables $( I_i^{\mathrm{ext}}, g_i^E, g_i^I)$ are rephrased as pools of ``pending kicks" $(H_i^{\mathrm{ext}},H_i^E, H_i^I)$. A single ``pending kick" quantifies the impact of a pre-synaptic spike that \textit{is yet to affect} the postsynaptic $v_i$. More specifically, when an external spike arrives, $[\frac{S^{\rm{ext}}_i}{a}]$ pending kicks are added to $H_i^{\mathrm{ext}}$. Each pending kick has an independent exponentially distributed waiting time $\mathbf{\tau} \sim \mathrm{Exp}(\tau^{\rm{ext}})$. When the waiting time is completed, a pending kick induces a state transition of $v_i$ (using rules described below), then removed from the pool. Therefore, as $M \to \infty$ and given an external spike train $\{t_{\mu_i^{\rm{ext}}}\}$, $aH_i^{\mathrm{ext}}(t) \to I_i^{\mathrm{ext}}(t)$ \textit{almost surely} for any $t$, in accordance with the strong law of large numbers. Similar procedures apply to $g_i^R$ as well, where $H_i^R$ is the corresponding pending kick pool ($R \in \{E, I\}$). $[\frac{S^{R}_{ij}}{a}]$ pending kicks are added to the pool whenever a type-R spike arrives, and the waiting time $\mathbf{\tau} \sim \mathrm{Exp}(\tau^{R})$. 

\heading{Approximating the Impact of Currents.} Leak currents, which induce exponential decay of $V_i$ in the absence of other currents, are represented by downward jumps to the next lower state for $m > 0$ (or upward for $m < 0$) at a rate of $g_L |m|$. When an ext-kick takes effect, $v_i$ increases by one state. On the other hand, The impact of recurrent pending kicks depends on $v_i$ and the reversal potentials. More specifically,
\begin{itemize}
    \item When an E-kick takes effect, $v_i$ jumps upward by $\delta^{E}(v_i) + u^E(v_i)$ states, where
\begin{align}
\label{Sect2.2-Eq3-BerE}
\delta^E(v_i) 
 = \left[\varepsilon^{E} - v_ia\right], \quad u^E(v_i) \sim \mathrm{Bernoulli}(q^E_{v_i}), \quad \textrm{and} \quad q^E_{v_i} &= \left\{\varepsilon^{E} - v_ia\right\};
\end{align}
\item When an I-kick takes effect, $v_i$ jumps downward by $\delta^I(v_i) + u^I(v_i)$ states, where
\begin{align}
\label{Sect2.2-Eq4-BerI}
\delta^I(v_i) = \left[v_ia - \varepsilon^{I}\right], \quad u^I(v_i) \sim \mathrm{Bernoulli}(q^I_{v_i}), \quad \textrm{and} \quad q^I_{v_i} &= \left\{v_ia - \varepsilon^{I} \right\}.
\end{align}
\end{itemize}
Here, $[\cdot]$ and $\{\cdot\}$ indicate the floor and sawtooth functions. $v_i$ should be bounded within the state space $\Gamma$. Should $v_i$ exceeds $M$ after an ext- or E-kick takes effect, it goes directly to $\mathcal{R}$ and neuron $i$ fires immediately. On the other hand, when $v_i$ approaches $-m^I$, $\delta^I(v_i)=0$, and $v_i$ never drops below $-m^I$. 

We stress that as $M \to \infty$ and given all external and internal spike trains, $v_i(t)\cdot\frac{V^{th}}{m^{th}}  \to V_i(t)$ \textit{almost surely} before neuron $i$ spikes.

\heading{Refractory Periods.} For a LIF neuron, the time that it already spent in the refractory period is a memory term. In a memoryless Markov model, we propose two solutions:
\begin{enumerate}
\item Assign an exponentially distributed waiting time with mean $\tau^{\rm{ref}}$ before $v_i$ exits $\mathcal{R}$.  
\item Assign $N_\mathcal{R}$ clocks with iid exponentially distributed waiting time with mean $\tau^{\rm{ref}}$. The neuron exits the refractory period when more than $N_\mathcal{R}(1-\frac{1}{e})$ waiting times are completed. 
\end{enumerate}
For solution 1, while the expectation of refractory period stays the same as the LIF neurons, the traces of $v_i$ and $V_i$ would be different after reset. Nevertheless, our numerical experiments suggest that the distribution of $v_i$ still yields a reasonable approximation to $V_i$ if all input spike trains are temporally stationary. Solution 2 is numerically more costly, but it guarantees that the total time spent in the refractory period converges to $\tau^{\rm{ref}}$ when $N_\mathcal{R}\to\infty$.

In summary, the state of neuron $i$ is described by the quartet $(v_i, H_i^E, H_i^I, H_i^{\mathrm{ext}})$. Given the spike train series received by neuron $i$, the dynamics of $v_i$ are Markovian. When the external spike trains are modeled as Poisson processes and internal spikes are generated by Markov processes, the entire network model is inherently Markovian.

\subsubsection{Setup of homogeneous populations}
Within each homogeneous population, synaptic coupling weights and external input rates are set as constants that only depend on the category of the source and target, i.e., in Eq.~\ref{Sect2.1-Eqn1-LIF},
\begin{itemize}
    \item $S^{\rm{ext}}_i = S^{Q\rm{ext}}$ if neuron $i$ is type-Q, and the number of pending kicks added by each external spike is denoted as $h^{Q\rm{ext}} = \left[\frac{S^{Q\rm{ext}}}{a}\right]$;
    \item $S^{R}_{ij} = S^{QR}$ if neuron $i$ is type-Q, and the corresponding number of pending kicks added to $H^{QR}$ is denoted as $h^{QR} = \left[\frac{S^{QR}}{a}\right]$
    \item The rate of Possion external input to neuron $i$ is $\lambda^Q$. 
\end{itemize}
Here, $Q,R\in\{E,I\}$.

To minimize the heterogeneity within each population, we introduce the setup of an annealed architecture by assigning the set of postsynaptic neurons of each spike \textit{on-the-fly}. Whenever a type-$R$ neuron spikes, the targeted postsynaptic neurons in the $Q$ populations 
are chosen independently with probabilities $p^{QR}$ ($Q,R\in\{E,I\}$). 
The motivation of this simplification is for analytical and computational convenience and is standard in many previous theoretical studies \cite{cai2006kinetic,brunel1999fast,wilson1972excitatory,buice2007field,cai2021model,gerstner2014neuronal}. Furthermore, previous numerical experiments \cite{wu2022multi} suggest that the dynamics produced by LIF networks with random quenched architecture is similar to those with homogeneous architecture when they have the same projection probabilities. Nevertheless, the Markovian approximation itself does not require the annealed architecture setup to work.

\vspace{0.15in}
To summarize, the state space of the Markov model is denoted as $\mathbf{\Omega}$. A network state $\omega \in \mathbf{\Omega}$ consists of $4N$ components ($N = N^E+N^I$)
\begin{align}
\label{Sect2.2-Eq5-networkState}
\omega=(v_1,\cdots,v_N,\quad H^E_1,\cdots,H^E_N,\quad H^I_1,\cdots,H^I_{N},\quad H^{\rm{ext}}_1,\cdots,H^{\rm{ext}}_{N}). 
\end{align}
In practice, the timescale of external input $\tau^{\rm{ext}}$ is sometimes set as 0.
This leads to the removal of the last $N$ components in Eq.~\ref{Sect2.2-Eq5-networkState} since $H^{\rm{ext}}_i$ will never accumulate.

\subsection{A Coarse-Grained Markov Model}
\label{Sect4.2-DimReduction}

The information gap between the CG-Markov system and the original LIF network is minimized by \textit{only} assuming the fast decorrelation of $H$ values and discretizing the dynamics of LIF neurons. In this subsection, we first state the assumption, then derive the CG-Markov model. 

\heading{Assumption 1:} \(H^R_i\) quickly decorrelates from its own history.

\subsubsection{Rationale of Assumption 1}
Assumption 1 is a natural consequence of the annealed architecture setup. For instance, consider two type-Q neurons \((i,j)\). Pending kicks are independently added to \(H^E_i(t)\) and \(H^E_j(t)\) for any E-spike produced by the rest of the network after time \(t\). This independence quickly erases their dependence on history as long as the network keeps producing firing events. This can be seen from 
$$
\mathrm{Var}[H^R_i(t')|H^R_i(t)] - \mathrm{Var}[H^R_i(t)] = \left(\frac{S^{QR}}{\tau^{R}}\right)^2p^{QR}(1-p^{QR})N^R\int_t^{t'}e^{-\frac{2(t'-s)}{\tau^R}}f^R(s)\,\mathrm{d}s,
$$
where \(f^R\) is the expected firing rate of each E-neuron. That is, the time that $\mathrm{Var}[H^R_i(t')|H^R_i(t)]$ reaches $k$ times of the original variance $\mathrm{Var}[H^R_i(t)]$ is \[ \tau^R\log\left(\frac{A}{A-k\mathrm{Var}[H^R_i(t)]}\right), \text{ where } A = \frac{(S^{QR})^2}{\tau^{R}}p^{QR}(1-p^{QR})N^Rf^R.\]

Under Assumption 1, \(H^E_i(t)\) and \(H^E_j(t)\) can be considered identically and independently distributed (IID) random variables. Similarly, \(H^E_i(t)\) is also independent of \(H^I_i(t)\) and \(v_i(t)\).
Therefore, any two neurons \((i,j)\) of the same type are indistinguishable if \(v_i = v_j\). In our simulations of the Markovian network dynamics, the correlation coefficient between $H^Q_i(t)$ and $v_i(t)$ concentrates around $10^{-2}$ and never goes beyond $0.2$.

We comment on the validity and necessity of Assumption 1. In this study, considering only AMPA and GABA-A synapses, \(\tau^{E,I}\) is constrained to a few milliseconds \cite{destexhe1998kinetic}. However, Assumption 1 may not hold if we consider synapses with much longer timescales (e.g., \(O(100\) ms) for NMDA synapses or \(O(1\) s)  for GABA-B synapses) and a low firing-rate regime (e.g., background activities \cite{pare1998impact}). Moreover, Assumption 1 is invalid if the population has a heterogeneous, quenched architecture that strongly favors a subset of type-Q neurons.

The goal of Assumption 1 is to coarse-grain the Markov model by considering only the distribution of neuron numbers in different states of potentials \(V\). This assumption is \textit{removable} if we are willing to consider a more complex state space that includes both \(V\) and \(H\). However, this would lead to a significantly larger number of random variables in the CG-Markov model and, consequently, more variables in the resulting ODE system.

\subsubsection{Deriving the Coarse-Grained Markov Model}
\label{Sect2.3.2-CGMarkov}
Assumption 1 ensures the interchangeability between neurons sharing the same state ($v$). Therefore, the coarse-grained Markov model only needs to account for the \textit{numbers} of neurons in each state rather than tracking the specific state of each neuron. Specifically, the voltage configuration of the entire network can be represented by two empirical distributions \(p^{E}(v)\) and \(p^{I}(v)\), where
\begin{align*}
    p^{E}(v) & = (n^E_{-m^I}, \cdots, n^E_{0}, ..., n^E_{m^{th}}, n^E_{\mathcal{R}}), \\
    p^{I}(v) & = (n^I_{-m^I}, \cdots, n^I_{0}, ..., n^I_{m^{th}}, n^I_{\mathcal{R}}).
\end{align*}
Here, \(n^Q_{m}\) is the number of type-Q neurons in state \(m\) (\(m \in \Gamma\)), and \(\sum_{m \in \Gamma} n^Q_{m} = N^Q\).

For the E/I-pending spike pools, it is sufficient to use four random variables \(H^{EE}, H^{EI}, H^{IE}, H^{II}\) to represent the pool sizes for each neuron, as \(H^R_i(t)\) and \(H^R_j(t)\) are IID for neurons of the same type (\(i,j\in Q \) and \(Q, R \in \{E, I\}\)).  According to the averaging for Markov chains \cite{pavliotis2008averaging}, only the expectations $\mu^{QR} = \mathbf{E}[H^{QR}]$ matters for CG dynamics if $H^{QR}$ evolves sufficiently fast and gets ergodic among all possible values before any significant change of $p^E(v)$ and $p^I(v)$. This is the case in the \textit{mean-driven regime} when the $D^{QR} = \mathbf{Var}[H^{QR}]$ is low. In the \textit{fluctuation driven regime,} to represent the noisy effect of recurrent E input on neurons, we propose the following state transitions including both upward and downward jumps:
\begin{align}
\label{Sect2.2-Eq6-StateTransE}
    \begin{cases}
        m\to m + \delta^{QE}_m &\text{at a rate of } \frac{1}{\tau^E}p^E_m\left(\mu^{QE}(t) + \frac{1}{2\tau^E}D^{QE}(t)\right) \\
        m\to m + \delta^{QE}_m+1 &\text{at a rate of } \frac{1}{\tau^E}q^E_m\left(\mu^{QE}(t) + \frac{1}{2\tau^E}D^{QE}(t)\right) \\
        m\to m^E_*+1 &\text{at a rate of }  \frac{1}{2(\tau^E)^2}p^E_mD^{QE}(t) \\
        m\to m^E_* &\text{at a rate of } \frac{1}{2(\tau^E)^2}q^E_mD^{QE}(t)
    \end{cases}
\end{align}
Here, $m^E_*$ and $m^E_*+1$ are the two states in which the neuron could jump to $m$ due to one E-kick (i.e., $m^E_* + \delta^E(m^E_*) + 1 = m$). The destination states are replaced by $\mathcal{R}$ when going beyond $m^{th}$. Also, $p^E_m = 1- q^E_m$. We argue that the downward jumps are to mimic the homogenization of Eq.~\ref{Sect2.1-Eqn1-LIF}, 

Therefore, the state of the CG-Markov model is
\begin{align}
\nonumber
    \omega =& \left(n^E_{-m^I}, \cdots, n^E_{0}, ..., n^E_{m^{th}}, n^E_{\mathcal{R}}, \quad n^I_{-m^I}, \cdots, n^I_{0}, ..., n^I_{m^{th}}, n^I_{\mathcal{R}}, \right.\\
\label{Sect2.3-Eqn7-CGfunction}
    & \left.\mu^{EE}, \mu^{EI}, \mu^{IE}, \mu^{II}, \quad D^{EE}, D^{EI}, D^{IE}, D^{II}\right).
\end{align}
The rates of all state transitions are parametrized by $\mu$'s, $D$'s, and parameters for leakage and external stimulus (taking the first solution for refractory periods): 
\begin{itemize}
    \item Leak: For \(m > 0\), \(n^Q_{m}\) decreases by one at a rate \(n^Q_{m} \cdot g^{Q\text{leak}} m\), accompanied by an increase of \(n^Q_{m-1}\) by one. The direction of the jump is reversed for \(m < 0\).
    \item External stimuli: \(n^Q_{m}\) decreases by one at a rate \(n^Q_{m} \cdot \lambda^Q\). If \(m + h^{Q\text{ext}} < m^{th}\), this is accompanied by an increase of \(n^Q_{m + h^{Q\text{ext}}}\) by one (since \(\tau^{\text{ext}} = 0\) and all external pending kicks take effect at once), otherwise, it results in a type-Q spike and an increase of \(n^Q_{\mathcal{R}}\) by one.
    \item Recurrent excitation: \(n^Q_{m}\) decreases by one at a rate \(\frac{1}{\tau^{E}}n^Q_m(\mu^{QE} + \frac{D^{QE}}{\tau^{E}})\) and the neuron is added to one of the four destination states in Eq.~\ref{Sect2.2-Eq6-StateTransE}. This increment is accompanied by a type-Q spike if the destination state exceeds \(m^{th}\).
    \item Recurrent inhibition: \(n^Q_{m}\) decreases by one at a rate \(\frac{1}{\tau^{I}}n^Q_m(\mu^{QI} + \frac{D^{QI}}{\tau^{I}})\) and the neuron is added to one of the four destination states: $m-\delta^I(m)$, $m-\delta^I(m)-1$, $m^I_*$, and $m^I_*-1$. 
    \item Reset: \(n^Q_{\mathcal{R}}\) decreases by one at a rate \(\frac{n^Q_{\mathcal{R}}}{\tau^{\text{ref}}}\), which is accompanied by an increase of \(n^Q_0\) by one.
    \item Increment of pending kicks: When a type-R spike occurs, \(H^{QR}\) increases by \(h^{QR}\) with probability \(p^{QR}\).
\end{itemize}
In practice, $\mu^{QR}$ and $D^{QR}$ can be collected by evolving $N^Q$ IID samples of $H^{R}_i$ following the rules described in Sect.~\ref{Sect2.2-MarkovIF}. Namely, a random sample would drop by one when a type-$R$ pending kick takes effect on a type-$Q$ cell, and a random selection of $p^{QR}N^{Q}$ samples increase by one when a type-$R$ spike takes place. 

We close this introduction of the CG-Markov model by pointing out that a spike can be induced by the fluctuations of I-kicks. Though counterintuitive, it is a well-known fact for fluctuation-driven dynamics.

\subsection{Parameters}
The choices of parameters are adopted from a large-scale spiking network model for Macaque primary visual cortex \cite{chariker2016orientation}, which are also used by previous studies \cite{li2019stochastic,cai2021model,wu2022multi}. For all SNN parameters used in the simulations, we list their definitions and values in Table \ref{Table1_Parameters}. Here, we remark that the range of projection probabilities $p^{QR}$ ($Q,R\in\{E,I\}$) are chosen to match the anatomical data in the macaque visual cortex, see \cite{chariker2016orientation} for reference. Also, $\tau^E<\tau^I$, since it is known that the Glu-AMPA receptors act faster than the GABA-GABA receptors, with both on a time scale of milliseconds \cite{koch1999biophysics}.


\begin{table*}[htbp]
\begin{center}
    \begin{tabular}{|l|c|l|l|l|}
      \hline
      Parameter Group  & Parameter        & Meaning                       & Standard Value & Tested Range\\ \hline
      Network          & $N^{I}$          & number of I cells             & 100  & 10-2000\\
      architecture     & $N^{E}$          & number of E cells             & 300  & $3N^I$ \\
                       & $P^{EE}$         & E-to-E coupling probability   & 0.8  & 0.2-0.8 \\ 
                       & $P^{EI}$         & I-to-E coupling probability   & 0.8  & 0.2-0.8 \\ 
                       & $P^{IE}$         & E-to-I coupling probability   & 0.8  & 0.2-0.8 \\ 
                       & $P^{II}$         & I-to-I coupling probability   & 0.8  & 0.2-0.8 \\
                       & $S^{EE}$         & E-to-E synaptic weight        & 0.95 & - \\ 
                       & $S^{EI}$         & I-to-E synaptic weight        & 2.71 & 2.3-3.1\\
                       & $S^{IE}$         & E-to-I synaptic weight        & 1.25 & - \\
                       & $S^{II}$         & I-to-I synaptic weight        & 2.45 & 1.1-2.6 \\\hline
      Neuronal         & $\tau^{\rm{ref}}$ & refractory period           & 4 ms & 0-10 ms\\
      physiology       & $\tau^E$         & E-synapse timescale           & 2 ms & 1-4 ms\\
                       & $\tau^I$         & E-synapse timescale           & 4.5 ms & -\\\hline
      External         & $S^{E\rm{ext}}$  & external-to-E synaptic weight & 1  & 1-4 \\ 
      input            & $S^{I\rm{ext}}$  & external-to-I synaptic weight & 1  & $S^{E\rm{ext}}$ \\
                       & $\lambda^E$      & external-to-E input rate      & 7 kHz & 3-10 kHz  \\
                       & $\lambda^I$      & external-to-I input rate      & 7 kHz & $\lambda^E$\\ \hline
    \end{tabular}
  
 \caption{Parameters regarding the network architecture (first row), individual neuronal physiology (second row), and external inputs (third row). Symbols, meanings, and values of relevant parameters are depicted.}
 \label{Table1_Parameters}
\end{center}
\end{table*}

\section{Deriving discrete-state ODEs from coarse-grained Markov models}
\label{Sect3-SDE}

In this section, we first derive a set of stochastic differential equations (SDE) for all variables in Eq.~\ref{Sect2.3-Eqn7-CGfunction} with Gaussian approximations of the recurrent pending kick pools ($H$s) and the flux of neurons between different neuron states. The Gaussian approximation requires large numbers of neurons in each state. To account for the finite-N effect, we then reduce this system to a lower-dimensional ODE system by combining consecutive voltage bins. This reduced ODE system is thus termed “discrete-state ordinary differential equations” (dsODEs).

\subsection{Discrete-state stochastic differential equations}
\label{Sect3.1-dsSDE-largeM}
We first need to ``continuize" the discrete $n$s and $H$s to obtain a differential system for temporal dynamics. Here, the SDE for neuron numbers can be regarded as the master equations for $n^Q_m$ when $N^{E,I}\to\infty$. Two ODEs trace the expectations and variances of each $H$, which in turn drive the SDEs for neuron numbers in different states. 

\subsubsection{Ordinary differential equations for pending spike pools}
We first derive differential equations of $H^{QE}$. In the CG-Markov model, $H^{QR}$ is a birth-death process subjected to the type-R spike series produced by the system. Therefore, in a small timestep $\delta t$:
\begin{align}
\label{Sect3.1-Eq8-HDiscrete}
    H^{QR}\left(t+\delta t\right) - H^{QR}(t) 
    = -\underbrace{\mathrm{Poisson}\left(\frac{H^{QR} \cdot \delta t}{\tau^E}\right)}_{\rm{\#\, of\, kicks\, consumed}} + \underbrace{h^{QR}\cdot\mathrm{Binomial}\left(F^R_{t,t+\delta t},p^{QR}\right)}_{\rm{\#\, of\, kicks\, added}},
\end{align}
where $F^R_{t,t+\delta t}$ is the number of type-R spikes produced by the system in $[t, t+\delta t)$. When $\delta t$ is small enough, $F^R_{t,t+\delta t} = N^Rf^R(t)\delta t$. Here $f^R(t)$ is the average firing rate of single type-R neurons.
We apply the Gaussian approximation to the two terms on the right-hand side of Eq.~\ref{Sect3.1-Eq8-HDiscrete}:
\begin{align*}
    &H^{QR}\left(t+\delta t\right) - H^{QR}(t) \\
    &= \left[-\frac{H^{QR}}{\tau^R}\delta t + \sqrt{\frac{H^{QR}}{\tau^R}}\delta W_1\right] + \left[h^{QR}N^Rf^R\cdot p^{QR}\delta t + h^{QR}\sqrt{N^Rf^Rp^{QR}(1-p^{QR})}\delta W_2\right],
\end{align*}
hence
\begin{align}
\label{Sect3.1-Eq9-HSDE}
    \ddt{H^{QR}}= \left(-\frac{H^{QR}}{\tau^R} + h^{QR}p^{QR}N^Rf^R\right)   
    + \left[ \sqrt{\frac{H^{QR}}{\tau^R}}\ddt{W_1}+ h^{QR}\sqrt{ N^Rf^R\cdot p^{QR}(1-p^{QR})}\ddt{W_2}\right], 
\end{align}
where $W_{1,2}$ are independent Wiener processes. Finally, $H^{QR}$ is a Gaussian random variable whose mean $\mu^{QR}$ and variance $D^{QR}$ respect the following two ODEs:
\begin{subequations}
    \label{Sect3.1-Eq10-H}
\begin{eqnarray}
\label{Sect3.1-Eq10EHode}
\ddt{\mu^{QR}} &=&  -\frac{\mu^{QR}}{\tau^R} + h^{QR}p^{QR}N^Rf^R \\
\label{Sect3.1-Eq10-VarHode}
\ddt{D^{QR}} &=&  \frac{1}{\tau^R}\left(-2D^{QR} + \mu^{QR}\right) + (h^{QR})^2N^Rf^R p^{QR}(1-p^{QR}).
\end{eqnarray}
\end{subequations}
We comment that, when $f^R(t)$ is fixed on $t\in[0,T]$, there exists
$$
\bm{u}^{QR}(t) = \lim_{a\to0} a\mu^{QR}(t)\quad\text{and }\bm{D}^{QR}(t) = \lim_{a\to0} a^2D^{QR}(t),
$$
since $h^{QR} = \left[\frac{S^{QR}}{a}\right]$.

\subsubsection{Stochastic differential equations for the number of neurons at each state}
For any state $m\in\Gamma$ and consider a small time interval $[t, t+\delta t)$ where at most one pending kick could take effect,
\begin{align}
\label{Sect3.1-Eq11-nDiscrete}
\Delta n^Q_m(t) = \Delta n^{Q\rm{leak}}_m(t) + \Delta n^{Q\rm{ext}}_m(t) + \Delta n^{QE}_m(t) + \Delta n^{QI}_m(t) + \Delta n^{Q\rm{reset}}_\mathcal{R}(t)\cdot \mathbf{1}_{0}(m),
\end{align}
the flux terms on the right-hand side are the total increment of neurons due to specific events listed in Sect.~\ref{Sect2.3.2-CGMarkov}. $\mathbf{1}_{0}(m) = 1$ only when $m = 0$, otherwise $\mathbf{1}_{0}(m) = 0$. When $m\neq \mathcal{R}$ and $m\neq0$
\begin{subequations}
\label{Sect3.1-Eq12-nflux}
\begin{eqnarray}
\label{Sect3.1-Eq12-nfluxLeak}
\Delta n^{Q\rm{leak}}_m &=& - \text{Poisson}\left(n^Q_m \cdot g^{Q\text{leak}} |m| \delta t\right) + \text{Poisson}\left(n^Q_{m+\mathrm{sgn}(m)} \cdot g^{Q\text{leak}} (|m|+1) \delta t\right), \\
\label{Sect3.1-Eq12-nfluxExt}
\Delta n^{Q\rm{ext}}_m &=& - \text{Poisson}(n^Q_m \cdot \lambda^Q \delta t) + \text{Poisson}(n^Q_{m - h^{Q\text{ext}}} \cdot \lambda^Q \delta t), \\
\nonumber
\Delta n^{QE}_m &=& - \text{Poisson}\left(\left(\mu^{QE} + \frac{D^{QE}}{2\tau^E}\right)\frac{n^Q_m\delta t}{\tau^E}  \right) \\
\label{Sect3.1-Eq12-nfluxRecE}
                   &&+ \delta a^{QE}(m^E_*) + \delta b^{QE}(m^E_*+1) + \delta c^{QE}(m+\delta^E(m)+1) + \delta d^{QE}(m+\delta^E(m)), \\
\nonumber
\Delta n^{QI}_m &=& - \text{Poisson}\left(\frac{n^Q_m(\mu^{QI}+D^{QI})}{\tau^I} \delta t \right) \\
\label{Sect3.1-Eq12-nfluxRecI}
                   &&+ \delta a^{QI}(m^I_*) + \delta b^{QI}(m^I_*-1) + \delta c^{QI}(m-\delta^I(m)-1) + \delta d^{QI}(m-\delta^I(m)).
\end{eqnarray}
\end{subequations}
Here, $\mathrm{sgn}(m) = \pm1$ and indicates the sign of $m$;  $(\delta a^{QE},\delta b^{QE})$ are the numbers of successful/unsuccessful Bernoulli tests in upward jumps of $v$ due to E-kicks; $(\delta c^{QE},\delta d^{QE})$ are the numbers of successful/unsuccessful Bernoulli tests in downward jumps of $v$ due to an E-kick; hence 
$$
\delta a^{QE}(m) + \delta b^{QE}(m) = \text{Poisson}\left(\left(\mu^{QE} + \frac{1}{2\tau^E}D^{QE}\right)\frac{n^Q_m\delta t}{\tau^E} \right); \quad \delta c^{QE}(m) + \delta d^{QE}(m) = \text{Poisson}\left(\frac{n^Q_mD^{QE}}{2(\tau^E)^2} \delta t \right);
$$
Similar definitions for $m^I_*$,  $(\delta a^{QI},\delta b^{QI})$ and $(\delta c^{QI},\delta d^{QI})$.  Specifically, for $m = 0$, the reset flux $\Delta n^{Q\rm{reset}}_m = \text{Poisson}\left(\frac{n^Q_{\mathcal{R}}}{\tau^{\text{ref}}} \delta t \right)$ is non-zero, and the inward leak flux come from both $m = \pm1$. We also note that, throughout Sect.~\ref{Sect3-SDE}, $n^Q_m = 0$ if $m\notin\Gamma$ to avoid flux from above $m^{th}$ or below $-m^I$.

As for $m = \mathcal{R}$, the leak flux is uninvolved, and recurrent fluxes are not taking neurons out of $\mathcal{R}$, hence
\begin{subequations}
    \label{Sect3.1-Eq13-nRflux}
\begin{eqnarray}    
\label{Sect3.1-Eq13-nRfluxExt}
\Delta n^{Q\text{ext}}_\mathcal{R} &=&  \sum_{k = m^{th}-h^{Q\rm{ext}}}^{m^{th}-1}\text{Poisson}(n^Q_{k+1} \cdot \lambda^Q \delta t)\,, \\
\label{Sect3.1-Eq13-nRfluxRecE}
\Delta n^{Q\rm{rec}}_\mathcal{R} &=&  \sum_{\ell = m^{th,E}_*}^{m^{th}}
\delta a^{QE}(\ell) + \sum_{\ell = m^{th,E}_*+1}^{m^{th}}
 \delta b^{QE}(\ell) + \sum_{\ell = m^{th,I}_*}^{m^{th}}
 \delta c^{QI}(\ell) + \sum_{\ell = m^{th,I}_*+1}^{m^{th}}
 \delta d^{QI}(\ell)\, , \\
\label{Sect3.1-Eq13-nRfluxReset}
\Delta n^{Q\text{reset}}_\mathcal{R} &=&  -\text{Poisson}\left(\frac{n^Q_{\mathcal{R}}}{\tau^{\text{ref}}} \delta t \right) \, ,
\end{eqnarray}
\end{subequations}
In Eq.~\ref{Sect3.1-Eq13-nRfluxRecE}, $m^{th,E}_*$ is the state farthest away from $m^{th}$ in which a neuron would fire with one E-kick taking effect.

We now apply Gaussian approximations to each of the flux terms. For $X\sim\mathrm{Poisson}(x\delta t)$, if $x$ is a constant, it is straightforward that $X\approx x\delta t + x^\frac12\delta W$, where $\delta W\sim\mathcal{N}(0,\delta t)$.
As for random variables in Eqs.~\ref{Sect3.1-Eq12-nfluxRecE}, \ref{Sect3.1-Eq12-nfluxRecI}, and \ref{Sect3.1-Eq13-nRfluxRecE}, they are Binomial random variables parametrized by Poisson random numbers, i.e., terms $\delta a$-$\delta d$ are Poisson as well. Therefore, 
\begin{align*}
    \delta a^{QE}(m) &\approx \upmu^{QE}_{a,m}\delta t +  \left(\upmu^{QE}_{a,m} \right)^\frac12\delta W^{QE}_{m\to m+\delta^E(m)+1}, \quad  \upmu^{QE}_{a,m} = \frac{q^E_mn^Q_m}{\tau^E} \left(\mu^{QE} + \frac{1}{2\tau^E}D^{QE}\right)\,, \\
    \delta b^{QE}(m) &\approx \upmu^{QE}_{b,m} \delta t + \left(\upmu^{QE}_{b,m} \right)^\frac12\delta W^{QE}_{m\to m+\delta^E(m)},\quad   
    \upmu^{QE}_{b,m} = \frac{p^E_mn^Q_m}{\tau^E} \left(\mu^{QE} + \frac{1}{2\tau^E}D^{QE}\right) \, .
\end{align*}
Here, 
the subscript of $\delta W$ indicates the source state and destination state of the flux, whereas the superscript indicates the cause of the flux. The fluxes $\delta c^{QE}(m)$ and $\delta d^{QE}(m)$ would be similar to the above. On the other hand, the $QI$-fluxes are completely analogical to the $QE$-fluxes. 

Finally, $M+1$ SDEs for $n^Q_m$ ($m\in\Gamma$) are derived by summing up all flux and taking $\delta t\to 0$. For $m>0$, 
\begin{subequations}
\label{Sect3.1-Eq14-nODEs}
\begin{eqnarray}   
    \nonumber
    \mathrm{d}n^Q_m &=& \left(-\mathrm{d}n^{Q\rm{leak}}_m  + \mathrm{d}n^{Q\rm{leak}}_{m+1} \right)  \,
    +\,  \left(-\mathrm{d}n^{Q\rm{ext}}_m + \mathrm{d}n^{Q\rm{ext}}_{m - h^{Q\text{ext}}} \right)   \\
    \nonumber
    && +\left[-\mathrm{d}a^{QE}(m) - \mathrm{d}b^{QE}(m) -\mathrm{d}c^{QE}(m) - \mathrm{d}d^{QE}(m)  \right. \\
    \nonumber
    && \left. + \mathrm{d}a^{QE}(m^E_*)+  \mathrm{d}b^{QE}(m^E_*+1) + \mathrm{d}c^{QE}(m+\delta^E(m)+1) + \mathrm{d}d^{QE}(m+\delta^E(m))\right] \\
    \nonumber
    && +\left[-\mathrm{d}a^{QI}(m) - \mathrm{d}b^{QI}(m) -\mathrm{d}c^{QI}(m) - \mathrm{d}d^{QI}(m)  \right. \\
    \label{Sect3.1-Eq14-nODEs-m}
    && \left. + \mathrm{d}a^{QI}(m^I_*)+  \mathrm{d}b^{QI}(m^I_*+1) + \mathrm{d}c^{QI}(m-\delta^I(m)-1) + \mathrm{d}d^{QI}(m-\delta^I(m))\right] \\
    \label{Sect3.1-Eq14-nODEs-R}
    \mathrm{d}n^Q_\mathcal{R} &= & -\mathrm{d}n^{Q\rm{reset}}_\mathcal{R} \,+\,N^Qf^Q(t)\mathrm{d}t, \\
    \nonumber
    N^Qf^Q(t)\mathrm{d}t  &= & \sum_{k = m^{th}-h^{Q\rm{ext}}}^{m^{th}-1}\mathrm{d}n^{Q\rm{ext}}_{k+1}  \\
    \label{Sect3.1-Eq14-nODEs-rate}
    && + \sum_{\ell = m^{th,E}_*}^{m^{th}}
\mathrm{d} a^{QE}(\ell) + \sum_{\ell = m^{th,E}_*+1}^{m^{th}}
 \mathrm{d} b^{QE}(\ell) + \sum_{\ell = m^{th,I}_*}^{m^{th}}
 \mathrm{d} c^{QI}(\ell) + \sum_{\ell = m^{th,I}_*+1}^{m^{th}}
 \mathrm{d} d^{QI}(\ell)\, , 
      \end{eqnarray}
\end{subequations}
Here, $\mathrm{d}n^{Q\textsc{cause}}_m$ indicates the Gaussian approximation of the corresponding Poisson flux \textit{leaving} state $m$:
\begin{align*}
    \mathrm{d}n^{Q\rm{leak}}_m &= \upmu^{Q\rm{leak}}_m \mathrm{d}t +\left(\upmu^{Q\rm{leak}}_m\right)^\frac12\mathrm{d}W^{Q\rm{leak}}_{m\to m-1}, \quad \upmu^{Q\rm{leak}}_m = g^{Q\text{leak}} mn^Q_m, \\
    \mathrm{d}n^{Q\rm{ext}}_m &= \upmu^{Q\rm{ext}}_m\mathrm{d}t + \left(\upmu^{Q\rm{ext}}_m\right)^\frac12\mathrm{d}W^{Q\rm{ext}}_{m\to m+h^{Q\text{ext}}}, \quad \upmu^{Q\rm{ext}}_m = \lambda^Q n^Q_m, \\
    \mathrm{d}n^{Q\rm{reset}}_\mathcal{R} &= \upmu^{Q\rm{reset}}_\mathcal{R} \mathrm{d}t  + \left(\upmu^{Q\rm{reset}}_\mathcal{R}\right)^\frac12 \mathrm{d}W^{Q\rm{reset}}_{\mathcal{R}\to0}, \quad \upmu^{Q\rm{reset}}_\mathcal{R} = \frac{n^Q_{\mathcal{R}}}{\tau^{\text{ref}}}.
\end{align*}
We also note that all white noises are labeled by their cause in the superscripts, the source and destination of the corresponding neuron flux in the subscript.   For $m = 0$, the reset flux $\mathrm{d}n^{Q\rm{reset}}_\mathcal{R} $ should be added to the right-hand side of Eq.~\ref{Sect3.1-Eq14-nODEs-m}, as well as the leak flux from both $m = \pm1$. The inward leaky flux would come from $m-1$ instead of $m+1$ for $m<0$.  The network firing rate $N^Qf^Q(t)$ accounts for all events that neurons jump above $m^{th}$.

\subsection{Dimension reduction with large voltage bin size}
\label{Sect3.2-dsSDE-Reduced}
The Markovian approximation to spiking network dynamics (and consequently, the dsSDE) depends on the choice of $M$. Usually, the approximation is better for larger $M$, but this presents several issues when applied to small neural circuits in the real brain. Most directly, it is unfeasible for analysis and simulations to rewrite an SNN with $O(10^{2-3})$ neurons into an SDE system with even higher dimensions. Additionally, the Gaussian approximation to the Poisson random numbers in neuron flux usually requires $n^Q_m > 10$. However, when $M \to \infty$, $n^Q_m$ would be 0 and 1 for most $m$, causing Eq.~\ref{Sect3.1-Eq14-nODEs} to deviate from the CG-Markov system significantly. In this subsection, we use well-established techniques from heterogeneous multiscale modeling (HMM) to reduce the SDE to a lower-dimensional deterministic ODE system.

\begin{figure}[htbp]
  \begin{center}
    \includegraphics*[bb=0in 3in 11in 15.2in,width=0.75\textwidth]{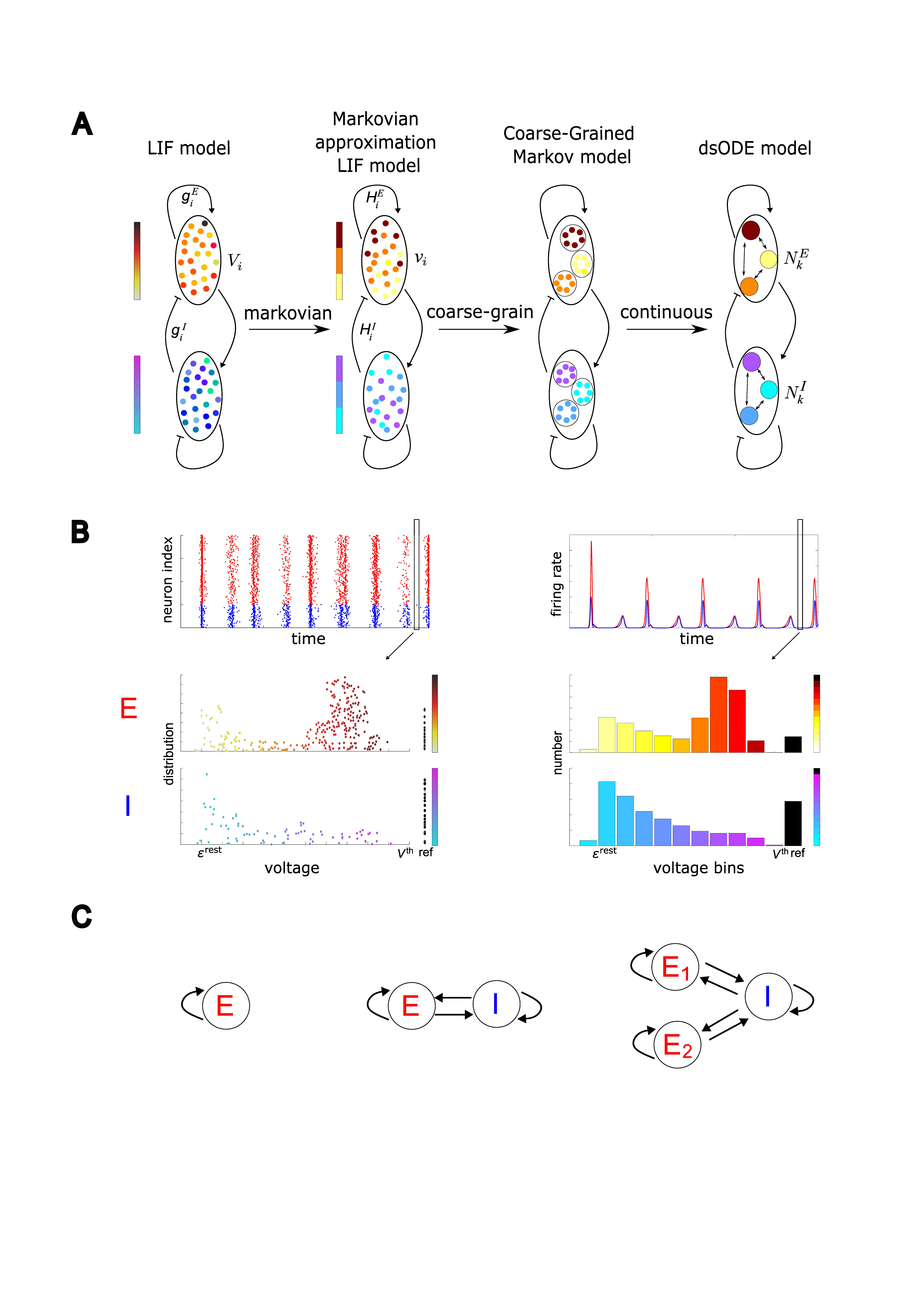}
    \caption{Schematic diagrams for model reductions and spiking network studied in this paper. \textbf{A.} A chain of model reductions. From left to right: leaky-integrate-fire (LIF) network with E \& I populations, where neuron states ($V$) are continuous (indicated by warm \& cold color codes); the Markovian spiking network, where neuron states ($v$) are discretized; the coarse-grained Markov network, where neurons in the same state are treated as interchangeable; and finally, the dsODE system, where the number of neurons in each state is treated as continuous. \textbf{B.} States of the LIF network and the corresponding dsODE system at a given time point. \textbf{Left:} the upper raster plot displays spiking events across all neurons within a time window. At a specific time point (marked by a vertical bar), the voltages of the E and I populations are presented as two distributions (lower plot), where the height reflects density. Black dots denote neurons in refractory states. \textbf{Right:} the dsODE system predicts firing rates that correspond to the LIF network dynamics (upper plot). At the same time point, the dsODE system tracks the number of neurons in each state rather than the detailed membrane potentials of individual neurons (lower plot). \textbf{C.} Three LIF network architectures on which the dsODE system is tested. From left to right: a single, self-connected E population; a network with both E and I populations; and a network with two competing E populations mediated by an I population.}
    \label{Fig1_dsODE_ideas}
  \end{center}
\end{figure}

\subsubsection{Dimension reduction of SDE  with heterogeneous multiscale methods}
While holding the two ODEs for $H$ (Eqs.~\ref{Sect3.1-Eq10EHode} and \ref{Sect3.1-Eq10-VarHode}), we propose to reduce Eq.~\ref{Sect3.1-Eq14-nODEs} by combining consecutive $L$ voltage bins. That is, the reduced system has $\mathbf{M} = \frac{M}{L}$ state variables and a larger voltage bin size $\ba = La$. Its state variables would be the total number of neurons in the reduced voltage state $\mbm$, i.e.,
\begin{align*}
    N^{Q}_\mbm = \sum_{m=\mbm L+1}^{(\mbm +1)L} n_m^Q, \quad\text{where }
    \mbm \in \mathbf{\Gamma} := \left\{ -\mbm^I, -\mbm^I+1, \ldots, \mathbf{-1}, \mathbf{0}, \mathbf{1}, \ldots, \mbm^{th}\right\} \cup \{\mathcal{R}\}
\end{align*}
Throughout this section and the rest of the paper, we use bold fonts to index the reduced voltage states to distinguish them from the voltage states in previous sections. 

The reduced dynamical system should account for the ``neuron flux" between the reduced voltage states. We start by focusing on all neurons in state $\mbm$ and trace the whereabouts of every neuron in a finite time step $\delta t$. In the CG-Markov system, this requires knowing the detailed voltage configuration $\{n^Q_m|[m/L] = \mbm\}$. Instead, inspired by the usual practice of HMM, we introduce an auxiliary variable $\bar{v}^{Q}_\mbm$, the average voltage of all neurons in state $\mbm$. When $a\to0$, it helps us infer the detailed voltage configuration as $\bm{n}^Q(v) = N^{Q}_\mbm P(v|\bar{v}^{Q}_\mbm)$ for $v\in I_\mbm = [\mbm\ba,(\mbm+1)\ba)$. Here, $\bm{n}^Q(v)$ represents the density of type-Q neurons at $V = v$.

There are indeed infinitely many ways to choose $P(v|\bar{v}^{Q}_\mbm)$. To accommodate the noise-induced jumps, we require $\bm{n}^Q(v)$ to be first-order differentiable on the boundaries of all $I_\mbm$.  For simplicity, we choose a piecewise linear scheme of $n^Q$ as a function of $v$. This scheme is made up by placing a ``turning point" in each  $I_\mbm$, whose specific location can be computed from $(N^{Q}_\mbm,\bar{v}^{Q}_\mbm)$ through a group of quadratic and cubic equations. The piecewise linear scheme is obtained by connecting all turning points consecutively (details in Appendix). Specifically, we set $\bm{n}^Q(v^{th}) = 0 $ to reflect the effect of noise in $H$s. This is because $D^{QR}\gg \mu^{QR}$ when $a\to0$, and neurons at $V^{th}$ would fire instantly due to noise in recurrent drives.

When we choose $\ba>S^{Q\rm{ext}}$, neuron flux can only take place between two neighboring combined states. Hence, following Eq.~\ref{Sect3.1-Eq14-nODEs}, we have for $\mbm\neq \mathcal{R}$
\begin{subequations}
\label{Sect3.2-Eq15-nflux_Redu}
\begin{eqnarray}
\label{Sect3.2-Eq15-nDiscrete_Redu}
\mathrm{d}N^Q_\mbm  &=& \mathrm{d} N^{Q\rm{leak}}_\mbm + \mathrm{d} N^{Q\rm{ext}}_\mbm + \mathrm{d} N^{QE}_\mbm + \mathrm{d} N^{QI}_\mbm + \mathrm{d} N^{Q\rm{reset}}_\mbm, \\
\label{Sect3.2-Eq15-nR_Redu}
\mathrm{d} N^{Q}_\mathcal{R} &=& \left(-\frac{N^{Q}_\mathcal{R}}{\tau^{\mathcal{R}}} + N^Qf^Q(t)\right)\cdot\mathrm{d}t, 
\end{eqnarray}
\end{subequations}
We now derive an asymptotic form of $\mathrm{d} N^{Q\textsc{cause}}_\mbm$ for $a\to0$, where $\textsc{cause}\in\{\mathrm{leak},\mathrm{ext},E,I\}$.
This can be done more conveniently by denoting the net upward flux through $v$ as $J^{Q\textsc{cause}}(v)$
, since $\mathrm{d} N^{Q\textsc{cause}}_\mbm = \left[J^{Q\textsc{cause}}(\mbm\ba) -  J^{Q\textsc{cause}}((\mbm+1)\ba)\right]\mathrm{d}t$.
Then, the reduced ODE system (dsODE) is formed by summing up the drift terms in these asymptotic forms and ignoring all the noise terms.
When taking $a\to0$, it is straightforward to obtain (details in Appendix):
\begin{subequations}
\label{Sect3.2-Eq16-J_Redu}
\begin{eqnarray}
\label{Sect3.2-Eq16-Jleak_Redu}
\mathrm{E}[J^{Q\rm{leak}}(v)] &=&-\lim_{a\to0} \int_{v}^{v+a} g^{Q\rm{leak}}\frac{x}{a}\bm{n}^Q(x)\, \mathrm{d}x = -g^{Q\rm{leak}} v\cdot \bm{n}^Q(v) \\
\label{Sect3.2-Eq16-Jext_Redu}
 \mathrm{E}[J^{Q\rm{ext}}(v)] &=& \int_{v-S^{Q\rm{ext}}}^{v} \lambda^Q \bm{n}^Q(x)\, \mathrm{d}x \\
 \label{Sect3.2-Eq16-JE_Redu}
\mathrm{E}[  J^{QE}(v)] &=& \left(\frac{\bm{u}^{QE}}{\tau^E} + \frac{\bm{D}^{QE}}{2(\tau^{E})^2}\right)\cdot(\varepsilon^{E}-v)\bm{n}^Q(v) - \frac{\bm{D}^{QE}}{2(\tau^{E})^2}\cdot(\varepsilon^{E}-v)^2\frac{\partial \bm{n}^Q}{\partial v} \\
 \label{Sect3.2-Eq16-JI_Redu}
\mathrm{E}[J^{QI}(v)] &=& -\left(\frac{\bm{u}^{QI}}{\tau^I} + \frac{\bm{D}^{QI}}{2(\tau^{I})^2}\right)\cdot(\varepsilon^{I}-v)\bm{n}^Q(v) - \frac{\bm{D}^{QI}}{2(\tau^{I})^2}\cdot(\varepsilon^{I}-v)^2\frac{\partial \bm{n}^Q}{\partial v}
\end{eqnarray}
\end{subequations}

As for the firing rate, we have 
\begin{align}
\label{Sect3.2-Eq17-fr_Redu}
         N^Qf^Q(t) = J^{Q\rm{ext}}(v^{th}) + J^{QE}(v^{th}) + J^{QI}(v^{th}) 
          = \mathrm{E}[J^{Q\rm{ext}}(v^{th})] + \mathrm{E}[  J^{QE}(v^{th})] + \mathrm{E}[J^{QI}(v^{th})] 
\end{align}

On the other hand, the change of $\bar{v}^{Q}_\mbm$ should account for the gain and loss from all neuron fluxes:
\begin{subequations}
\label{Sect3.2-Eq18-vflux-Redu}
    \begin{align}
\nonumber
&\frac{\mathrm{d}\left[N^Q_\mbm\bar{v}^Q_\mbm\right]}{\mathrm{d}t}= \\
\label{Sect3.2-Eq18-vflux-Redu-diffu}
& \mbm\ba\sum_{\rm{cause\neq ext}}J^{Q\textsc{cause}}(\mbm\ba)  -   (\mbm+1)\ba\sum_{\rm{cause\neq ext}}J^{Q\textsc{cause}}((\mbm+1)\ba) \\
\label{Sect3.2-Eq18-vflux-Redu-jump}
 & + \int_{\mbm\ba-S^{Q\rm{ext}}}^{\mbm\ba} \lambda^Q \bm{n}^Q(x)\cdot(x+S^{Q\rm{ext}})\, \mathrm{d}x - \int_{(\mbm+1)\ba-S^{Q\rm{ext}}}^{(\mbm+1)\ba} \lambda^Q \bm{n}^Q(x)\cdot x\, \mathrm{d}x \\
\label{Sect3.2-Eq18-vflux-Redu-Inside}
 & + N^Q_\mbm\left[g^{Q\rm{leak}}(\varepsilon^{\mathrm{rest}}-\bar{v}^Q_\mbm) + \frac{\bm{u}^{QE}}{\tau^E}(\varepsilon^{E}-\bar{v}^Q_\mbm)+\frac{\bm{u}^{QI}}{\tau^I}(\varepsilon^{I}-\bar{v}^Q_\mbm)\right] + \lambda^QS^{Q\rm{ext}}\int_{\mbm\ba}^{(\mbm+1)\ba-S^{Q\rm{ext}}} \bm{n}^Q(x)\, \mathrm{d}x
    \end{align}
\end{subequations}
In Eqs.~\ref{Sect3.2-Eq18-vflux-Redu-diffu} and \ref{Sect3.2-Eq18-vflux-Redu-jump}, all the flux terms $J$ are replaced by their expectations, same as Eq.~\ref{Sect3.2-Eq15-nflux_Redu}. Eq.~\ref{Sect3.2-Eq18-vflux-Redu-Inside} approximates the change of voltages for neurons staying in the same bin.

Together with the recurrent drive terms $\bm{u}^{QR}$ and $\bm{D}^{QR}$, Eqs.~\ref{Sect3.2-Eq15-nflux_Redu}, \ref{Sect3.2-Eq17-fr_Redu} and \ref{Sect3.2-Eq18-vflux-Redu} form our dsODE. It is a closed, deterministic ODE system despite the integration and derivative terms of $\bm{n}^Q(v)$ in $J(v)$. This is due to the linearity $\bm{n}^Q(v)$ around the boundaries of $I_\mbm$. In general, when $N>100$, we find that dsODEs accurately capture the LIF network dynamics by taking $\mathbf{M}= 5-20$ states (see Sect.~\ref{Sect4-Simulation}).

In addition to the dsODE system, if one wishes to recover dynamical features related to the randomness in the spiking network, we propose to account for noise in the fluxes producing firing events, while ignoring noise in other fluxes. This results in a ``dsODE+noise" scheme, and is termed the dsSDE system. Our rationale is that randomness not directly related to firing events is not contributing to the recurrent drives in the next time step, thus is less relevant to the transient dynamics and can be averaged out when the system evolves. Therefore, the the dsSDE system only modifies $N^{Q}_{\mbm^{th}}$, $N^{Q}_{\mathcal{R}}$, and $\bar{v}^Q_\mbm$ in the corresponding dsODE system. More specifically, the terms
\begin{align}
          \label{Sect3.2-Eq19-frNoise_Redu}
(\mathrm{E}[J^{Q\rm{ext}}(v^{th})])^\frac12 \frac{\mathrm{d}W^{Q\rm{ext}}}{\mathrm{d}t} + \mathrm{E}[J^{QE}(v^{th})])^\frac12 \frac{\mathrm{d}W^{QE}}{\mathrm{d}t} + (\mathrm{E}[ J^{QI}(v^{th})])^\frac12 \frac{\mathrm{d}W^{QI}}{\mathrm{d}t}
\end{align}
are added to the right-hand-side of Eq.~\ref{Sect3.2-Eq17-fr_Redu}, and the terms
\begin{align}
\label{Sect3.2-Eq19-vfluxNoise-Redu}
    -v^{th}\cdot\left(\left\{\mathrm{E}[J^{QE}(v^{th})]\right\}^\frac12 \frac{\mathrm{d}W^{QE}}{\mathrm{d}t} + \left\{\mathrm{E}[ J^{QI}(v^{th})]\right\}^\frac12 \frac{\mathrm{d}W^{QI}}{\mathrm{d}t}\right) - \left\{\int_{v^{th}-S^{Q\rm{ext}}}^{v^{th}} \lambda^Q \bm{n}^Q(x)\cdot x^2 \, \mathrm{d}x\right\}^\frac12\mathrm{d}W^{Q\rm{ext}}
\end{align}
are added to the right-hand-side of Eq.~\ref{Sect3.2-Eq18-vflux-Redu} for $\mbm = \mbm^{th}$ (note that $(\mbm^{th}+1)\ba = v^{th}$).

\subsubsection{An efficient numerical implementation of dsODEs}
The evolution of the dsODEs (Eqs.~\ref{Sect3.2-Eq15-nflux_Redu}, \ref{Sect3.2-Eq17-fr_Redu}, and \ref{Sect3.2-Eq18-vflux-Redu}) using the basic Euler scheme (or Euler-Maruyama scheme for dsSDE) requires the calculation of “turning points” at each time step.
To streamline this process and reduce computational complexity, we propose a more efficient numerical implementation of the dsODEs. The primary idea behind this approach is to approximate the detailed voltage configuration $\bm{n}^Q(v)$ as a uniform distribution centered around $\bar{v}^{Q}_\mbm$. At each time step $\delta t$, Gaussian convolutions are applied to the voltage configurations, with the changes in $N^{Q}_\mbm$ and $\bar{v}^{Q}_\mbm$ driven by the movement of neurons into and out of state $\mbm$. Meanwhile, we omit the leak terms during parameter exploration (see Sect.~\ref{Sect4-Simulation}) to further accelerate computation, as these terms have minimal impact on the dynamics of spiking networks. The full dynamics, including leaky terms, are reinstated in Sect.\ref{Sect5-Comparison}, where we compare the performance of the dsODE system with other population methods. Further technical details are provided in the Appendix.

It is important to note that this efficient numerical scheme does not converge to an ODE system as $\delta t\to 0$. The non-differentiability of $\bm{n}^Q(v)$ at state boundaries causes the recurrent flux $J^{QR}(v)$ to become infinitely large. In spite of this, when using $\delta t = 0.1$ ms, the efficient implementation produces numerical results that closely approximate those of the dsODEs.

\section{dsODEs approximate simulations of spiking networks}
\label{Sect4-Simulation}
We perform a systematic comparison between the dynamics of the LIF network and the dsODE system. This section demonstrates that, when coupled with the same parameter choices, dsODEs accurately capture the initial transient dynamics, long-term behaviors, and finite-N fluctuations of the LIF network dynamics. Additionally, dsODEs successfully predict the bifurcation points in the LIF network dynamics across all parameter investigations, even during transitions between temporal homogeneity, synchrony, and oscillations across multiple frequency bands. Furthermore, dsODEs quantitatively predict the transition waiting times when the LIF network dynamics exhibit metastability.

Throughout the remainder of the paper, we fix $\mathbf{M} = 20$ states. While this section primarily focuses on the dsODE system, we also test the dsSDE scheme in a few examples to recover critical noise-induced dynamical features of spiking networks. The dsSDE scheme is tested more systematically when we perform bifurcation analysis in Sect.~\ref{Sect4.5-Bifurcation}, and when we investigate finite-N fluctuations in Sect.~\ref{Sect4.6-FiniteN}.

\begin{figure}[htbp]
  \begin{center}
  \begin{subfigure}{.5\textwidth}
  \textbf{A}\\
  \includegraphics*[bb=2.5in 0.1in 34.5in 19.5in,width=\textwidth]{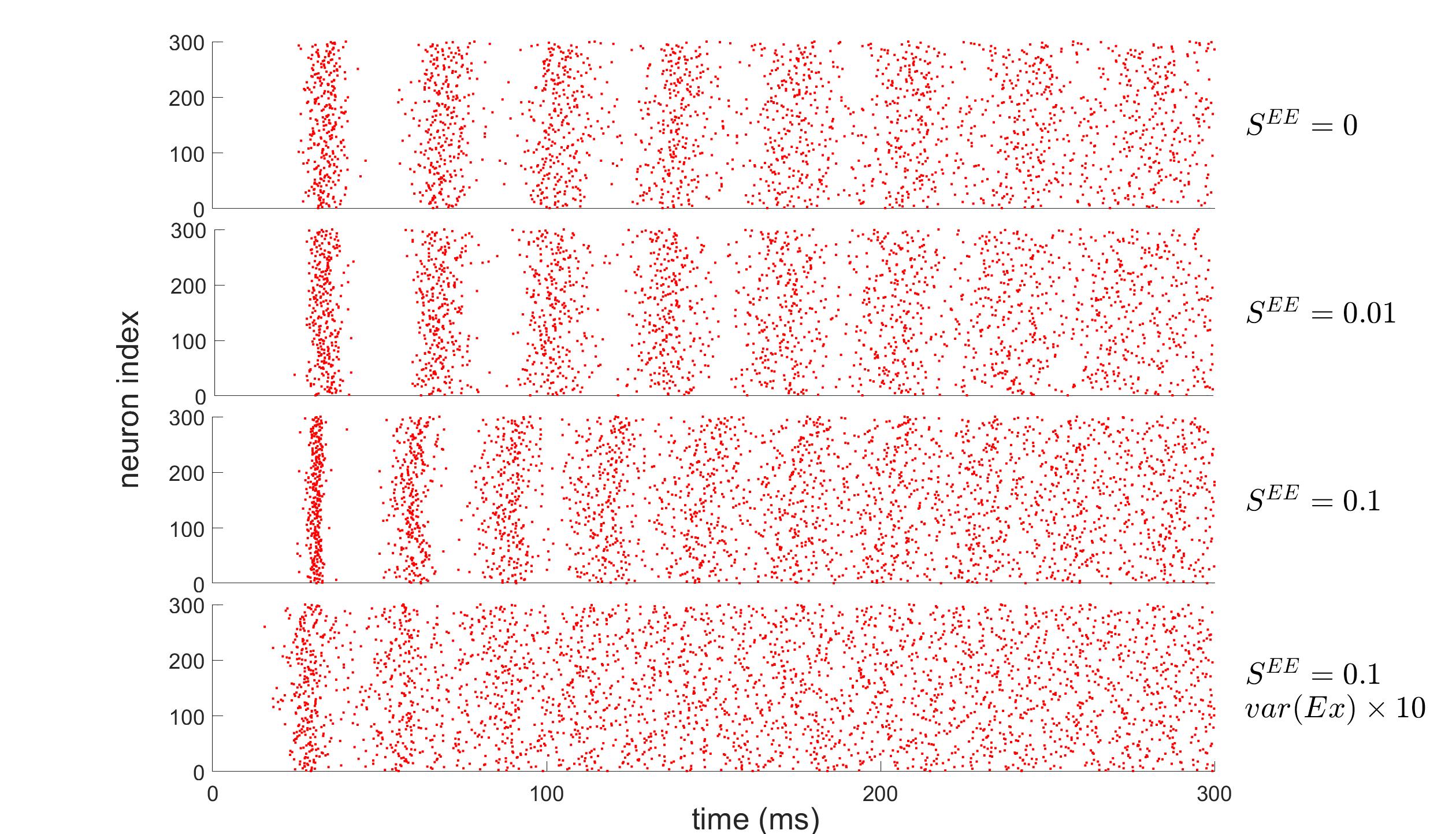}
  \end{subfigure}%
  \begin{subfigure}{.5\textwidth}
  \textbf{B}\\
  \includegraphics*[bb=2.5in 0.1in 34.5in 19.5in,width=\textwidth]{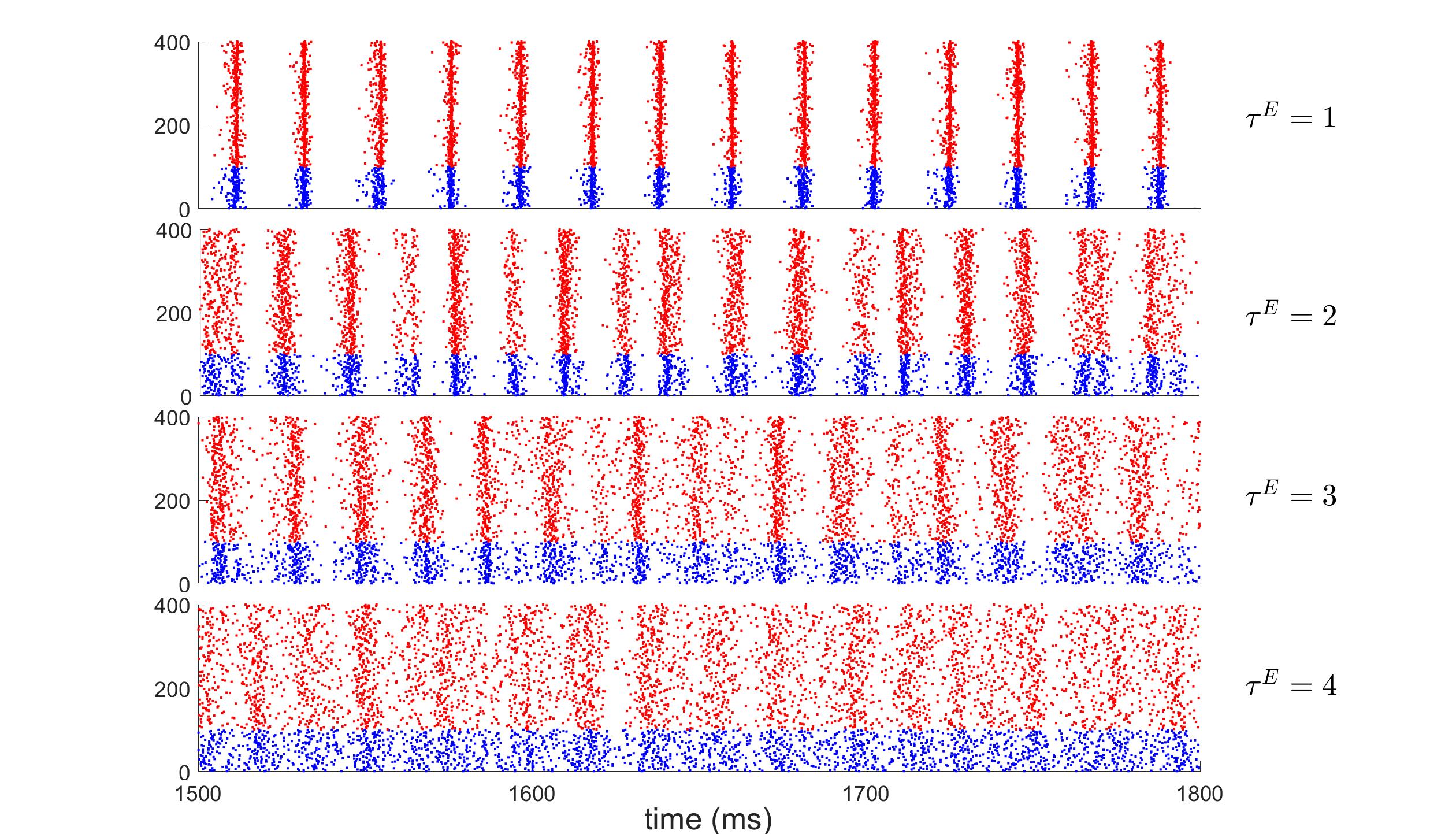}
  \end{subfigure}  \\[1ex]
    \begin{subfigure}{.5\textwidth}
  \textbf{C}\\
  \includegraphics*[bb=2.5in 0.1in 34.5in 19.5in,width=\textwidth]{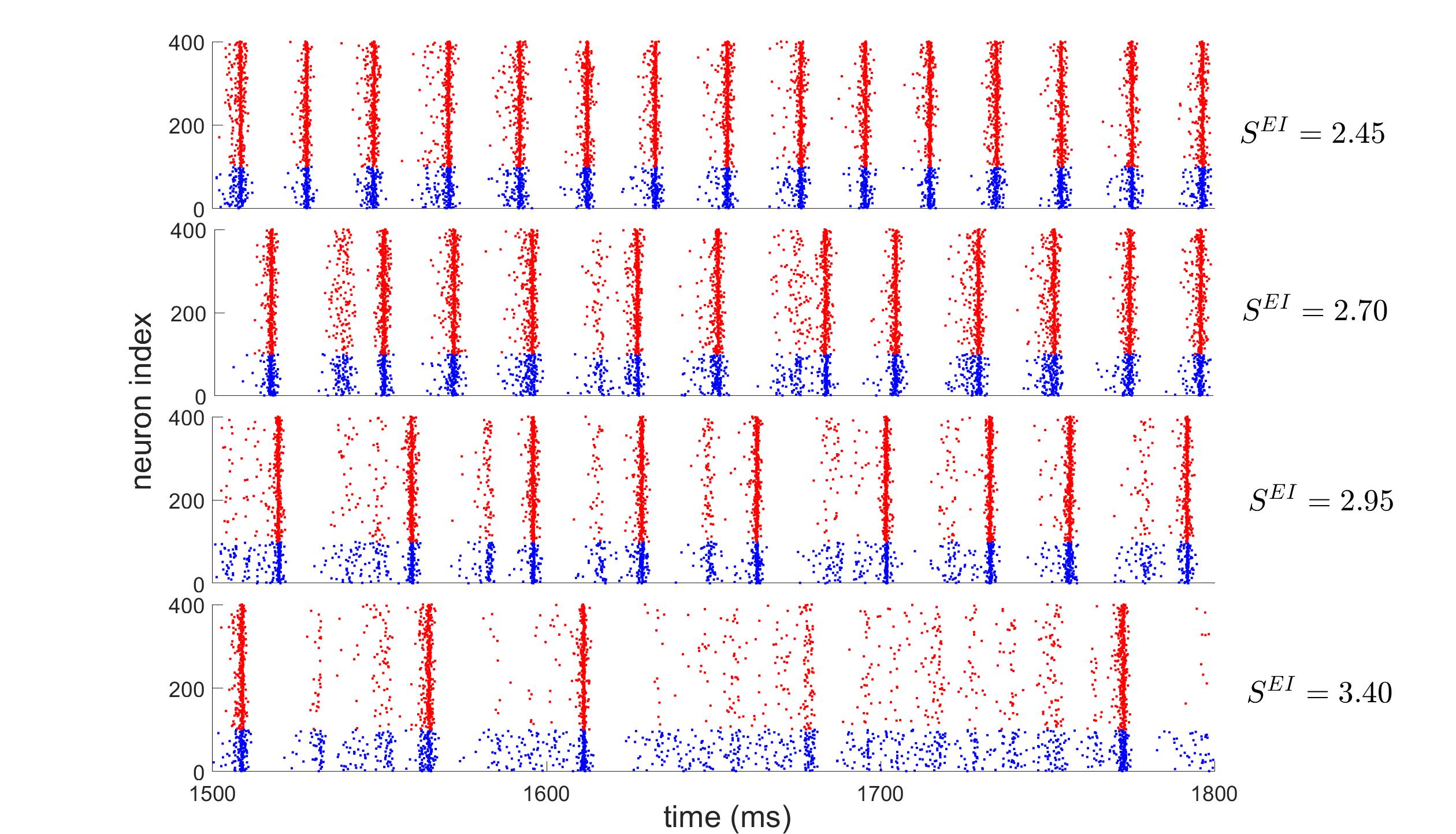}
  \end{subfigure}%
  \begin{subfigure}{.5\textwidth}
  \textbf{D}\\
  \includegraphics*[bb=2.5in 0.1in 34.5in 19.5in,width=\textwidth]{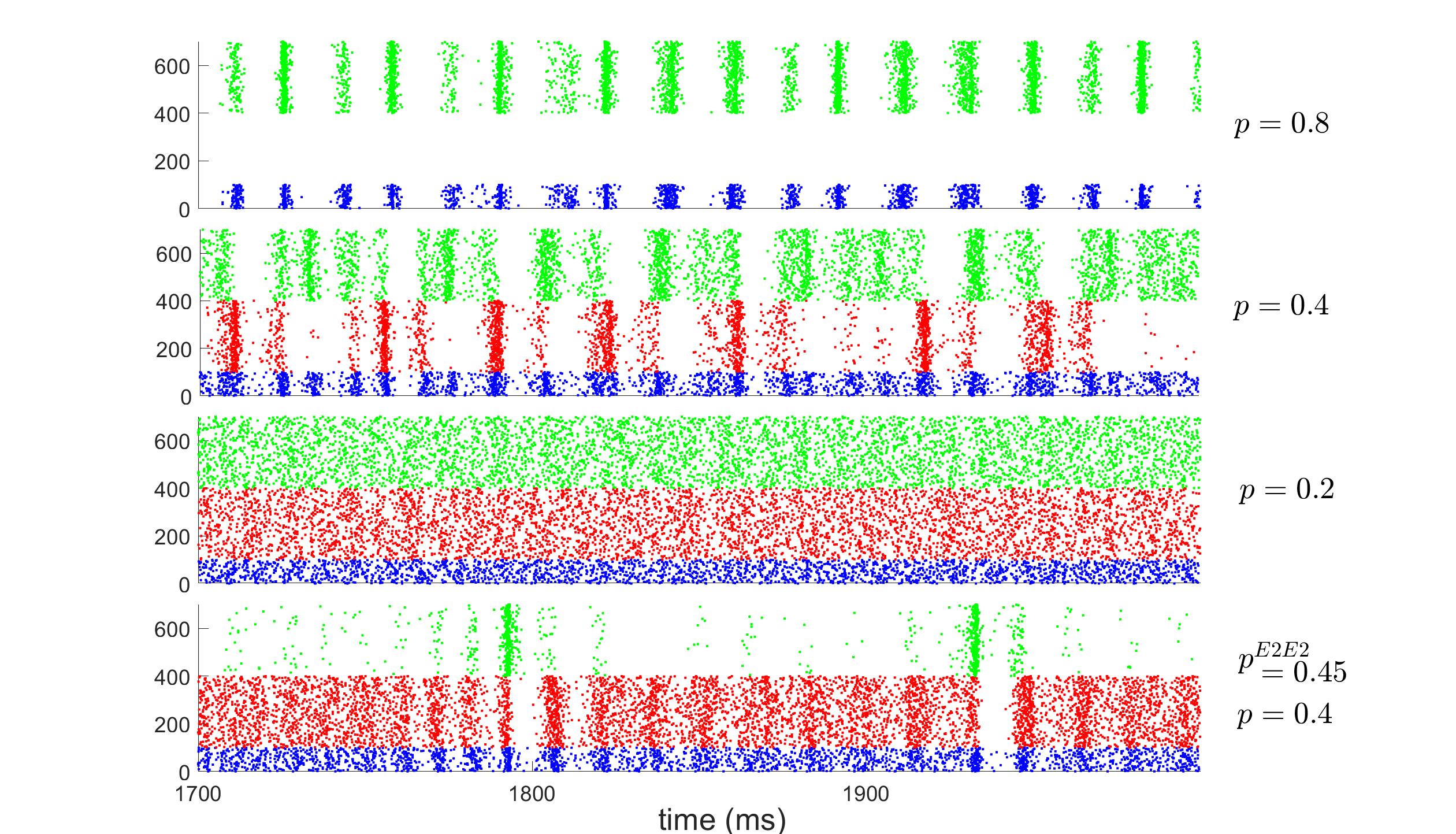}
  \end{subfigure} 
    \caption{Raster plots presenting examples of LIF network dynamics. Neuron $i$ firing at time $t$ is represented by a colored dot at $(t,i)$.
\textbf{A.} Initial transient dynamics produced by a population of 300 excitatory (E) neurons. The synaptic coupling weights are labeled next to the raster plots. In the bottom panel, the variance of external input is increased tenfold compared to the setup in the third panel.
\textbf{B-C.} Different types of dynamics produced by a 400-neuron LIF network (300 E and 100 inhibitory (I) neurons). Only $\tau^E$ and $S^{EI}$ are varied, while all other parameters are set to the standard values from Table~\ref{Table1_Parameters}.
\textbf{D.} Different types of dynamics produced by a 700-neuron LIF network (two E populations with 300 neurons each, and 100 I neurons) by varying the projection probabilities within the same E population. $p^{E_1E_1}$ and $p^{E_2E_2}$ are kept the same, except in the bottom panel, where the biased activation duration is produced. All other parameters are set to standard values in Table~\ref{Table1_Parameters} (e.g., $p^{E_1I} = p^{E_2I} = p^{EI} = 0.8$).}
    \label{Fig2_LIF_dynamics}
  \end{center}
\end{figure}

\subsection{Tested spiking networks and dynamics}
We tested dsODEs with three LIF networks consisting of different homogeneous population configurations: (a) one excitatory (E) population, (b) one E and one inhibitory (I) population, and (c) two competing E populations mediated by one I population (Fig.~\ref{Fig1_dsODE_ideas}C). SSNs with one E and one I population are known to generate a rich spectrum of temporal behaviors \cite{brunel1999fast,rangan2013emergent}. Additionally, SSNs with two competing E-populations can exhibit jumps between multiple stable states, making them useful for modeling multi-stable perceptions and decision-making processes \cite{wang2002probabilistic,mazzucato2015dynamics}. A subset of the SNN dynamics we investigated is shown in the raster plots in Fig.~\ref{Fig2_LIF_dynamics}, where each colored dot represents a spiking event produced by a specific neuron at a given time. 

Fig.~\ref{Fig2_LIF_dynamics}A presents four examples of transient dynamics generated by a single E population of 300 neurons. Initially, all neurons start at the resting potential $v_i = 0$ and receive independent external stochastic stimuli. Due to the synchronized membrane potentials at $t = 0$, a few spiking clusters emerge before the network dynamics stabilize into temporal homogeneity. While recurrent coupling strengths can increase the synchronicity of the initial spiking clusters (Fig.~\ref{Fig2_LIF_dynamics}A, $S^{EE} = 0-0.01$), adding more independent noise to the external inputs effectively desynchronizes the population dynamics (Fig.~\ref{Fig2_LIF_dynamics}A, bottom).

Fig.~\ref{Fig2_LIF_dynamics}B and C depict examples of rich dynamics produced by a pair of coupled E and I populations. Fig.~\ref{Fig2_LIF_dynamics}B displays oscillatory behaviors of the SNNs in the Gamma band (or ``Gamma oscillations", 30-80 Hz). In addition, as the ratio $\tau^E/\tau^I$ increases, we observe a transition of the Gamma oscillations from temporal synchrony to homogeneity. This phenomenon has been well studied previously \cite{borgers2003synchronization,KeeleyEtAl2019}. On the other hand, Fig.~\ref{Fig2_LIF_dynamics}C shows a different type of phase transition as the recurrent inhibitory input to E cells ($S^{EI}$) increases. When $S^{EI}$ is low ($=2.45$), the oscillatory temporal dynamics consist of regularly occurring spiking clusters of similar size within the gamma band (30-80 Hz). However, as $S^{EI}$ increases, smaller spiking clusters appear, and a stable alternation between strong and weak clusters emerges ($S^{EI} = 2.95$), leading to oscillations in another frequency band (Beta, 15-30 Hz). More interestingly, further increasing $S^{EI}$ to 3.4 results in a bi-stable stochastic transition between synchrony and homogeneity (or weak synchrony). These phenomena were first reported in our recent study \cite{wu2022multi}.

Fig.~\ref{Fig2_LIF_dynamics}D illustrates the switching dynamics between two competing E-populations, $E_1$ and $E_2$, mediated by an I population that suppresses both E populations. The bistability can be explained as follows: when $E_2$ is completely suppressed by the I population, the network behaves similarly to the E-I network shown in the top panel of Fig.~\ref{Fig2_LIF_dynamics}B, producing stable, temporally oscillatory or homogeneous dynamics in $E_1$ and I. An example of this can be seen in the first panel ($p^{E_1E_1} = p^{E_2E_2} = 0.8$). However, the other E-population can regain dominance (a ``switch" from $E_1$ to $E_2$), which may occur due to noise or insufficient excitation. For example, when $p^{E_1E_1} = p^{E_2E_2} = 0.4$, after a large firing cluster of $E_1$, the recurrent excitation within $E_1$ is insufficient to prevent $E_2$ from taking over. This results in regular, unbiased switching dynamics between both E-populations. Furthermore, reducing $p^{E_1E_1} = p^{E_2E_2}$ to 0.2 causes both E populations to struggle to maintain dominance. However, when $p^{E_1E_1}$ and $p^{E_2E_2}$ differ slightly (bottom panel), the network exhibits a strong bias toward switching: $E_1$ dominates most of the time, but it may lose dominance after its dynamics settle after a few firing clusters. 

\subsection{One E-population}
\label{Sect4.2-1E}
\begin{figure}
  \begin{center}
    \includegraphics*[bb=2in 0in 35in 20.5in,width=0.65\textwidth]{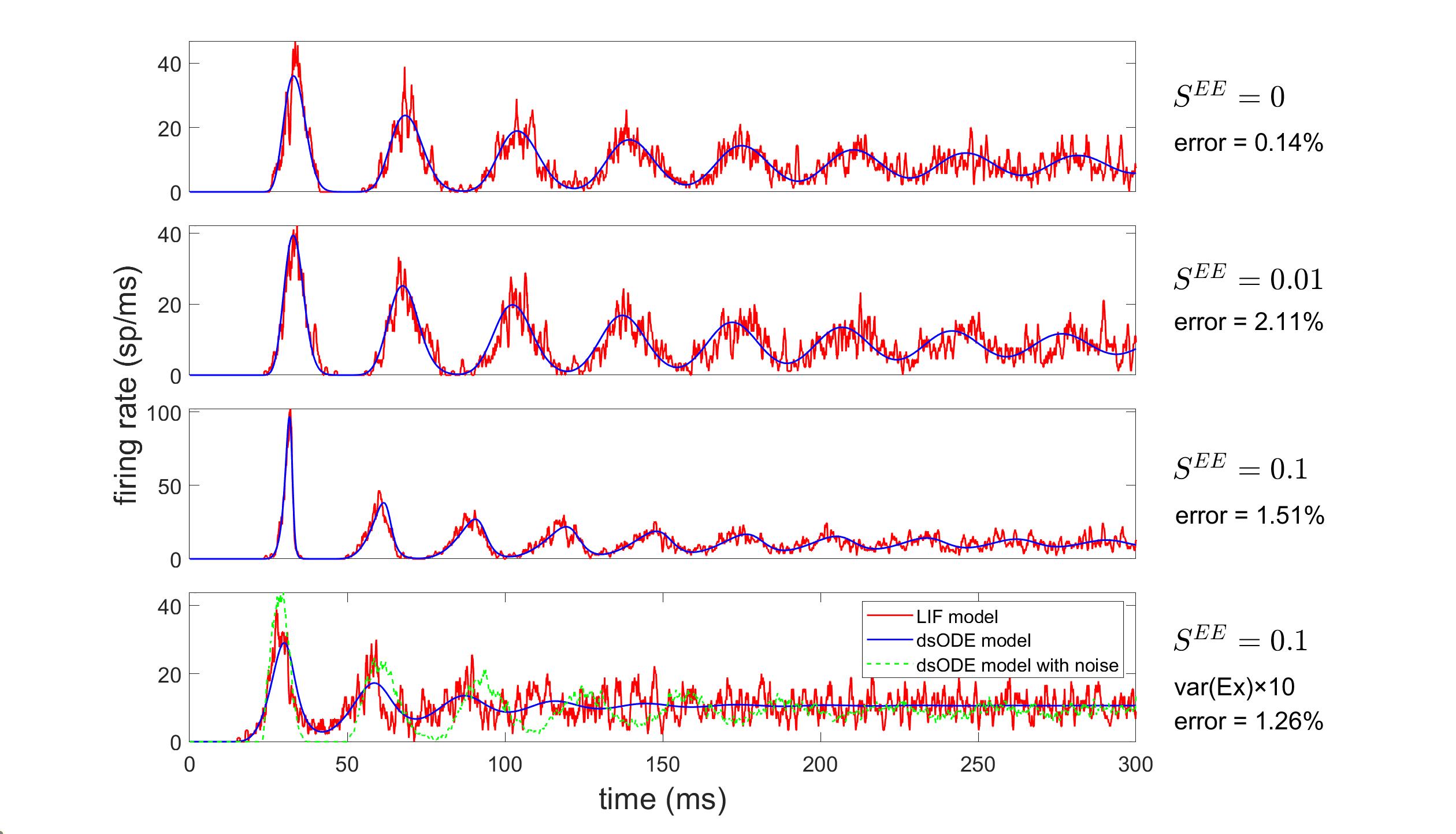}
    \caption{The dsODE model predicts dynamics produced by single LIF E-populations. The four panels correspond to the panels in Fig.~\ref{Fig2_LIF_dynamics}A with the same parameter choices and time windows. Firing rates produced by dsODEs and LIF networks are shown by red and blue curves, respectively.}
    \label{Fig3_1E}
  \end{center}
\end{figure}

Our dsODE accurately captures the initial transient dynamics produced by single E populations, including the timing, amplitudes, and shapes of the first few peaks in the firing rate curves. These peaks correspond to the firing clusters in Fig.~\ref{Fig2_LIF_dynamics}A before the dynamics settle down. However, the LIF network exhibits fluctuations in spike timings, which are reflected by deviations from the dsODE predictions. These deviations become more pronounced when fluctuations in the external input are amplified (bottom panels of Fig.~\ref{Fig2_LIF_dynamics}A and Fig.~\ref{Fig3_1E}). Nevertheless, by reintroducing the noise terms to the dsODEs (see details in Sect.~\ref{Sect3.2-dsSDE-Reduced}), we observe similar amplitudes of deviation (green curve, bottom panel of Fig.~\ref{Fig3_1E}). Additionally, the errors in the averaged firing rates predicted by the dsODE remain within approximately 3\%.

\subsection{A pair of coupled E/I-populations}
\label{Sect4.3-1E1I}
To test dsODEs predictions on LIF networks with one E and one I-population, we perform sweeps of different types of parameters. We include the synaptic time scale ($\tau^E$), synaptic coupling weights ($S^{EI}$ and $S^{II}$), projection probabilities ($p^{EI}$ and $p^{II}$), the length of refractory periods ($\tau^{\mathcal{R}}$), and the noise in external input. Here, we choose to present results of varying $\tau^E$ and $S^{EI}$ (as shown in Fig.~\ref{Fig2_LIF_dynamics}B-C) in detail, leaving the rest to the bifurcation analysis in Sect.~\ref{Sect4.5-Bifurcation}. For convenience, we show only the dynamics of the E-population for both models in this section.

\begin{figure}[htbp]
    \begin{subfigure}{0.7\textwidth}
      {\bf A}\\
      \includegraphics*[bb=2in 0in 35in 20in,width=\textwidth]{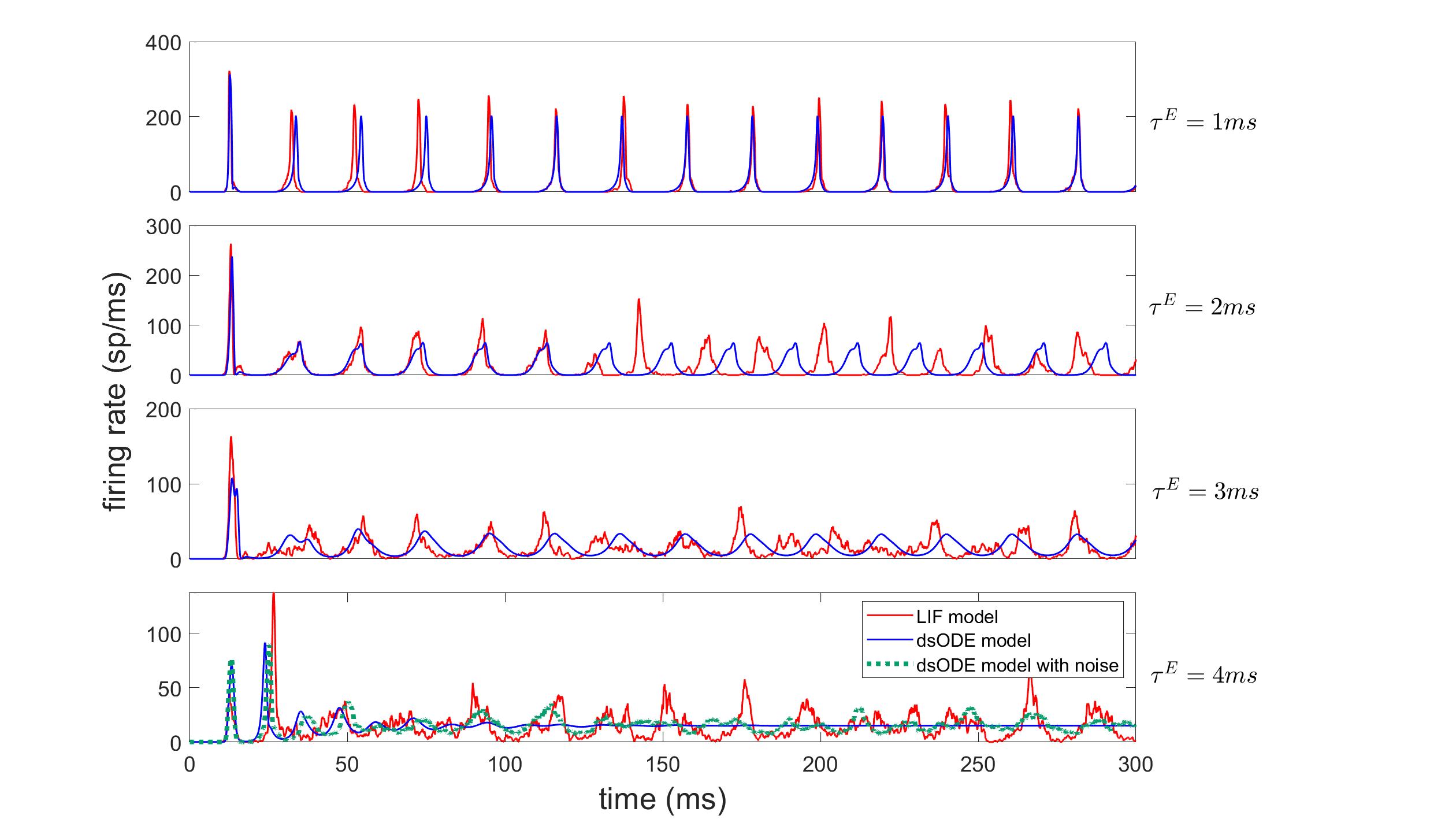}
    \end{subfigure} \\[1ex]%
    \begin{subfigure}{\textwidth}
      \begin{subfigure}{.24\textwidth} 
        {\bf B}\\
        \includegraphics*[bb=8.5in 1in 29.2in 20in,width=\textwidth]{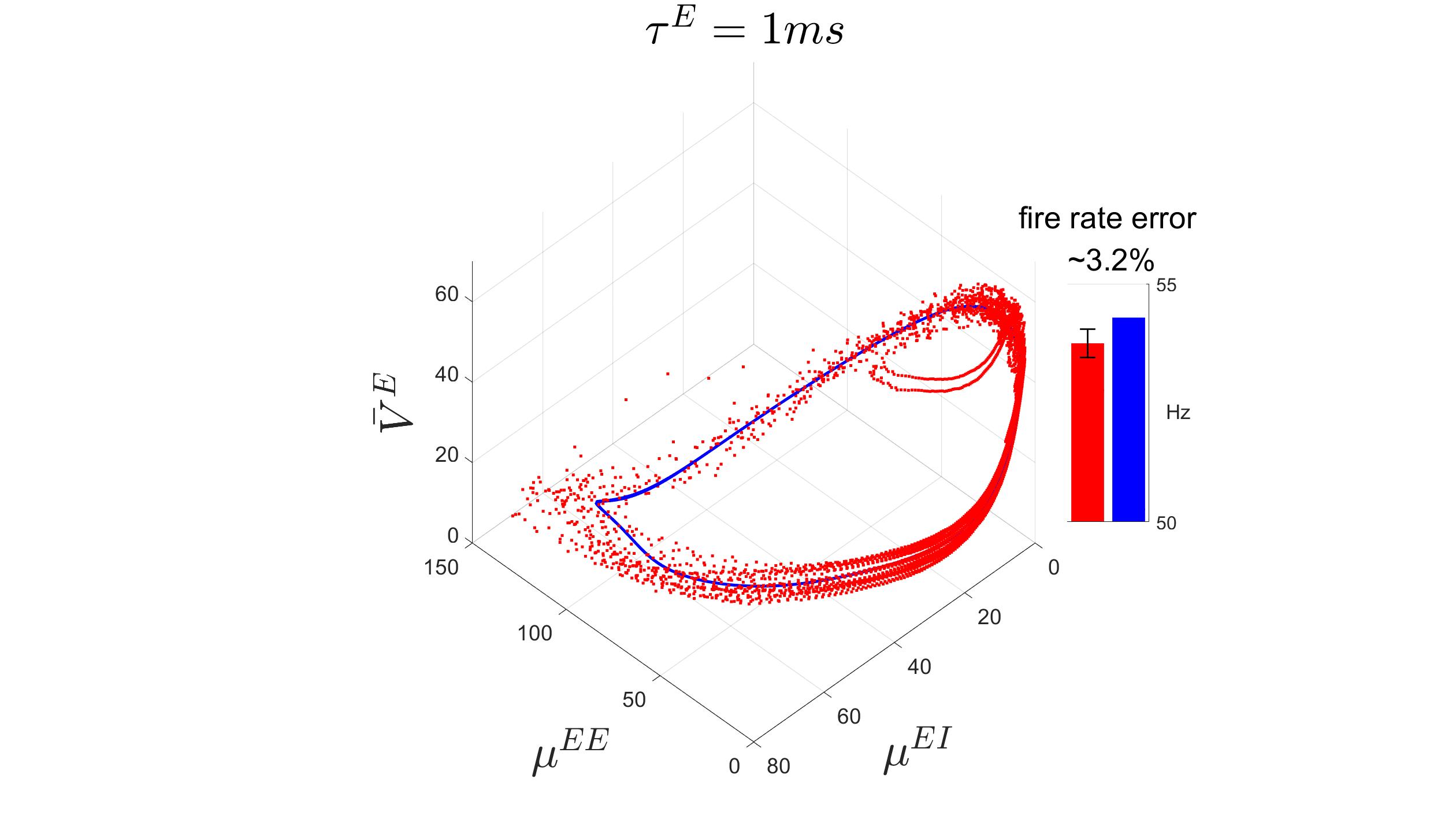}
      \end{subfigure}%
      \begin{subfigure}{.24\textwidth}
        {\bf C}\\
        \includegraphics*[bb=8.5in 1in 29.2in 20in,width=\textwidth]{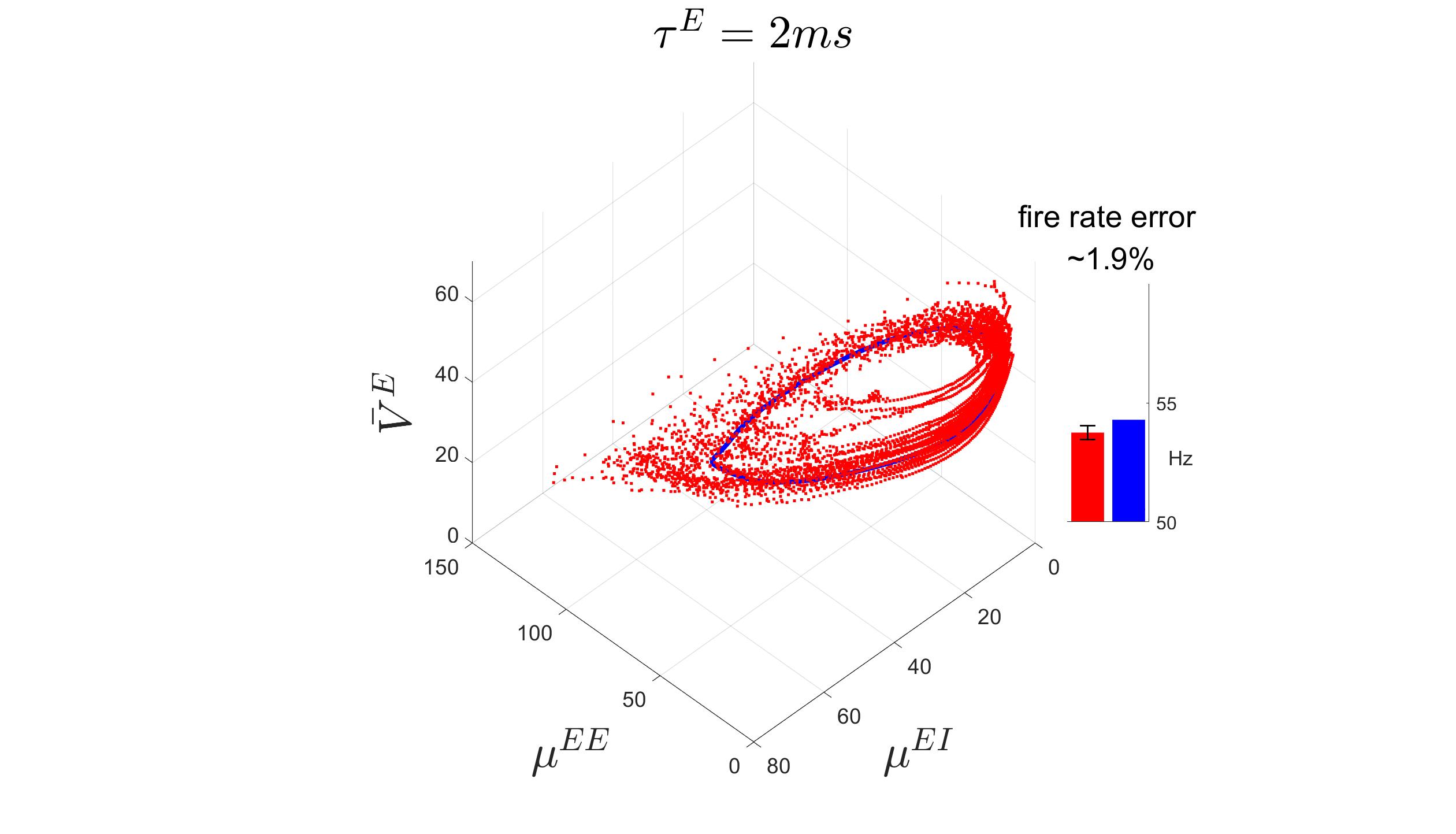}
      \end{subfigure}
      \begin{subfigure}{.24\textwidth}
        {\bf D}\\
        \includegraphics*[bb=8.5in 1in 29.2in 20in,width=\textwidth]{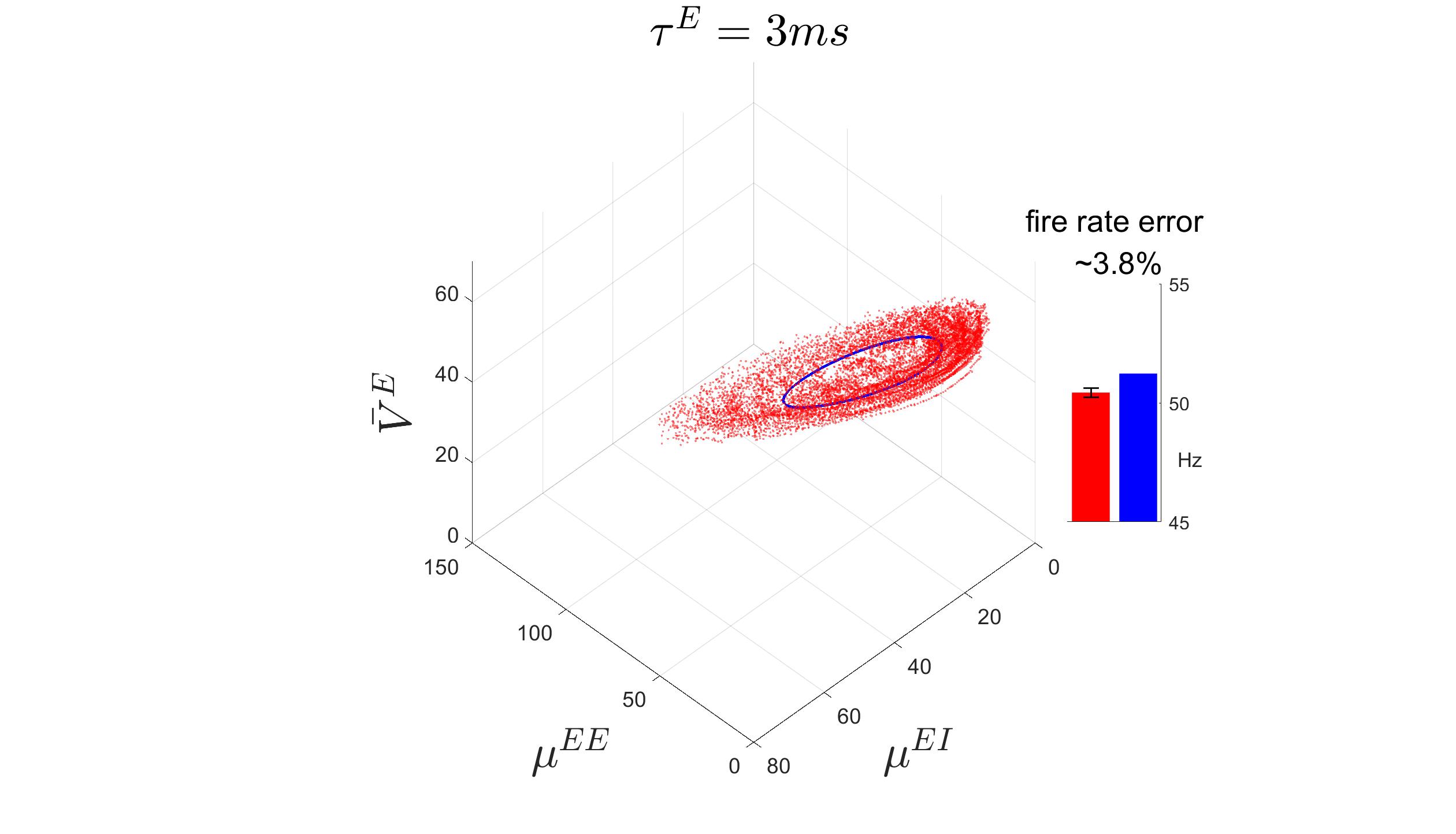}
      \end{subfigure}%
      \begin{subfigure}{.24\textwidth}
        {\bf E}\\
        \includegraphics*[bb=8.5in 1in 29.2in 20in,width=\textwidth]{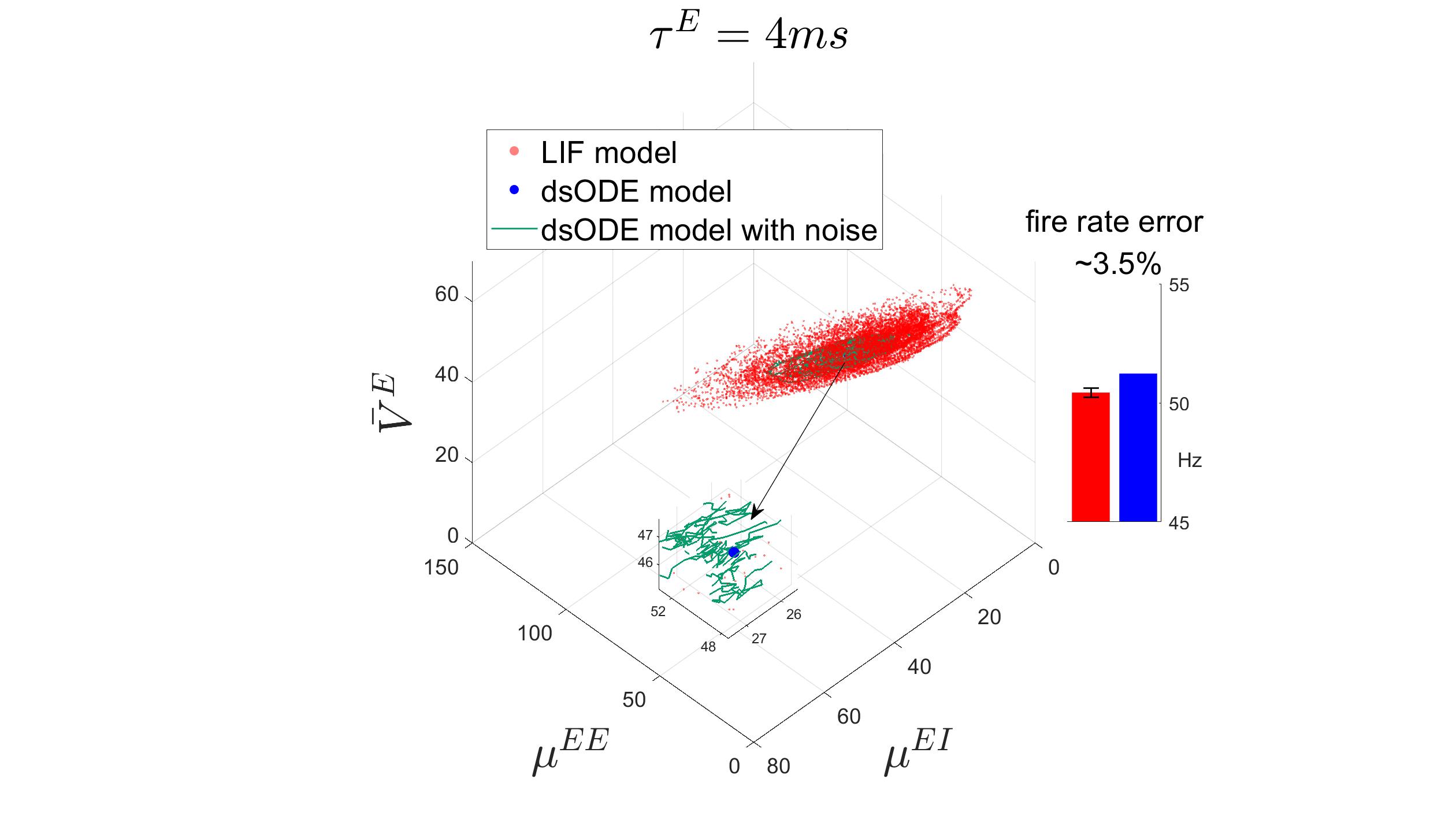}
      \end{subfigure}
    \end{subfigure}
    \caption{The dsODE model predicts synchronicity within spiking networks. A pair of coupled E/I-populations are investigated by varying excitatory synaptic timescale $\tau^E$. \textbf{A.} dsODEs produce faithful initial transient dynamics for $t<100$ms. The four panels correspond to those in Fig.~\ref{Fig2_LIF_dynamics}B. Weak oscillations are also recovered by adding noise back to the dsODE when $\tau^E = 4$ ms (green dashed curve, bottom panel). \textbf{B-E.} For $\tau^E = 1-4$ ms, respectively, trajectories of dsODEs (blue curves) and LIF networks (red dashed curves) for $t\in[1000, 2000]$ ms are projected onto the phase space of $(\bar{V}^E, \mu^{EE}, \mu^{EI})$. In \textbf{E}, the point attractor (blue dot) is shown in the inset. The comparison of averaged firing rates is presented by the phase plots.}
    \label{Fig4_1E1I_tau}
\end{figure}

When varying $\tau^E$, dsODEs not only capture the initial transient dynamics of the LIF network for $t < 100$ms (Fig.~\ref{Fig4_1E1I_tau}A), but also quantitatively predict the Gamma oscillations in the long run (Fig.~\ref{Fig4_1E1I_tau}B-E). To visualize the dynamics of dsODEs and LIF networks, we choose a 3D phase space consisting of $(\bar{V}^E, \mu^{EE}, \mu^{EI})$ and focus on the behavior of the E-population. Here, $\bar{V}^E$ is the average membrane potential of all E-neurons. When projecting the LIF dynamics to a reduced phase space, the oscillatory dynamics are represented by cycles in the trajectories. We find that when $\tau^E = 1$, 2, and 3ms, the dsODE system admits limit cycles (blue curves, Fig.~\ref{Fig4_1E1I_tau}B-D) whose sizes, shapes, and frequencies are close to the trajectories of LIF networks (red dashed curves, Fig.~\ref{Fig4_1E1I_tau}B-D). On the other hand, when $\tau^E = 4$, the dsODE system only admits a point attractor (the blue dot, inset of Fig.~\ref{Fig4_1E1I_tau}E) while the LIF network still exhibits weak oscillations in the gamma band. Nevertheless, we find that the weak oscillation is partially recovered when adding noise terms back to the dsODE system.

\begin{figure}[htbp]
    \begin{subfigure}{0.7\textwidth}
      {\bf A}\\
      \includegraphics*[bb=2in 0.5in 32.5in 20in,width=\textwidth]{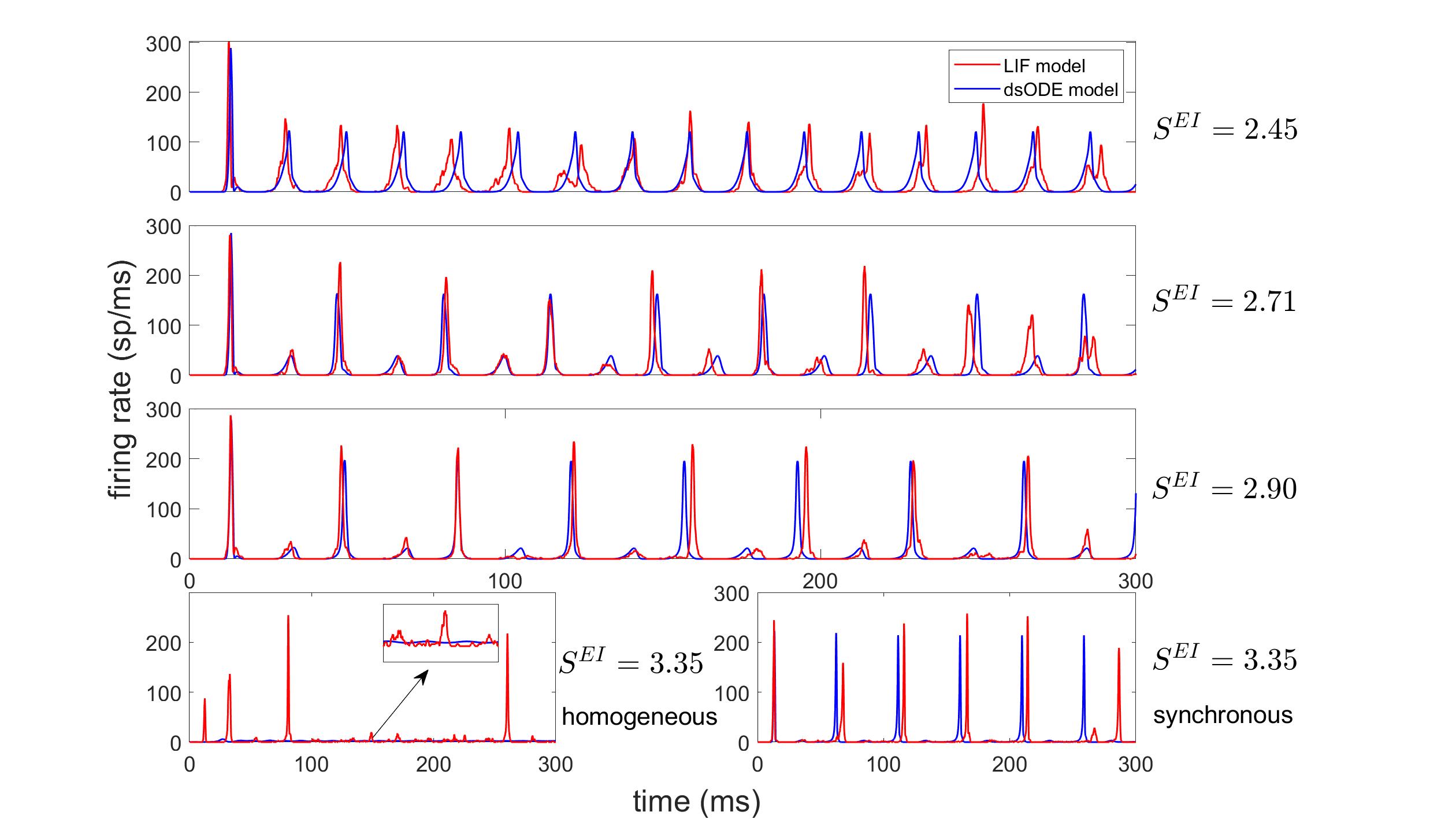}
    \end{subfigure} \\[1ex]%
    \begin{subfigure}{\textwidth}
      \begin{subfigure}{.24\textwidth} 
        {\bf B}\\
        \includegraphics*[bb=9.5in 1in 29.2in 20in,width=\textwidth]{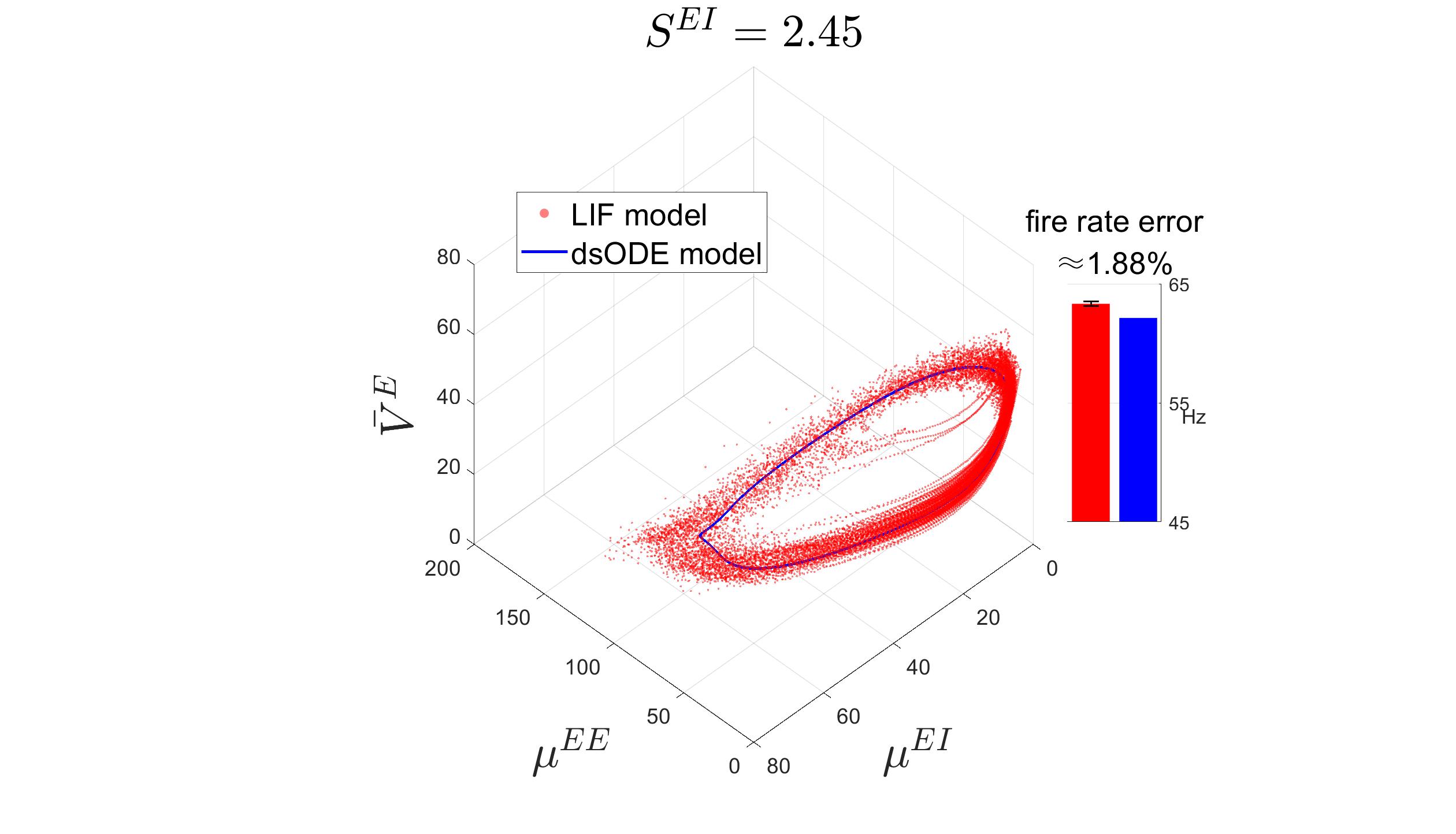}
      \end{subfigure}%
      \begin{subfigure}{.24\textwidth}
        {\bf C}\\
        \includegraphics*[bb=9.5in 1in 29.2in 20in,width=\textwidth]{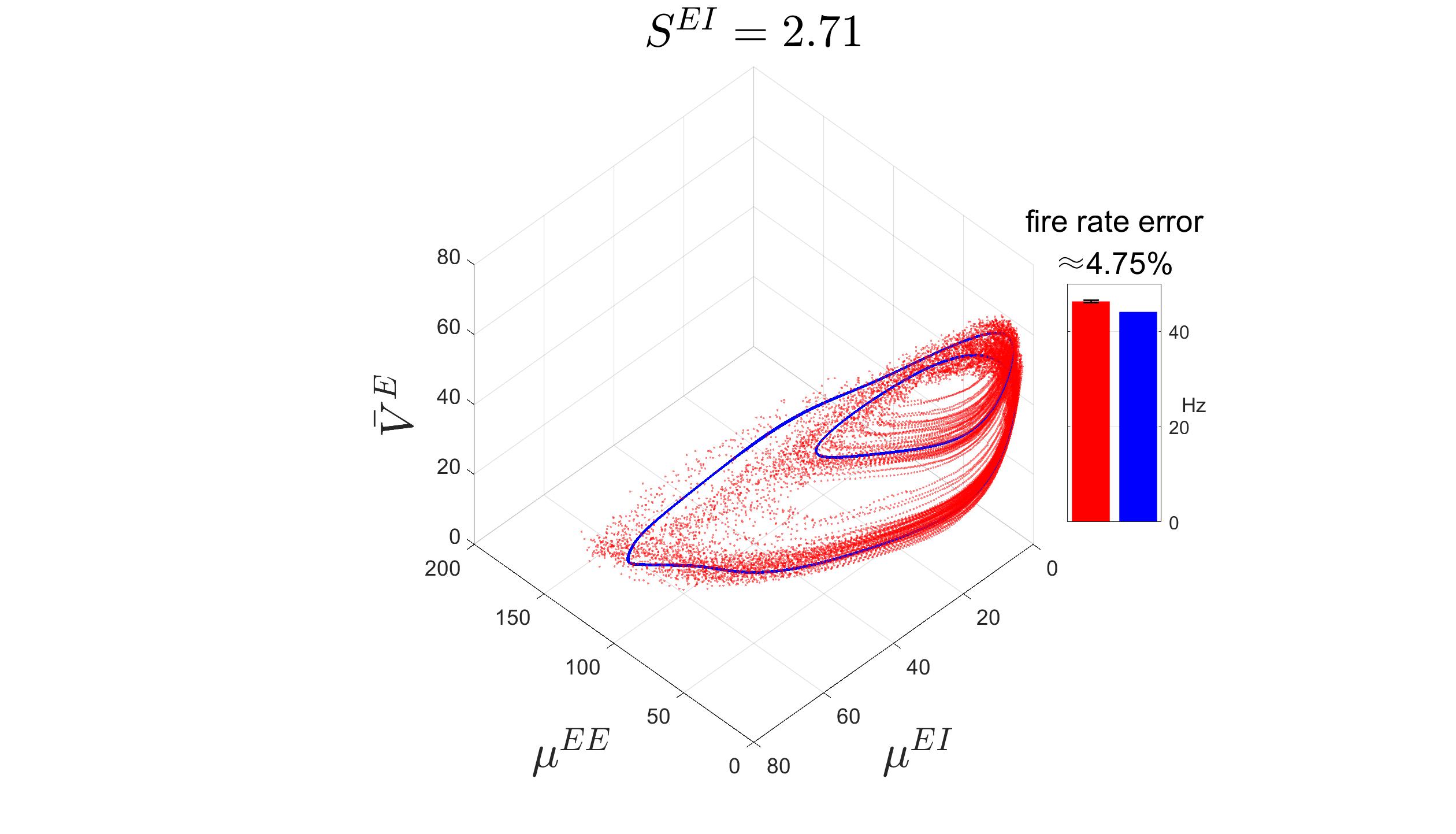}
      \end{subfigure}
      \begin{subfigure}{.24\textwidth}
        {\bf D}\\
        \includegraphics*[bb=9.5in 1in 29.2in 20in,width=\textwidth]{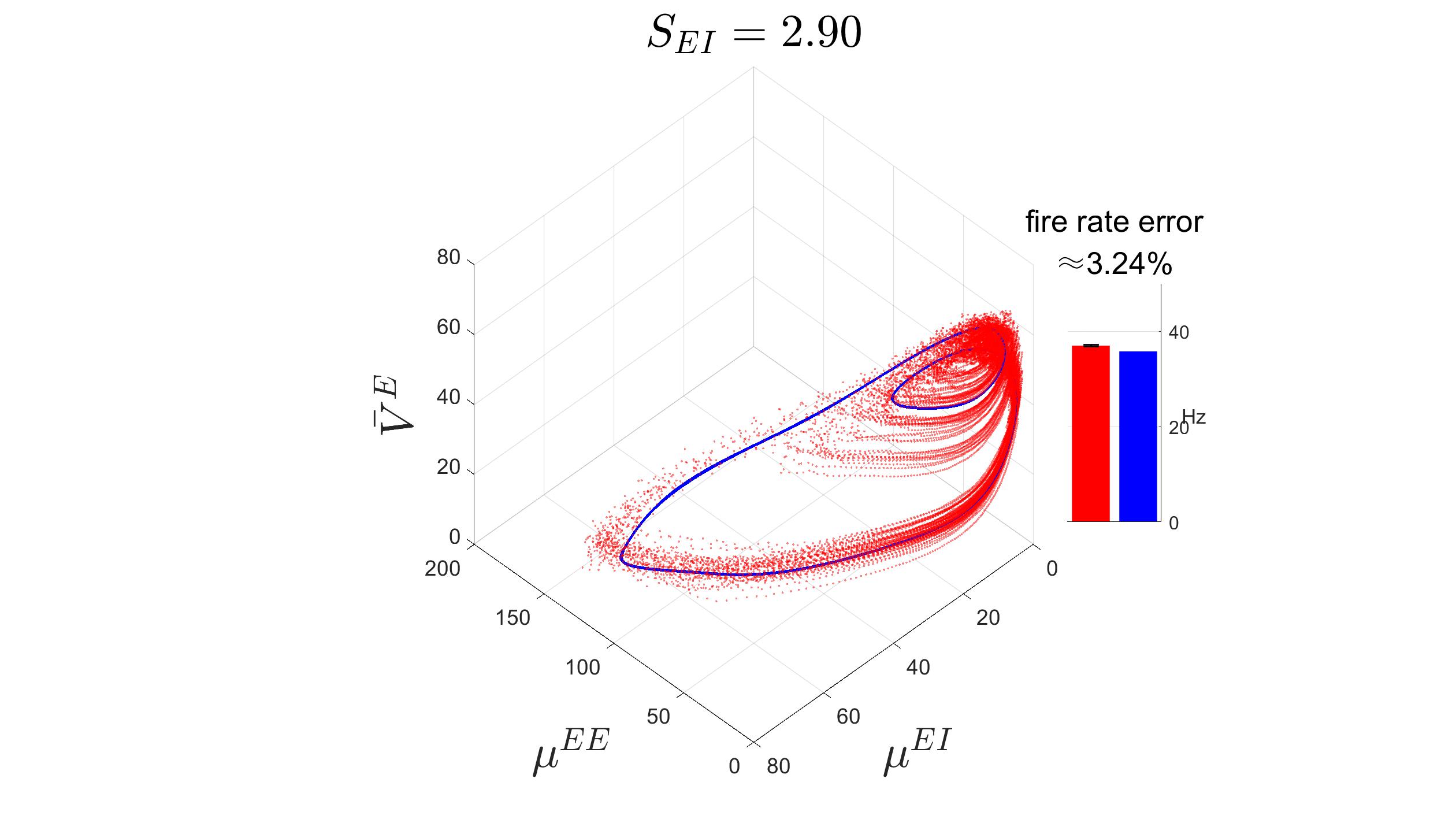}
      \end{subfigure}%
      \begin{subfigure}{.24\textwidth}
        {\bf E}\\
        \includegraphics*[bb=9.5in 1in 29.2in 20in,width=\textwidth]{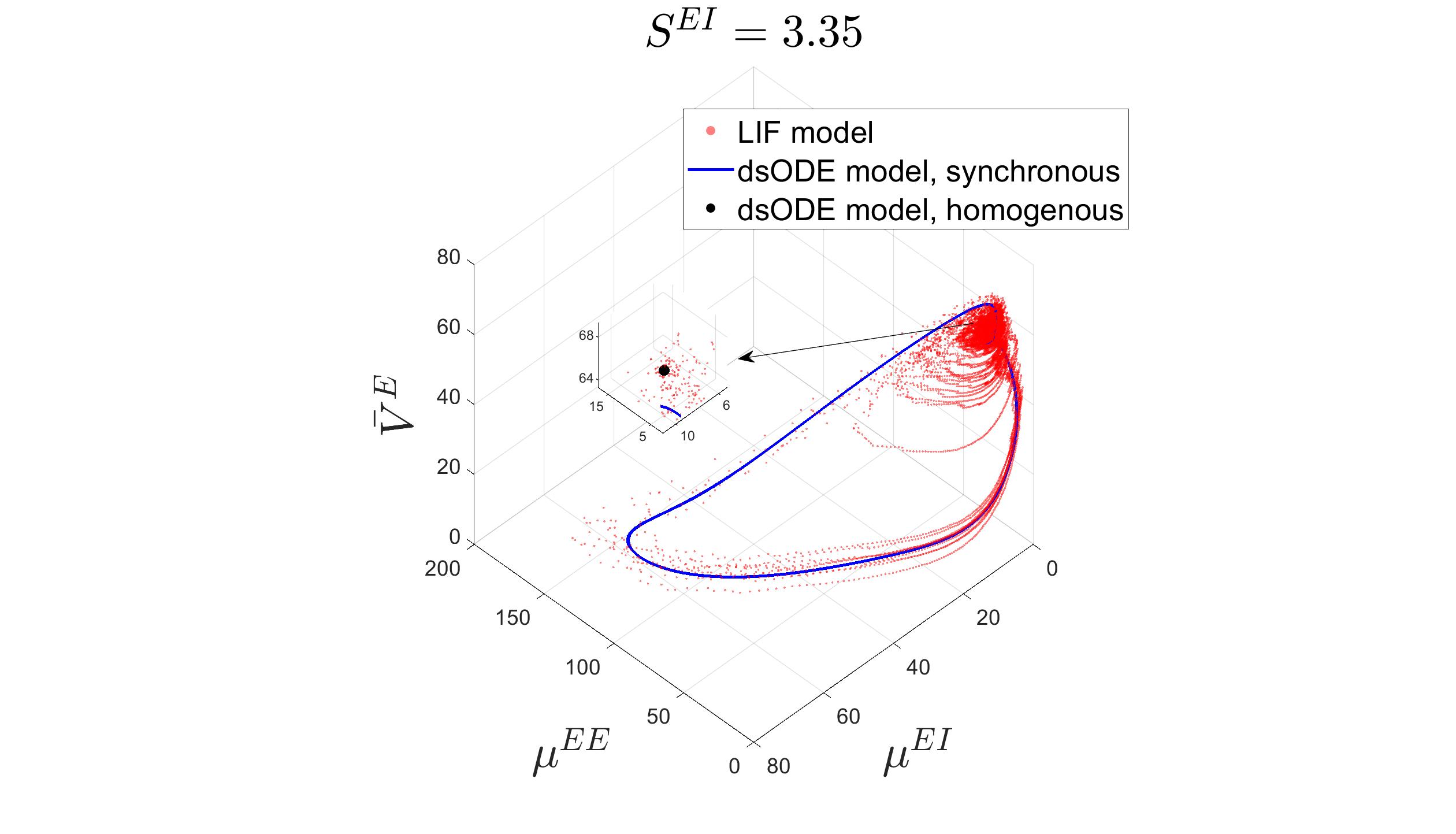}
      \end{subfigure}
    \end{subfigure}
    \caption{The dsODE model predicts the beats of oscillatory dynamics and the coexistence of multiple attractors. A pair of coupled E/I-populations are investigated by varying I-to-I synaptic coupling weight $S^{EI}$. \textbf{A.} For $S^{EI} = $2.45, 2.71, and 2.90, dsODEs accurately capture the initial transient dynamics for $t<100$ms (rows 1-3). For $S^{EI} = 3.35$, two distinct dsODE trajectories converging to different attractors are compared to LIF dynamics (row 4). The four panels correspond to Fig.~\ref{Fig2_LIF_dynamics}C. \textbf{B-E.} For different values of $S^{EI}$, trajectories of dsODEs (blue curves) and LIF networks (red dashed curves) for $t\in[1000, 2000]$ ms are projected onto the same phase space as in Fig.~\ref{Fig4_1E1I_tau}. In \textbf{E}, the attractive point is displayed in the inset along with part of the other attractor, i.e., the limit cycle. The comparison of averaged firing rates is presented through phase plots for \textbf{B-D}.}
    \label{Fig5_1E1I_SEI}
\end{figure}

The performances of dsODE are as good when varying $S^{EI}$ in the aspects of initial transient dynamics (Fig.~\ref{Fig5_1E1I_SEI}A), long-term behaviors, and average firing rates (Fig.~\ref{Fig5_1E1I_SEI}B-E). In the phase space, dsODE presents the alternation between the strong \& weak clusters as double cycles (blue curves, Fig.~\ref{Fig5_1E1I_SEI}CD). On the other hand, the bi-stable dynamics at $S^{EI}= 3.35$ is represented by the coexistence of two attractors in the same dsODE system: one limit cycle standing for synchronized, oscillatory dynamics, and one point attractor standing for the homogeneous dynamics (inset, Fig.~\ref{Fig5_1E1I_SEI}E). Adding noise terms back to the dsODEs realizes the stochastic transition between the two attractors. In the LIF model, the average duration of the synchronous state is 121 ms (95\% confidence interval, or CI: 52-457 ms), 
and the average duration of the homogeneous state is 95 ms (95\% CI: 72-416 ms). 
Correspondingly, the dsSDE model (dsODE plus the noise terms Eq.~\ref{Sect3.2-Eq19-frNoise_Redu} and Eq.\ref{Sect3.2-Eq19-vfluxNoise-Redu}) yields average duration as 254 ms (95\% CI: 52-624 ms) and 182 ms (95\% CI: 113-468ms). Though the ranges of dwell time are comparable, we speculate that the difference of average duration comes from the stronger randomness LIF model induced by its discrete nature, hence the shorter transition waiting time. 

Overall, for the 400-neuron LIF network tested, dsODE predictions yield a small relative error ($<6\%$) of the average firing rate for all choices of parameters (side plots in Fig.~\ref{Fig4_1E1I_tau}B-E and Fig.~\ref{Fig5_1E1I_SEI}B-E). A more systematic comparison of the prediction errors of firing rates between dsODE and previous theories is shown in Sect.~\ref{Sect5-Comparison}.

\subsection{Two competing E-populations mediated by one I-population}
\label{Sect4.4-2E1I}
\begin{figure}[tbp]
    \begin{subfigure}{0.5\textwidth}
      {\bf A}\\
          \includegraphics*[bb=2.5in 0in 33in 19in,width=.98\textwidth]{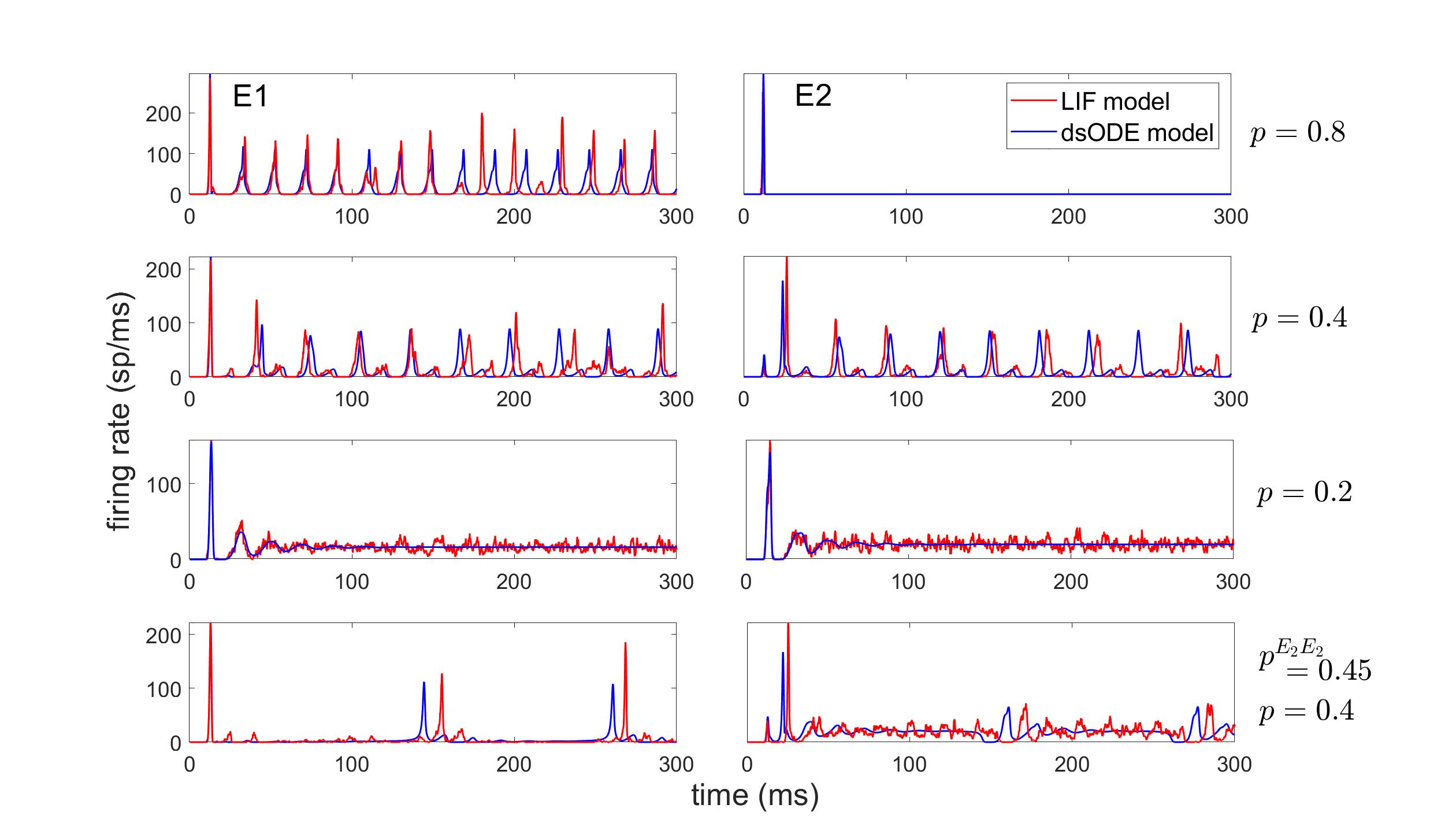} 
    \end{subfigure}
    \begin{subfigure}{0.5\textwidth}
      {\bf B}\\
      \includegraphics*[bb=2.5in 0in 33in 19in,width=.98\textwidth]{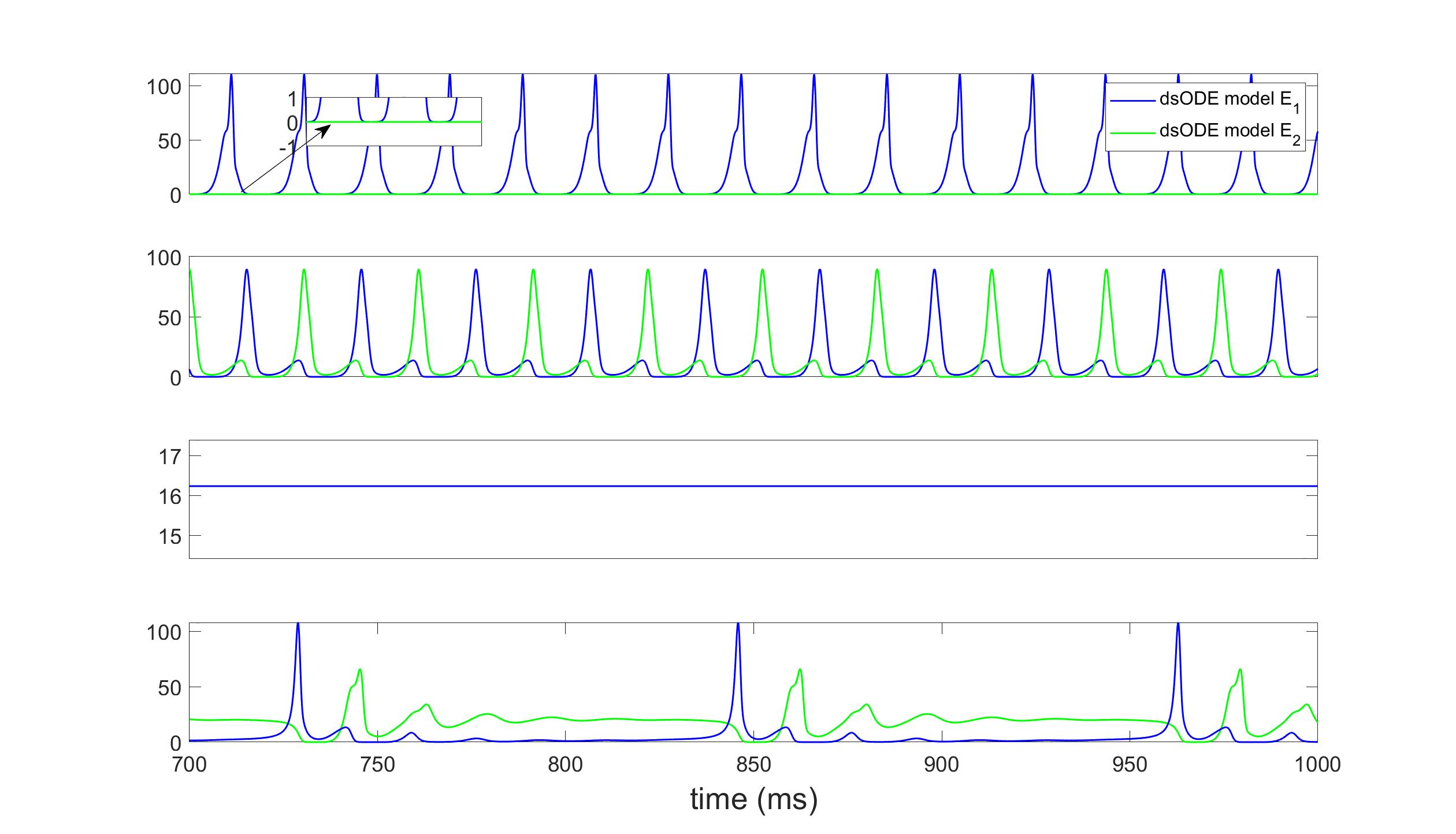}
    \end{subfigure}\\[1ex]%
 \begin{subfigure}{\textwidth}
      \begin{subfigure}{.24\textwidth} 
        {\bf C}\\
        \includegraphics*[bb=8.2in 1in 29.2in 20in,width=\textwidth]{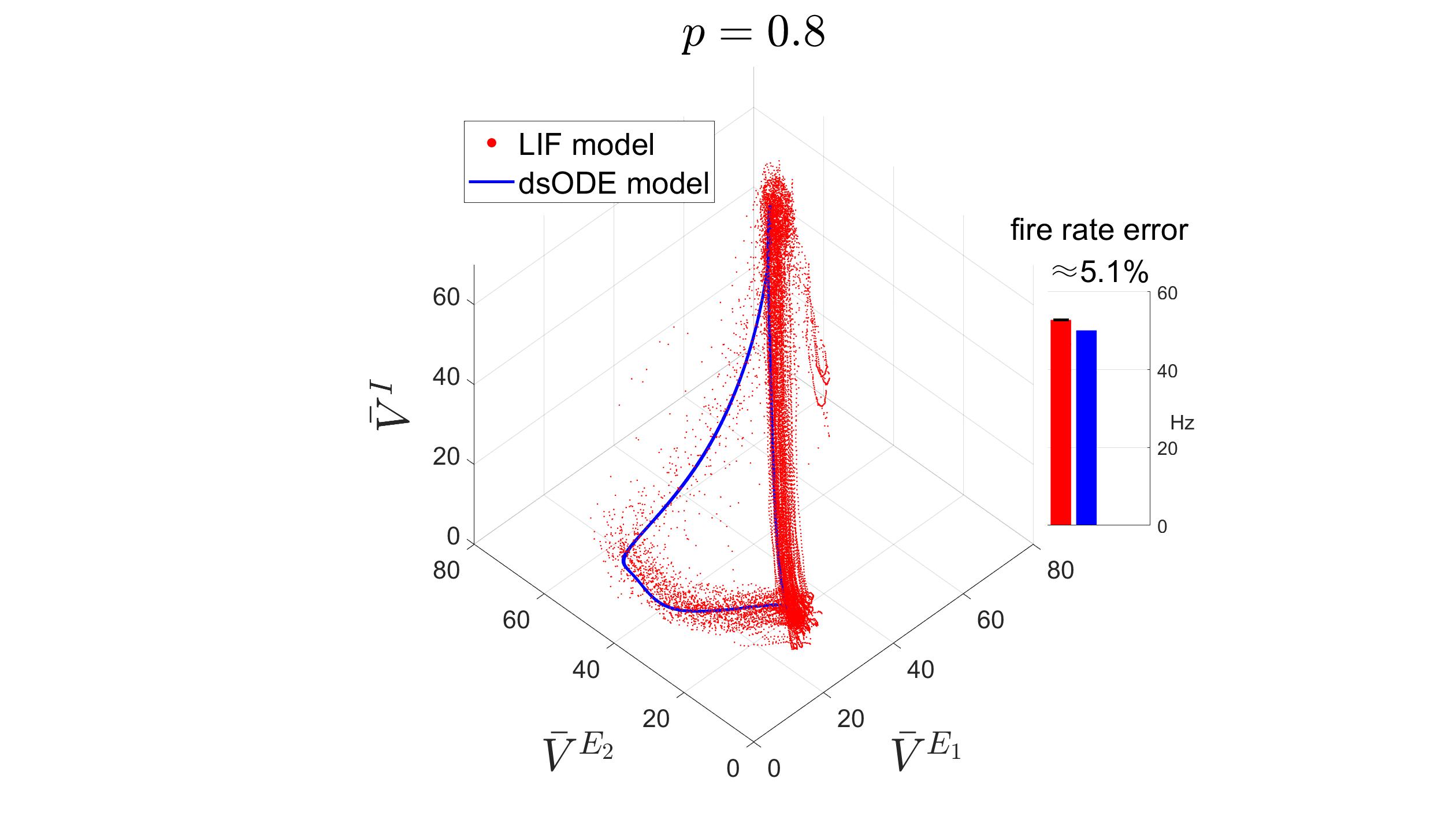}
      \end{subfigure}%
      \begin{subfigure}{.24\textwidth}
        {\bf D}\\
        \includegraphics*[bb=8.2in 1in 29.2in 20in,width=\textwidth]{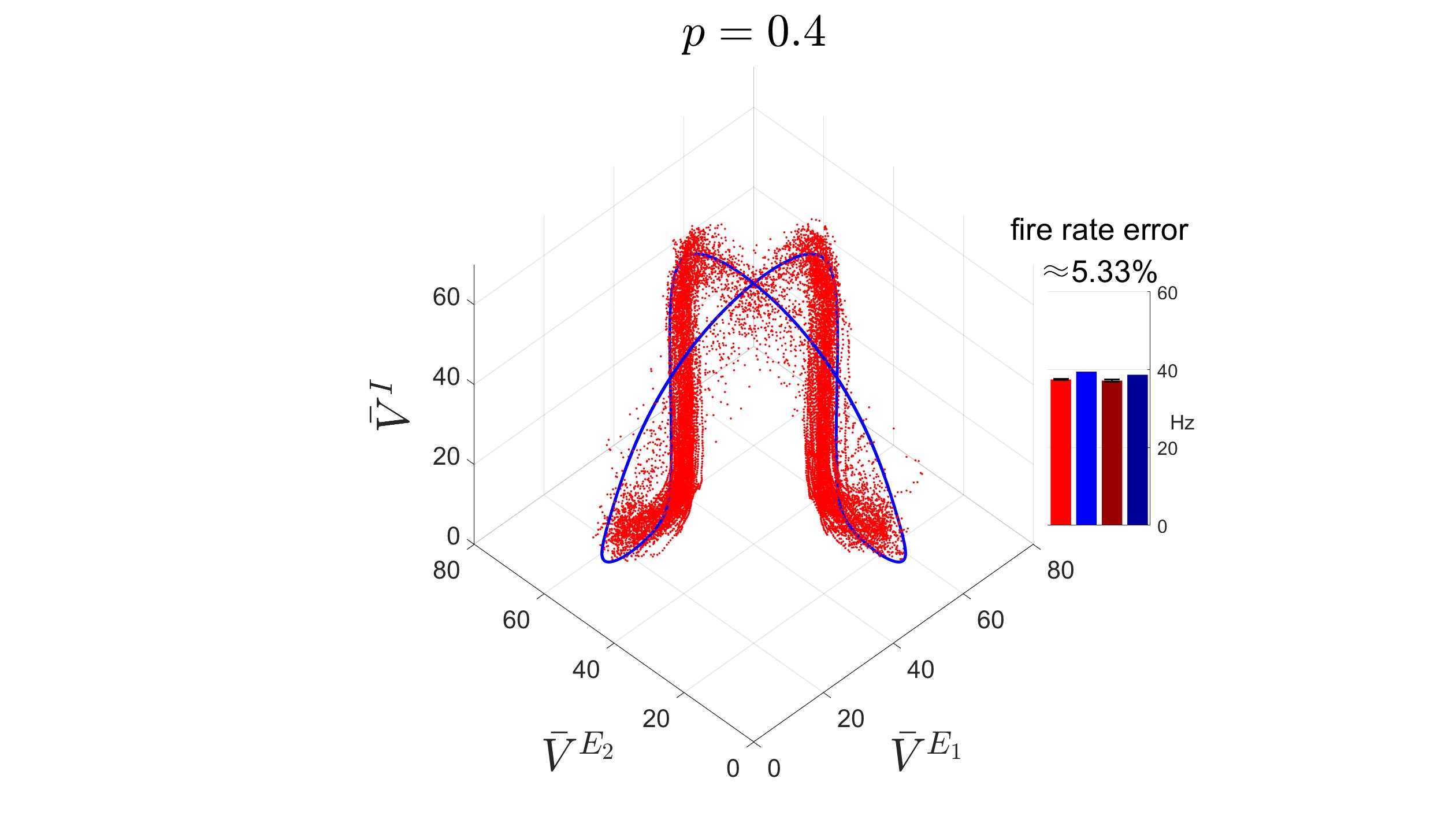}
      \end{subfigure}
      \begin{subfigure}{.24\textwidth}
        {\bf E}\\
        \includegraphics*[bb=8.2in 1in 29.2in 20in,width=\textwidth]{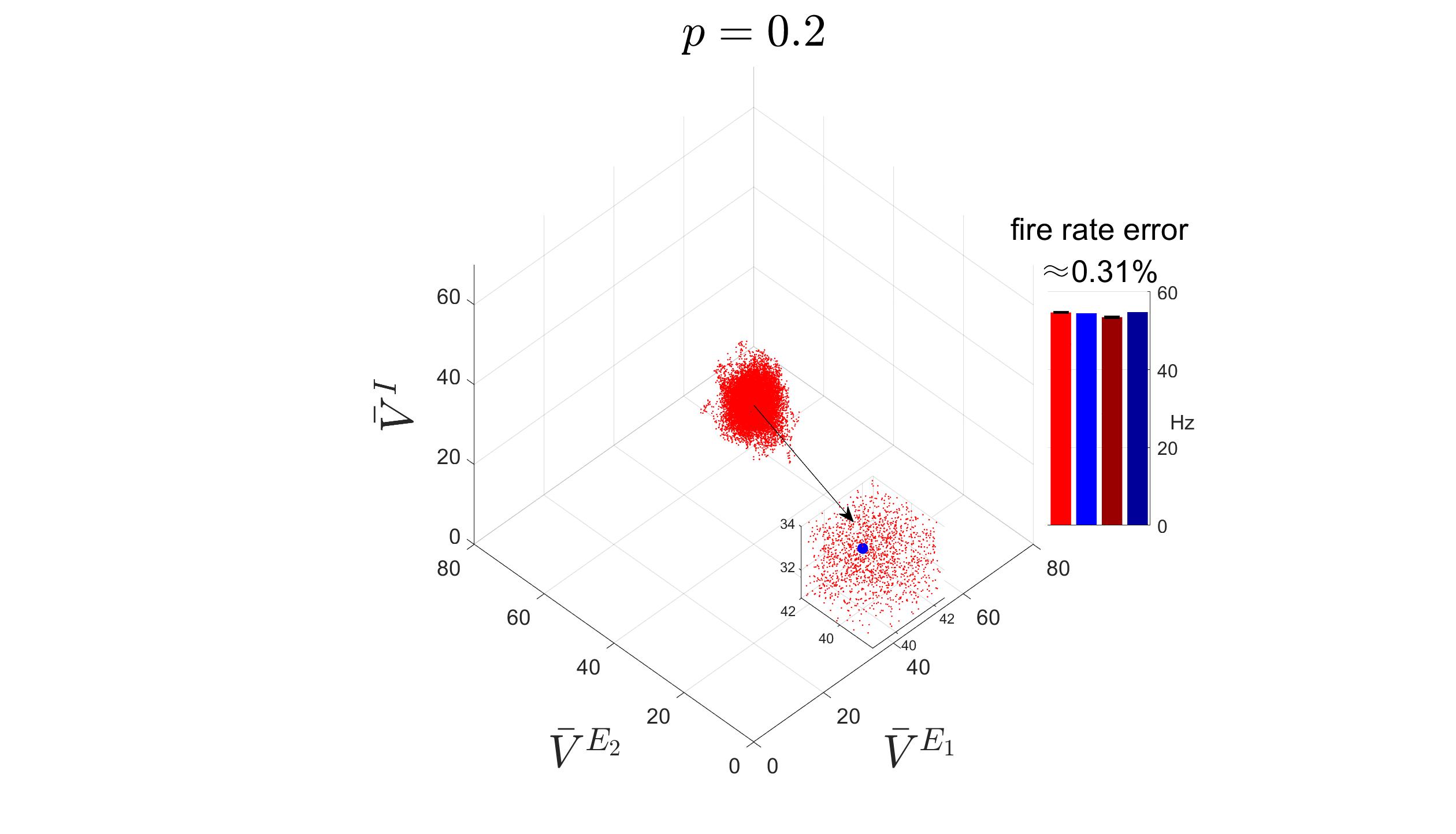}
      \end{subfigure}%
      \begin{subfigure}{.24\textwidth}
        {\bf F}\\
        \includegraphics*[bb=8.2in 1in 29.2in 20in,width=\textwidth]{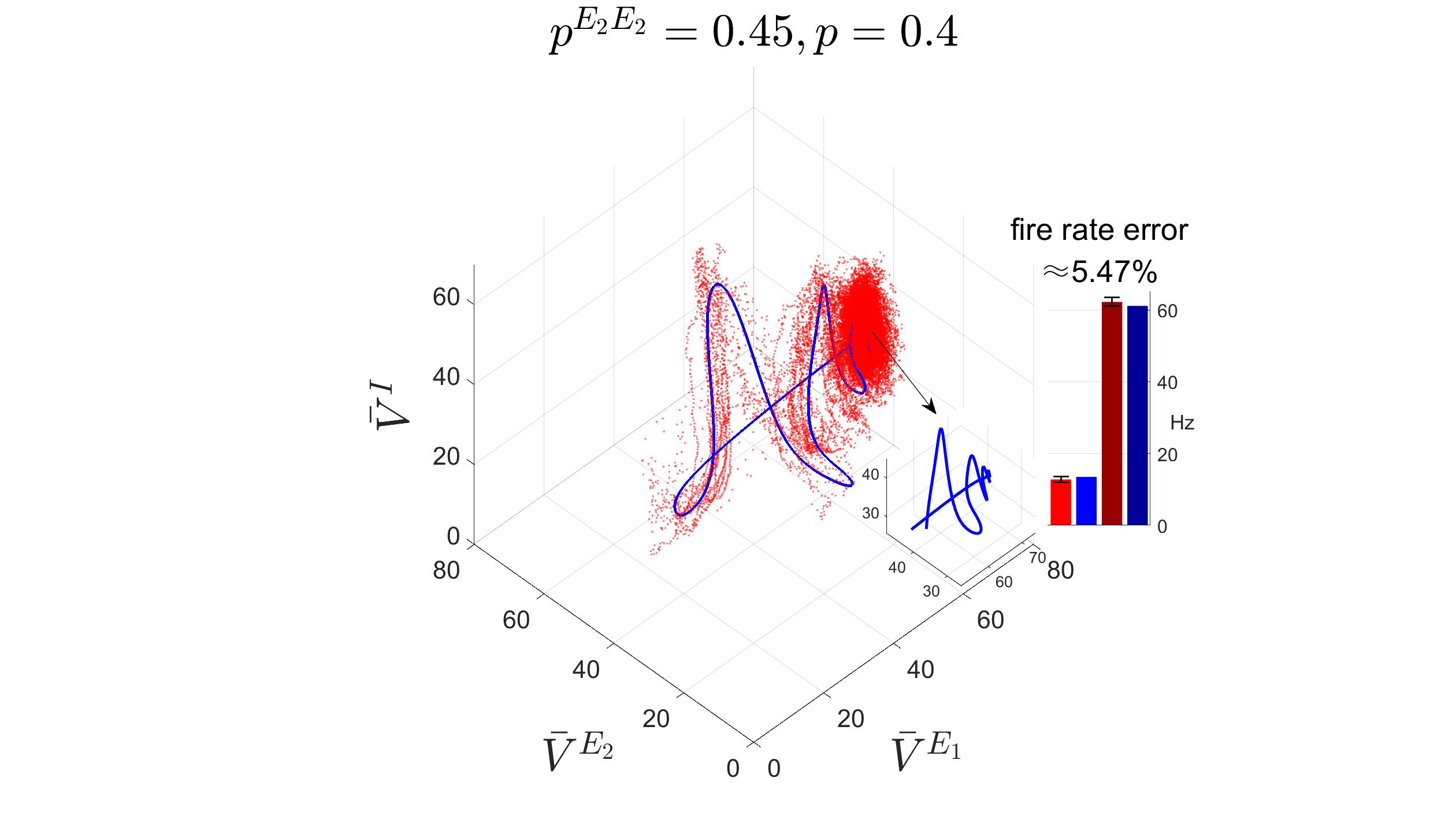}
      \end{subfigure}
    \end{subfigure}
    \caption{The dsODE model predicts the competition between multiple E-populations. Two competing E-populations mediated by one I-population are investigated by varying the E-to-E projection probabilities within each E-population. \textbf{A.} Comparison of initial transient dynamics for both E-populations ($E_1$: left panels; $E_2$: right panels; red curves: LIF dynamics; blue curves: dsODE predictions). \textbf{B.} Long-term dynamics of both E-populations are displayed. \textbf{C-F.} For each choice of E-to-E projection probabilities, trajectories of dsODEs (blue curves) and LIF networks (red dashed curves) for $t\in[1000,2000]$ ms are projected onto a subspace defined by $(\bar{V}_{E_1}, \bar{V}_{E_2}, \bar{V}_{I})$. Insets in \textbf{E} and \textbf{F} show detailed views of the attractors. The dsODE predictions of temporally averaged E-firing rates are shown in the bar maps (red: $E_1$ LIF; blue: $E_1$ dsODE; dark red: $E_2$ LIF; dark blue: $E_2$ dsODE). }
    \label{Fig6_2E1I_p}
\end{figure}

The dsODE system can also accurately replicate the competition between multiple E populations. For the four examples in Fig.~\ref{Fig2_LIF_dynamics}D, the dsODEs successfully capture both the initial transient dynamics for $t<100$ ms (Fig.~\ref{Fig6_2E1I_p}A) and the long-term behavior of both E populations (Fig.~\ref{Fig6_2E1I_p}B-F). In addition to simpler cases where one or neither of the E populations dominates all the time (first and third rows, Fig.~\ref{Fig6_2E1I_p}AB), the dsODEs also quantitatively predict the details of the switching dynamics (second and fourth rows, Fig.~\ref{Fig6_2E1I_p}AB).

When projected onto a 3D phase space consisting of $(\bar{V}_{E_1}, \bar{V}_{E_2}, \bar{V}_I)$ (Fig.~\ref{Fig6_2E1I_p}C-F), the attractors produced by the dsODEs (blue) closely follow the trajectories of the corresponding SNNs (red dashed lines). Furthermore, the dsODE attractors reveal crucial dynamic characteristics that are obscured by the randomness in the LIF networks. In the case where $p^{E_1E_1}$ is slightly larger than $p^{E_2E_2}$ (Fig.~\ref{Fig6_2E1I_p}F), the dsODE attractor shows that the switch from $E_1$ to $E_2$ is induced by a saddle point (inset). Specifically, after the first and largest spiking clusters of $E_1$, the trajectory approaches the saddle along its stable submanifold in a spiral. Subsequently, the network moves into the unstable submanifold, leading to the switch to $E_2$’s dominance. In contrast, the cause of the switch is much harder to discern from the LIF network trajectories.

In addition to small prediction errors in the average firing rates of both E populations ($<5$\%), a more striking result is that dsODEs also accurately predict the expected waiting time for transitions between the dominance of the E populations. For $p = 0.4$, the transition waiting time in the LIF model is $19.3\pm 6.2$ ms, while the dsODE predicts $15.4$ ms. For biased transitions, $t_{E_1\to E_2} = 13.3\pm 2.3$ and $t_{E_2\to E_1} = 132.2\pm 43.9$ ms, while the dsODE predicts 16.4ms and 100.4ms.

\subsection{Bifurcation analysis}
\label{Sect4.5-Bifurcation}
An ultimate test for the dsODE system is to perform a bifurcation analysis for the corresponding spiking neural networks (SNNs). Bifurcation analysis is a powerful tool in nonlinear dynamical systems, revealing how qualitative changes in a system’s behavior arise from small variations in parameters. However, this analysis is particularly challenging for dsODEs to detect the drastic changes in SNN dynamics that occur around bifurcation points, including multi-stability, oscillatory behaviors, or even chaotic dynamics, as shown by the examples in Sect.~\ref{Sect4.3-1E1I} and below. Therefore, capturing such delicate phenomena requires a mathematical theory that accurately reflects the underlying system dynamics, rather than one merely relying on data fitting. 

We perform bifurcation analysis on SNNs composed of a pair of E/I populations, focusing on a dynamical feature called the “beat index.” During oscillatory dynamics in SNNs, the size and duration of spiking clusters may vary over time. The beat index refers to the number of spiking clusters between the recurrence of two clusters with similar sizes and durations (see more details of the beat index in \cite{wu2022multi}). For example, in Fig.~\ref{Fig2_LIF_dynamics}C, the beat index is 1 (or “1-beat”) for $S^{EI} = 2.45$, but increases to 2-beat for $S^{EI} = 2.95$, where strong and weak clusters alternate. Similarly, $S^{EI} = 2.70$ produces 3- or 4-beat dynamics. Accurately predicting the beat index requires capturing SNN dynamics across different time scales—both within and between spiking clusters—making the corresponding bifurcation analysis a true test of the robustness and accuracy of dsODEs.

The visualization of beat indices in bifurcation maps has been detailed in our previous study \cite{wu2022multi}. Briefly, the size and duration of each spiking cluster are determined by the state of the SNN at the onset of the cluster, denoted by $t_i$. A key indicator of the SNN state is $\bar{m}(t_i) = \bar{V}_E(t_i) - \bar{V}_I(t_i)$, which represents the difference between the average membrane potentials of the E and I populations at time $t_i$. Intuitively, a large $\bar{m}$ implies an advantage of the E population in the E-I competition, leading to a larger spiking cluster, and vice versa. For visualization purpose, we track $\Delta \bar{m}(t_i) = \bar{m}(t_{i+1}) - \bar{m}(t_i)$, the difference between consecutive $\bar{m}$ values. The beat index is then represented by the number of peaks in the distribution of $\Delta \bar{m}$. For example, in 1-beat dynamics, $\Delta \bar{m} \approx 0$, while 2-beat dynamics alternate between two distinct $\bar{m}$ values, resulting in $\Delta \bar{m} = \pm |\bar{m}(t_{i+1}) - \bar{m}(t_i)|$, and so on. 

\begin{figure}[htbp]
\begin{subfigure}{\textwidth}
    \begin{subfigure}{0.5\textwidth}
    {\bf A}\\
        \includegraphics*[bb=2.5in 0in 33in 20in,width=.98\textwidth]{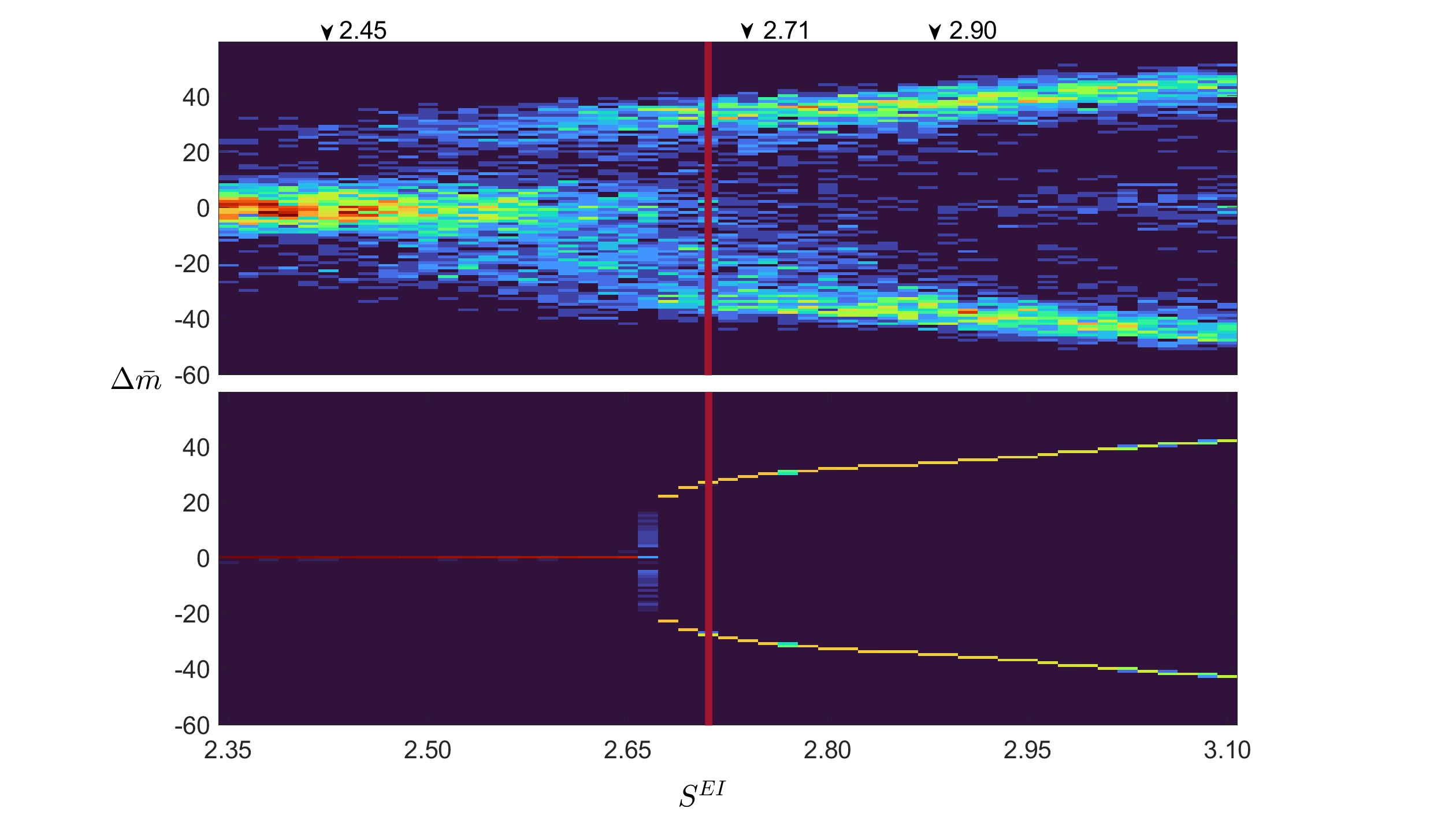}
    \end{subfigure}
    \begin{subfigure}{0.5\textwidth}
    {\bf B}\\
        \includegraphics*[bb=2.5in 0in 33in 20in,width=.98\textwidth]{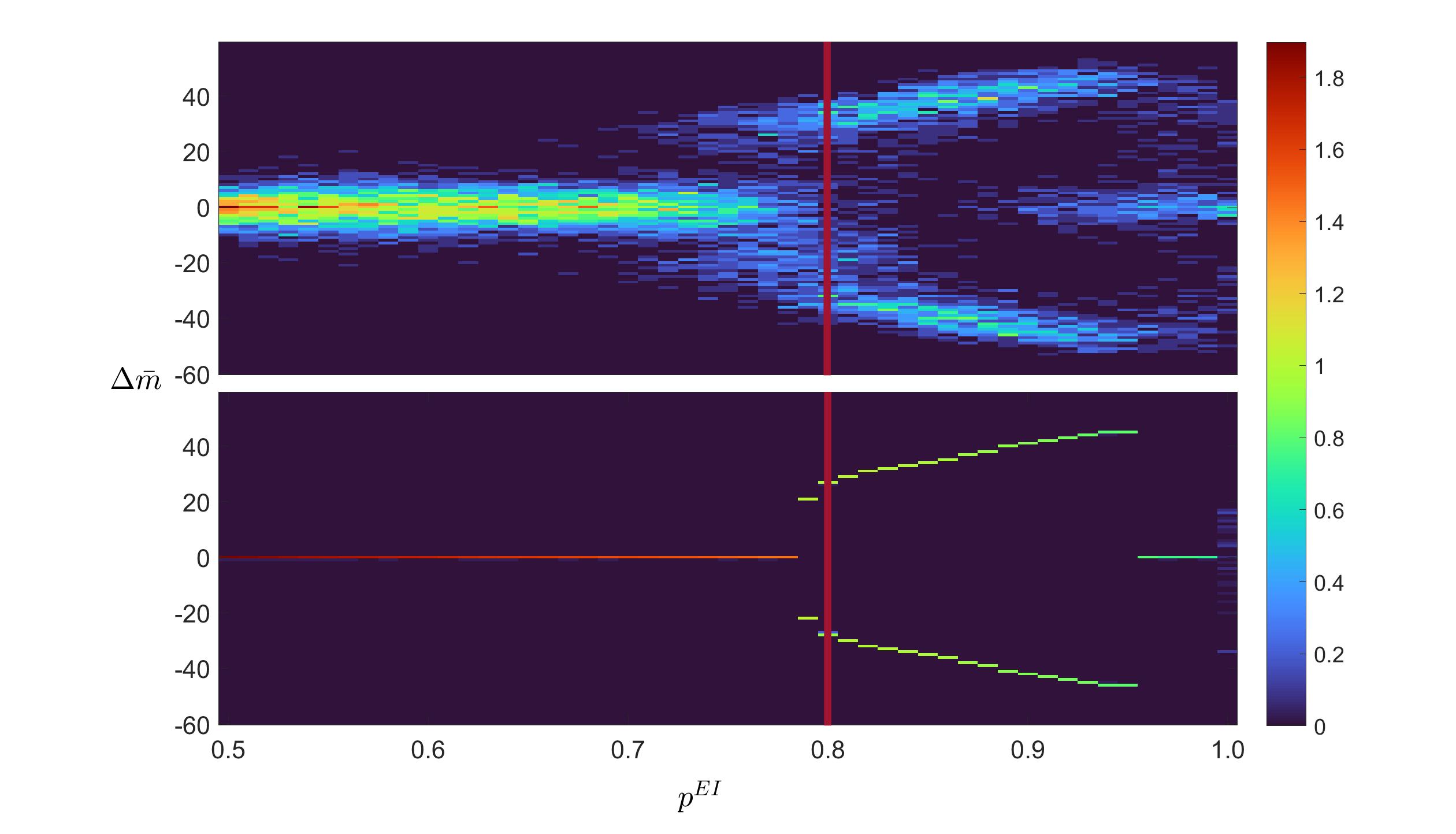}
    \end{subfigure}
\end{subfigure}\\[1ex]
    \begin{subfigure}{0.5\textwidth}
    {\bf C}\\
        \includegraphics*[bb=2.5in 0in 33in 19in,width=.98\textwidth]{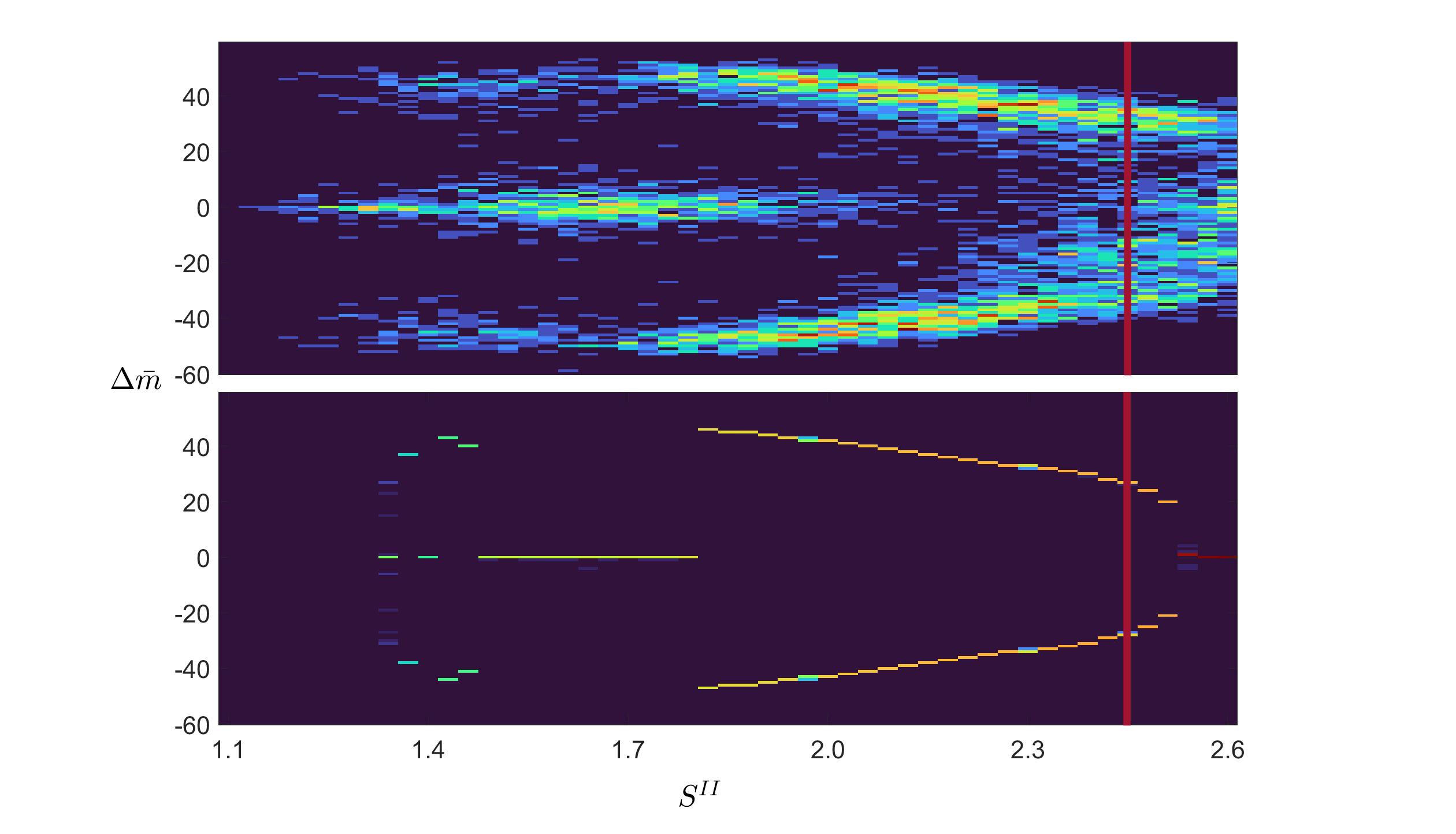}
    \end{subfigure}
    \begin{subfigure}{0.5\textwidth}
    {\bf D}\\
        \includegraphics*[bb=2.5in 0in 33in 19in,width=.98\textwidth]{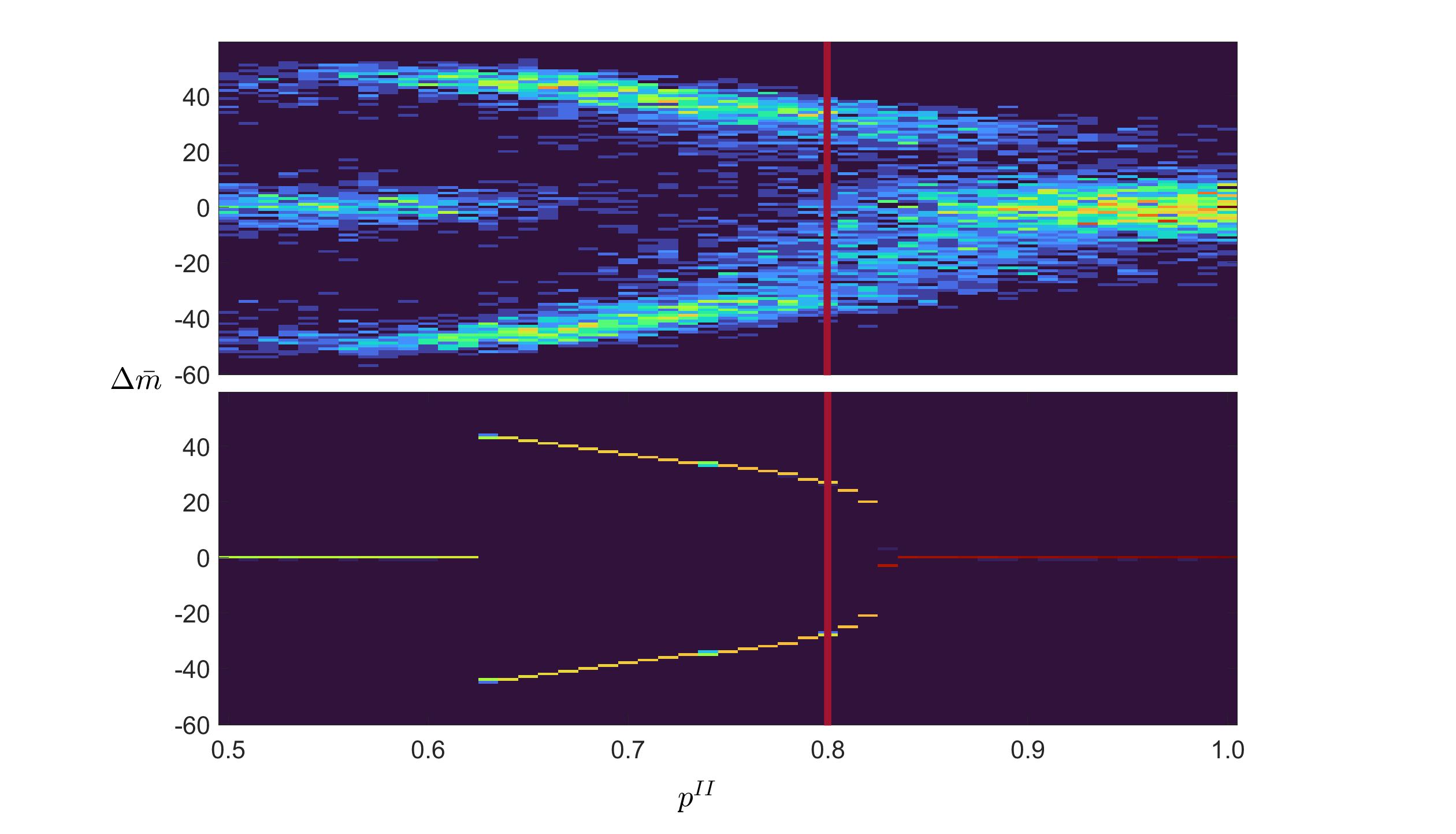}
    \end{subfigure}\\[1ex]
    \begin{subfigure}{0.5\textwidth}
    {\bf E}\\
        \includegraphics*[bb=2.5in 0in 33in 20in,width=.98\textwidth]{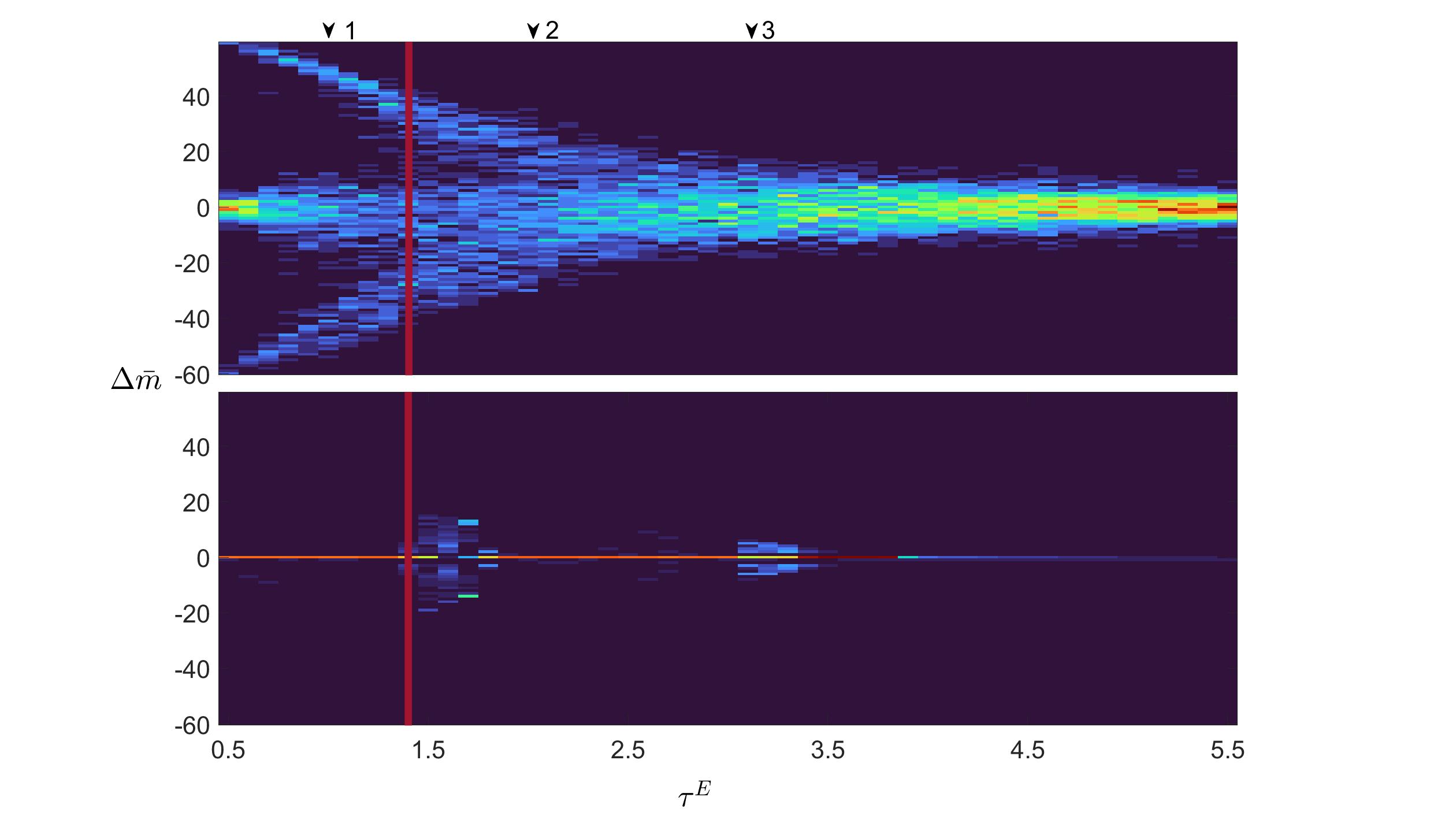}
    \end{subfigure}
    \begin{subfigure}{0.5\textwidth}
    {\bf F}\\
        \includegraphics*[bb=2.5in 0in 33in 20in,width=.98\textwidth]{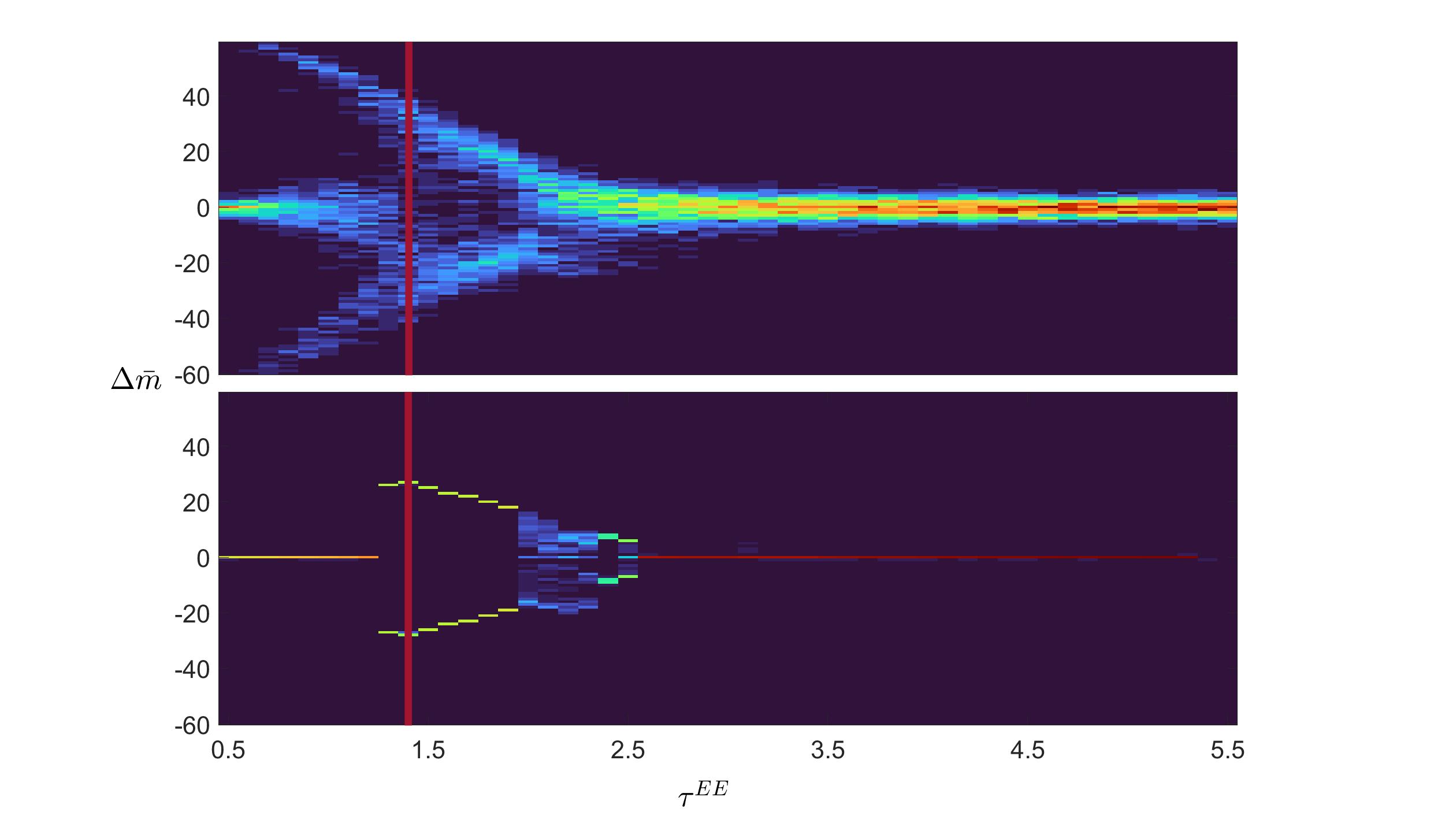}
    \end{subfigure}\\[1ex]
    \begin{subfigure}{0.5\textwidth}
    {\bf G}\\
        \includegraphics*[bb=2.5in 0in 33in 19in,width=.98\textwidth]{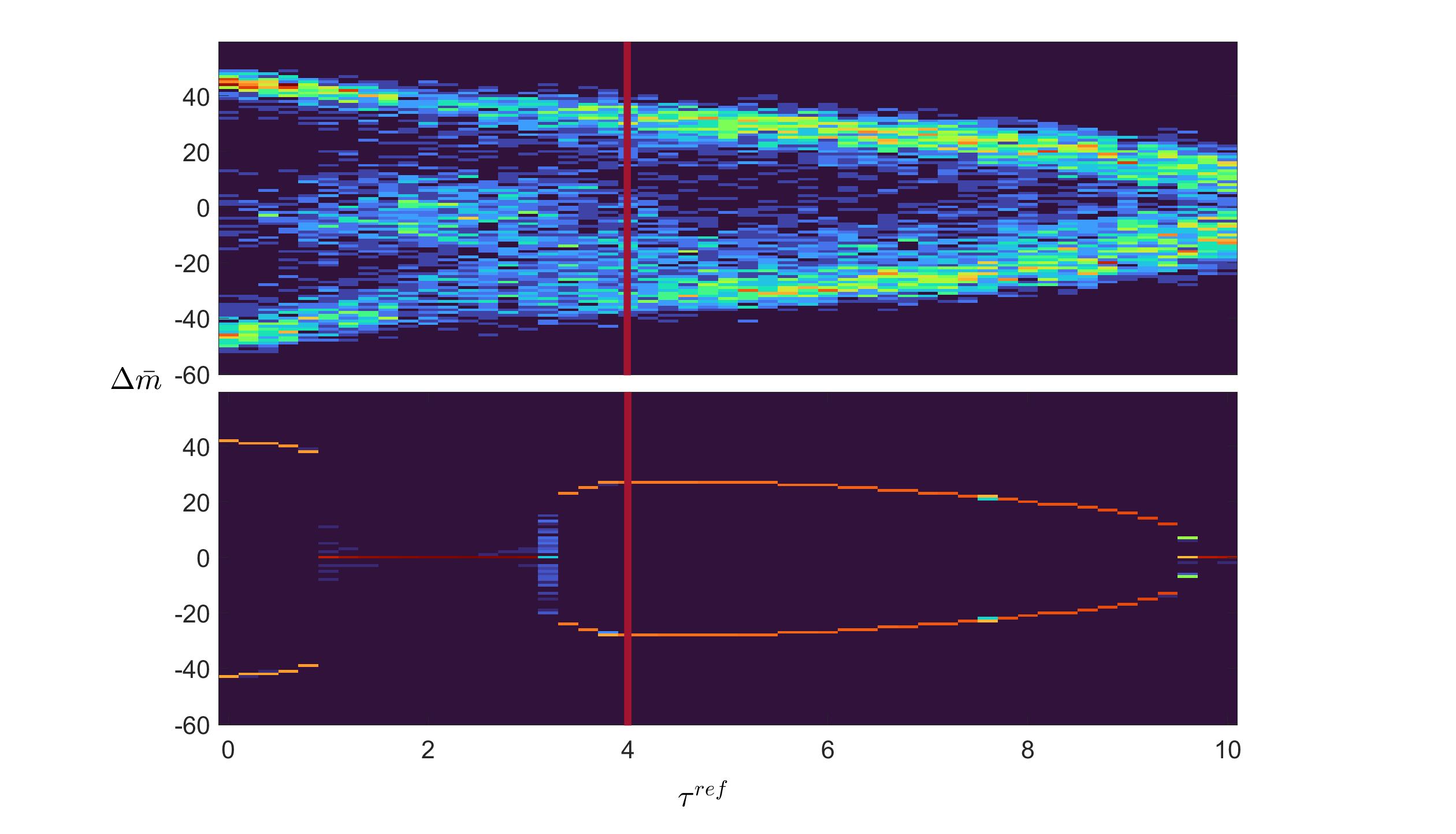}
    \end{subfigure}
    \begin{subfigure}{0.5\textwidth}
        {\bf H}\\
          \includegraphics*[bb=2.5in 0in 33in 19in,width=.98\textwidth]{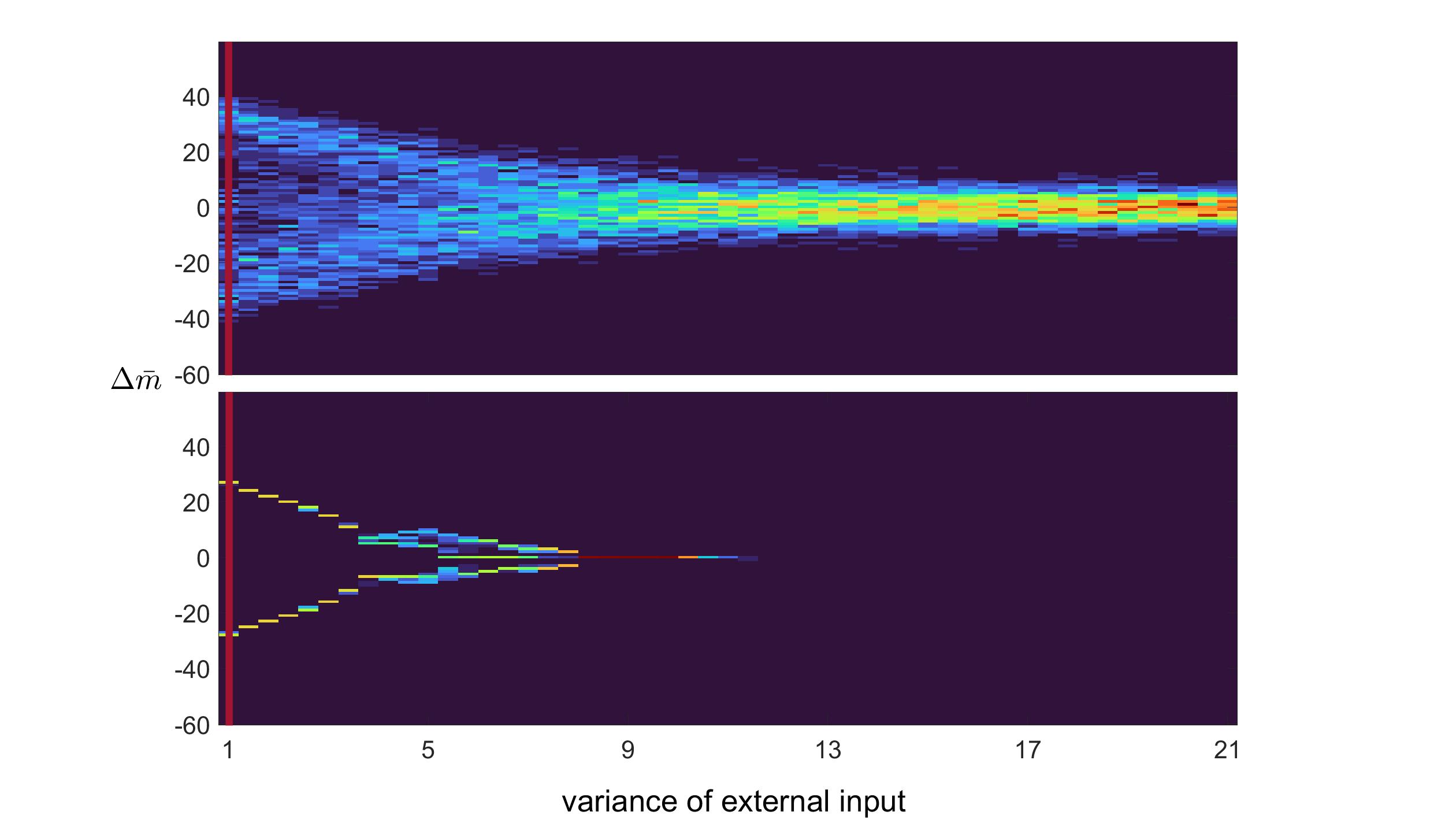}
    \end{subfigure}
    
    \caption{Comparing the bifurcation maps of spiking networks and dsODEs. The colormap represents the distribution of spiking clusters at different $\Delta \bar{m}$, which measures the change in the mean membrane potential difference across the onsets of consecutive spiking clusters. Red bars highlight the standard parameter setup when a particular parameter is varied. \textbf{A}: Varying $S^{EI}$, with the first three cases from Fig.~\ref{Fig5_1E1I_SEI} indicated by arrows. \textbf{B}: Varying $p^{EI}$. \textbf{C}: Varying $S^{II}$. \textbf{D}: Varying in $p^{II}$. \textbf{E}: Varying $\tau^{E}$, with the first three cases from Fig.~\ref{Fig4_1E1I_tau} indicated by arrows. \textbf{F}: Varying $\tau^{EE}$, i.e., only the synaptic timescale between E neurons is changed, while the timescale for E-to-I projections is fixed. \textbf{G}: Varying $\tau^{\rm ref}$. \textbf{H}: Varying the variance of external input (measured by the ratio to the original variance).}
    \label{Fig7_bifurcations}
\end{figure}

Figure~\ref{Fig7_bifurcations} presents bifurcation maps for eight key parameters of the spiking network, encompassing synaptic coupling weights, projection probability, and physiological timescales. Each map shows the density of spiking clusters, characterized by the difference in initial population voltage $\Delta \bar{m}$ from one cluster to the next. The colormaps, normalized to the standard Gamma frequency (40 Hz), indicate the frequency of spiking clusters. That is, the more intense red areas represent higher frequencies, and vice versa. Notably, regions with no density signify convergence of dsODEs to point attractors, implying the absence of spiking clusters in the temporal dynamics (e.g., for low $S^{II}$ and high variance in external inputs). In each bifurcation map, one parameter is varied while all others remain at their standard values as detailed in Table~\ref{Table1_Parameters}. A red vertical bar marks the standard value of the parameter being examined.

Overall, dsODEs (the bottom map in each panel) quantitatively predict the locations of bifurcation points and the number of branches for spiking networks (the top map in each panel), accurately reflecting the correct beat numbers. However, due to noise, the LIF network dynamics may deviate from the attractors predicted by dsODEs, leading to less regular bifurcation maps or even additional peaks in the $\Delta \bar{m}$ distribution. Examples include $S^{EI} < 2.65$ in Fig.~\ref{Fig7_bifurcations}A, $S^{II} > 2.45$ in Fig.~\ref{Fig7_bifurcations}C, and $p^{II} \in (0.50, 0.62)$. Additionally, branches that are close together in the dsODE predictions may also be blurred by noise in the LIF dynamics, such as $p^{EI} \in (0.5, 0.55)$ in Fig.~\ref{Fig7_bifurcations}B and $\tau_E \in (2.5, 3.1)$. Another interesting finding is the chaos-like behavior of dsODEs when varying $\tau^{EE}$, the synaptic timescale between E neurons, while keeping the synaptic timescale for E to I projections fixed ($\tau^{EE} \in (1.75, 2.40)$). 

Despite the discrepancies above, we stress that the bifurcation maps predicted by the dsSDE scheme are strikingly similar to LIF networks (Fig.~\ref{FigS1_bifurcations_of_dsODE_noise}). We find that adding noise terms back to the dsODE system successfully recovers the extra branches induced by the noise, as well as the blurring effect.

\subsection{Finite-N effect}
\label{Sect4.6-FiniteN}

\begin{figure}[htbp]
  \begin{center}
    \begin{subfigure}{\textwidth}
      {\bf A}\\
          \includegraphics*[bb=1.5in 0.5in 34in 19.2in,width=.98\textwidth]{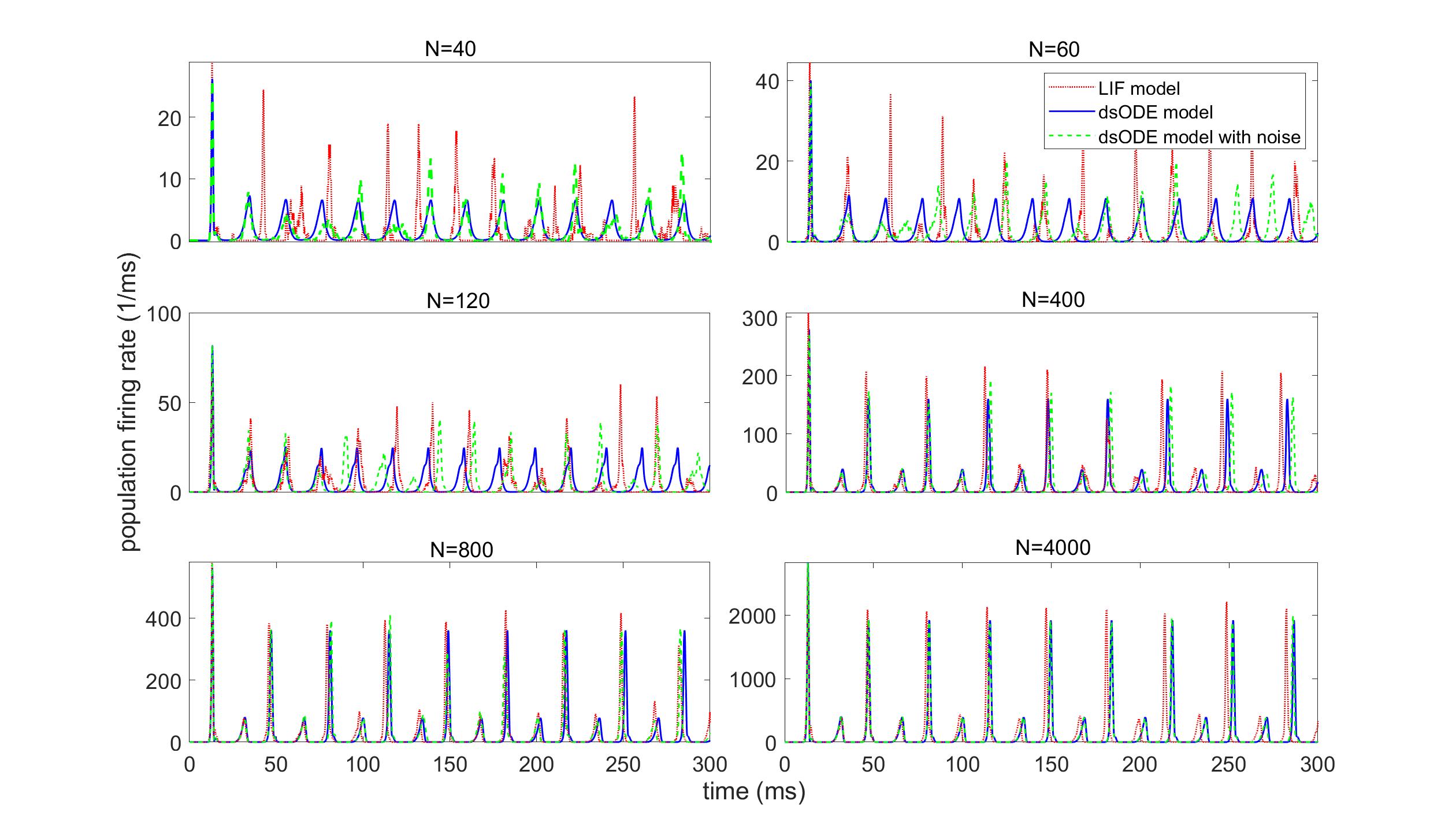} 
    \end{subfigure}\\[1ex]%
    \begin{subfigure}{\textwidth}
      \begin{subfigure}{.5\textwidth} 
        {\bf B}\\
        \includegraphics*[bb=3in 10in 33in 19.2in,width=\textwidth]{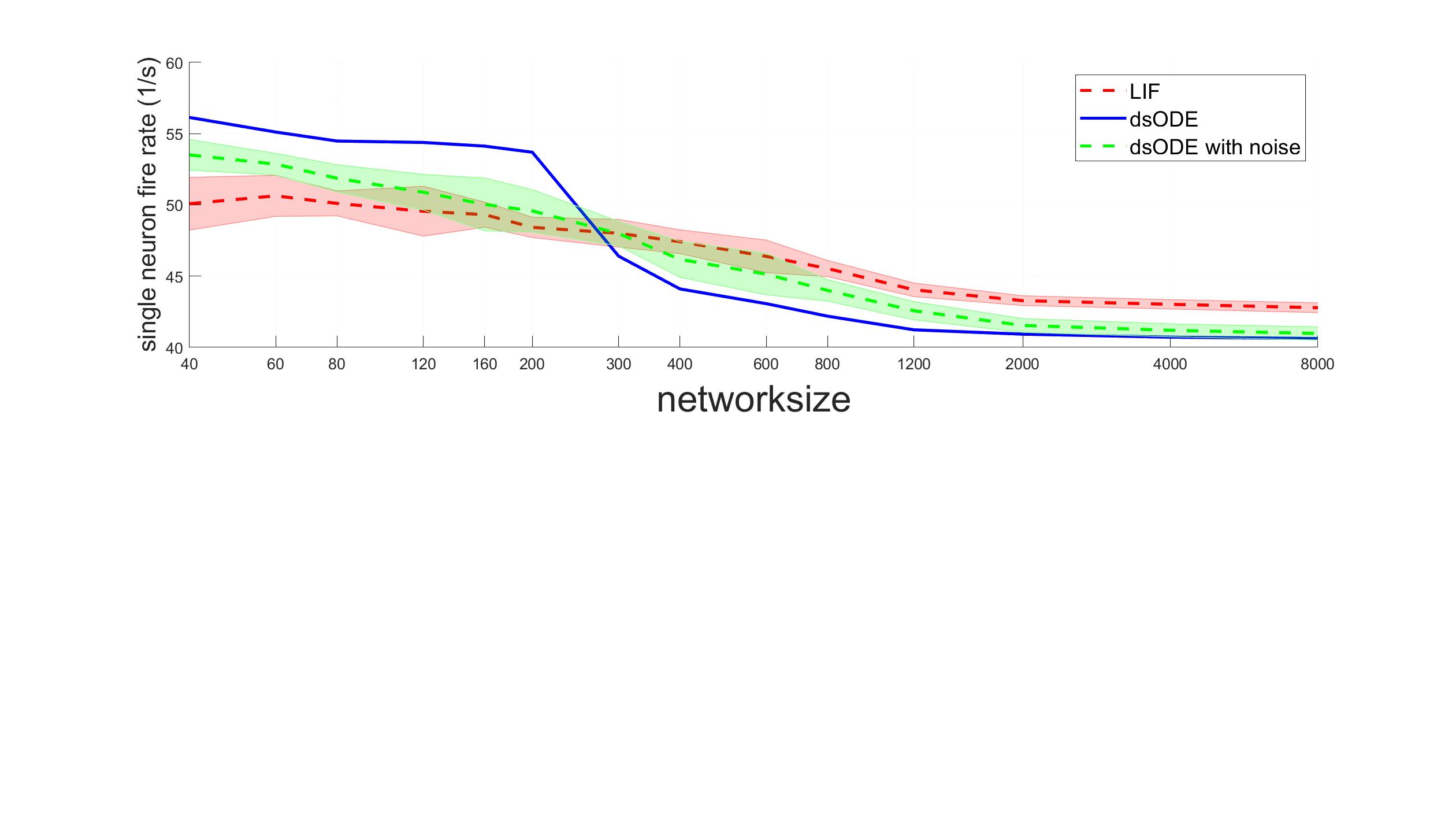}
      \end{subfigure}%
      \begin{subfigure}{.5\textwidth}
        {\bf C}\\
        \includegraphics*[bb=3in 0.5in 33in 9.7in,width=\textwidth]{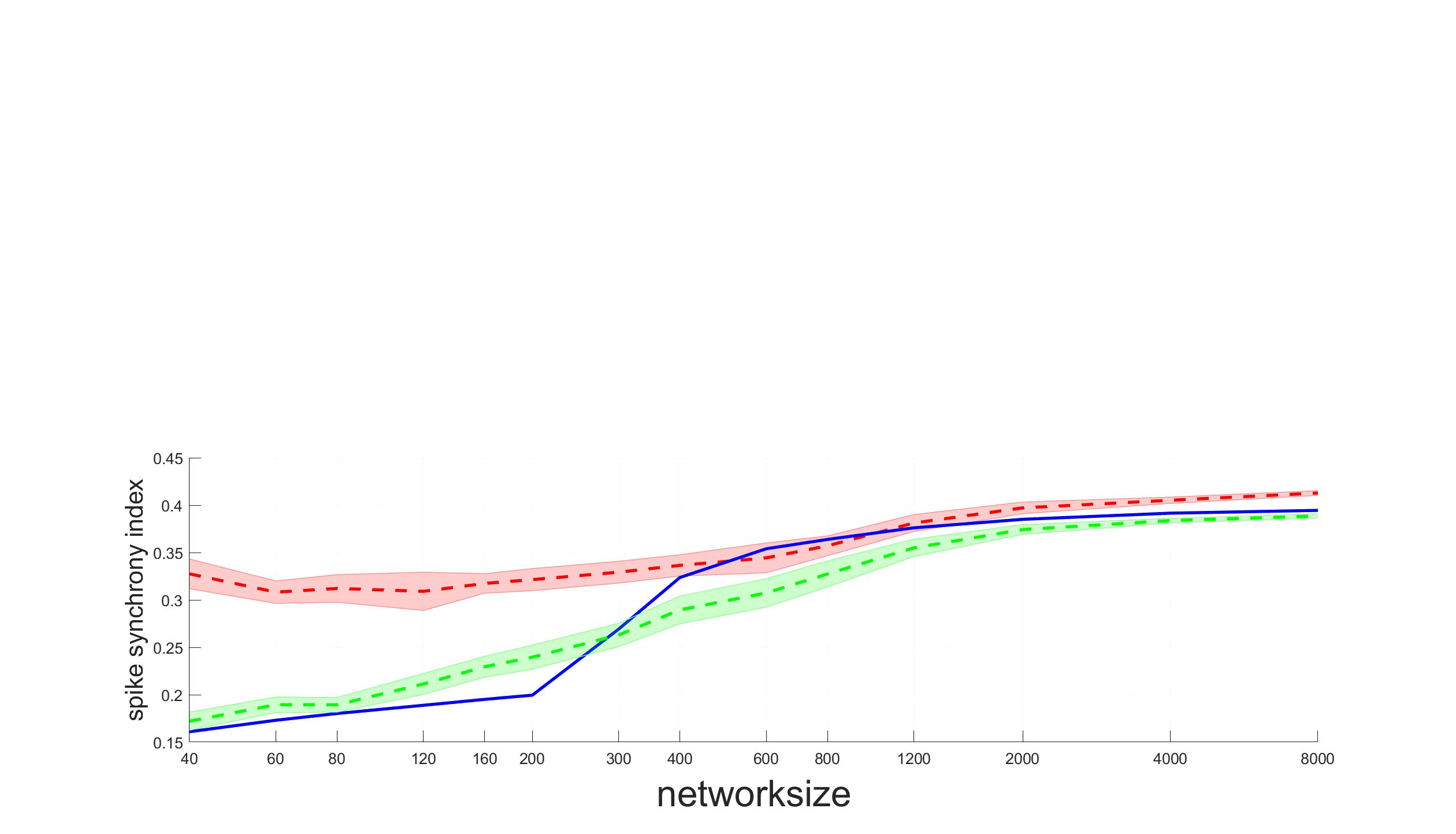}
      \end{subfigure}
    \end{subfigure}
    \caption{Adding noise to dsODE helps recover the finite-neuron effect. \textbf{A.} Predictions of the dsODE model (blue curves) and dsSDE model (green dashed curves) across different network scales. \textbf{B.} Single-neuron firing rates produced by LIF networks, dsODE, and dsSDE as the network scale varies. \textbf{C.} Changes in the Spike synchrony index with the network scale. In both \textbf{B} and \textbf{C}, the dsSDE model provides more accurate predictions compared to the dsODE model.} 
    \label{Fig8_finiteN}
  \end{center}
\end{figure}

Neural networks with finite numbers of neurons can exhibit intrinsic fluctuations and variability \cite{schwalger2017towards}. These finite-neuron effects are critical for accurately capturing the dynamics of real neural systems (e.g., minicolumns in the cerebral cortex consisting of only $O(10-10^3)$ neurons).

Traditional mean-field approaches often assume $N \to \infty$ and weak coupling, neglecting the stochastic fluctuations present in finite systems \cite{brunel2000dynamics,cai2006kinetic,MontbrioEtAl2015}. In previous studies, finite-neuron effects were typically explored through perturbations of mean-field theory \cite{el2009master,bressloff2010stochastic,trousdale2012impact,buice2010systematic,buice2013dynamic}. However, this approach often results in oversimplification. First, finite-neuron effects may induce synchronization properties and oscillatory behavior that are absent in the corresponding mean-field equations. Second, finite-size fluctuations become more pronounced around critical points in parameter space, potentially leading to transitions between different types of dynamics, as shown in Sect.~\ref{Sect4.3-1E1I} and \ref{Sect4.4-2E1I}. Moreover, case studies in \cite{joglekar2019case} also demonstrated that scaling down a large SNN without parameter adjustment can lead to catastrophic results.

The finite-neuron effect arises from the discrete nature of neurons and spiking events. The impact of spikes is “quantum” rather than continuous. Therefore one spike alone may significantly alter the network dynamics, and the variability of individual neurons is not negligible. Generally, any mathematical theory of SNNs (including dsODE) based on differential equations must violate this discrete nature to some extent. This violation can sometimes be mitigated by adding noise terms back to the equations, such as the $\sqrt{\frac{f(t)}{N}}$ white noise used in previous studies, where $f(t)$ is the average firing rate of single neurons. However, the effectiveness of this modification depends on two conditions: 1) whether the mathematical theory accurately captures the nonlinear network dynamics and the non-white stochasticity in $f(t)$ itself, and 2) whether the number of neurons is “small but large enough” for the Gaussian approximation to hold for the randomness in spike production.

In dsODE theory, the neuron number N plays a role in two key aspects. First, when $f^R(t)$ is constant, Eq.~\ref{Sect3.1-Eq10-H} gives $D^{QR}/\mu^{QR} = \frac{1}{2} + h^{QR}(1 - p^{QR})$. To maintain the network firing rate as $N \to \infty$, we set $h^{QR} \sim O(\frac{1}{N^R})$ while fixing $p^{QR}$. As a result, $\mu^{QR}$ converges to a non-zero constant, and there is an $O(\frac{1}{N^R})$ deviation in the variance of recurrent input from the standard Ornstein-Uhlenbeck process. Second, the noise terms in Eq.~\ref{Sect3.2-Eq19-frNoise_Redu} and \ref{Sect3.2-Eq19-vfluxNoise-Redu} are related to \(N^Q_{\mbm^{th}}\), the number of neurons in bin \(\mbm^{th}\), which is proportional to $N^Q$. When $N^Q$ is very small, the Gaussian approximations for these noise terms break down.

To assess how well dsODE captures the finite-neuron effect, we conducted a case study using parameter settings that produce the 2-beat dynamics in Fig.~\ref{Fig2_LIF_dynamics}C ($S^{EI} =$ 2.95 and $N =$ 400). When scaling the network size, we kept $S^{QR}_N = S^{QR}\cdot 400/N$. We found that the pure ODE system predictions for dynamics, firing rates, and synchrony (quantified as the spike synchrony index, see definition in \cite{rangan2013emergent}) were satisfactory for $N > 400$ (blue curves in Fig.~\ref{Fig8_finiteN}). Notably, the average firing rate per neuron was not constant for different $N$, making these results non-trivial. However, for very small N (e.g., $N = 40$ or 60), the amplitude of oscillatory dynamics and firing rates predicted by the ODE system were significantly different from the corresponding SNNs. Nevertheless, these discrepancies were partially mitigated by the dsSDE model, i.e., adding the noise terms from Eq.~\ref{Sect3.2-Eq19-frNoise_Redu} and \ref{Sect3.2-Eq19-vfluxNoise-Redu} back to dsODEs (green dashed curves in Fig.~\ref{Fig8_finiteN}).

A comprehensive investigation of the dsODE system’s performance with respect to finite-neuron effects is beyond the scope of this paper, though our case study shows promising results and can be potentially extended to other parameter regimes. Nonetheless, our results suggest that the CG-Markov model is a strong candidate for retaining finite-neuron effects, as it does not violate the discrete nature of neurons and spikes, warranting further in-depth analysis.

\section{Comparison between dsODEs and previous population methods}
\label{Sect5-Comparison}
In this section, we discuss the relationships and differences between dsODEs and several classical approaches. We begin by comparing dsODEs with rate models (e.g., the Wilson-Cowan equations, \cite{wilson1973mathematical}) and Fokker-Planck (FP) equations \cite{brunel2000dynamics,cai2006kinetic,vinci2023self}, two of the most widely used population-level methods for modeling spiking neural networks. We will also examine the dsODE framework in contrast with the refractory density method (RDM, \cite{schwalger2017towards}), a recent breakthrough in the field. Similar to dsODEs, RDM does not rely on the assumptions of an infinite number of neurons or weak coupling weights, making it the first effective mathematical theory for modeling local neural circuits under realistic conditions. 

We further compare the predictive power of dsODEs, RDM, and a recent variant of FP  equations that incorporates finite-neuron effects \cite{vinci2023self}. For simplicity, we focus on the long-term average firing rate
\begin{align}
\label{Sect5-Eqn20-FRate}
    \bar{f} = \frac1T\lim_{T\to\infty}\int_{0}^{T}f(t)\,\mathrm{d}t.
\end{align}
which is the most fundamental dynamical statistics and a prerequisite for any mathematical theory aiming to capture the detailed dynamics of spiking networks. Since the LIF equations in \cite{schwalger2017towards,vinci2023self} are formulated slightly differently, we choose to adapt our dsODEs. See simulation details of LIF networks, dsODEs, RDM, and FP equations in Appendix.

\subsection{Comparison with rate models}
Starting from the seminal work by Wilson and Cowan \cite{wilson1972excitatory,wilson1973mathematical}, rate models have become perhaps the most widely used population method among brain and neuronal network modelers. Rate models are typically employed to describe the average firing rates of neuronal populations without accounting for the precise timing of individual spikes. For example, a typical rate model for a network with a pair of E/I populations can be formulated as:
\begin{subequations}
    \label{Sect5.1-Eqn22-RateModel}
    \begin{eqnarray}
        \tau^E\ddt{f^E} &=& -f^E + F^E(w^{EE}f^E + w^{EI}f^I+ I^E) \, , \\
        \tau^I\ddt{f^I} &=& -f^I + F^I(w^{IE}f^E + w^{II}f^I + I^I)\, ,
    \end{eqnarray}
\end{subequations}
where $f^{E,I}(t)$ denote the firing rates of E/I populations, $I^{E,I}(t)$ represent external inputs to both populations, and $w^{QR}$ are the coupling strengths between populations. Due to their lack of consideration for the detailed physiological processes of individual neurons, the interaction kernels for firing rates, $F^{E,I}$, are often assumed for analytical convenience or chosen to reflect specific dynamical regimes (e.g., rectified linear, sigmoid functions, and so on). Moreover, the reaction timescales, $\tau^{E,I}$, are typically chosen arbitrarily, even though in an SNN they should depend on many intrinsic features of the E/I populations (e.g., network architecture, neuronal adaptation, etc.) and may vary significantly under different dynamic regimes. Overall, rate models provide a macroscopic view of neural population activities. They are efficient for large-scale simulations and can offer rich analytical insights, but they are not designed to capture the fine temporal dynamics in SNNs and physiological details of single neurons.

To illustrate this limitation, we perform a 1D parameter scan of $w^{EI}$, which is analogous to the I-to-E synaptic coupling weight $S^{EI}$ in Table~\ref{Table1_Parameters}. As shown in Fig.~\ref{FigS2_Rate_Model}, the dynamics of Eq.~\ref{Sect5.1-Eqn22-RateModel} undergo two Hopf bifurcations as $w^{EI}$ varies, transitioning first from a stable fixed point to a one-beat oscillation, and then returning to a stable fixed point. In contrast to the example shown in Fig.~\ref{Fig2_LIF_dynamics}C, the rate model does not exhibit more complex dynamical structures, such as 2-beat oscillations or transitions between homogeneous and synchronous states. This limitation arises from the low dimensionality of the rate model: the “2-beat” dynamics are represented by trajectories with two overlapping cycles (Fig.~\ref{Fig5_1E1I_SEI}CD), which require at least three dimensions in an ODE system. Due to its formulation, Eq.~\ref{Sect5.1-Eqn22-RateModel} is inherently restricted to two dependent variables, limiting its ability to represent higher-dimensional dynamics.

\subsection{Comparison with the Fokker-Planck method}
Using principles from statistical physics, the Fokker-Planck (FP) equation describes the temporal evolution of the probability density function for the states of neurons, typically under the assumption $N \to \infty$. For example, the FP equation for a single E-population of LIF neurons can be formulated as follows (by removing the inhibitory terms from Eq.~\ref{Sect2.1-Eqn1-LIF}):
\begin{align}
\nonumber
    \partial_t \rho(v,g^E,t) 
    =\, &  \partial_v\left\{\left[g^{E\rm{leak}}v + g^E\left(v-\varepsilon^{E}\right)\right]\rho\right\} + \lambda^E\left[\rho\left(v-S^{E\rm{ext}},g^E,t\right) - \rho(v,g^E,t)\right] \\
    \label{Sect5-eq21-FP}
    & + \partial_{g^E}\left(\frac{g^E}{\tau^E}\rho\right) + pf^E(t)N^E\left[\rho\left(v,g^E-\frac{S^{EE}}{\tau^E},t\right) - \rho(v,g^E,t)\right],
\end{align}
where the firing rate is determined by the probability flux crossing $V^{th}$: 
$$f^E(t) = \int_0^\infty \mathrm{d}g^E\cdot\left\{-\left[g^{E\rm{leak}}v + g^E\left(v-\varepsilon^{E}\right)\right]\rho + \lambda^E\int_{V^{th}-S^{E\rm{ext}}}^{V^{th}}\mathrm{d}v\cdot\rho\left(v-S^{E\rm{ext}},g^E,t\right)\right\}.$$

FP equations and their variants (such as kinetic theories) have been highly successful in making quantitative predictions for various regimes of neural dynamics. However, these methods are constrained by the assumption $N \to \infty$, which is necessary to justify the probability density function $\rho(v,g^E,t)$. This leads to two main limitations. First, Eq.~\ref{Sect5-eq21-FP} requires $S^{EE} \to 0$, so that the impact of individual spikes becomes negligible (a condition often referred to as the “weak-coupling” assumption in many studies), thereby enabling the derivation of a differential equation. Second, Eq.~\ref{Sect5-eq21-FP} requires a low degree of synchrony, so that the recurrent input can be approximated as a Poisson process. Otherwise, the equation would blow up when a significant fraction of the population fires within a short time window, such as in the spiking clusters observed in Fig.~\ref{Fig2_LIF_dynamics}. In fact, analytical work has shown that Eq.~\ref{Sect5-eq21-FP} may blow up when $\tau^E$ is small \cite{caceres2014beyond}. Furthermore, implementing the fire-and-reset physiology necessitates complex boundary conditions. When $\tau^{\rm{ref}} = 0$, the probability flux crossing $V^{th}$ must be directly added back to the rest potential $\varepsilon^{\mathrm{rest}} = 0$, and for non-zero refractory periods, this can introduce additional analytical challenges.

The dsODE system shares several conceptual similarities with the FP method and its variants but differs in two key aspects. First, dsODE does not require $N \to \infty$ and, consequently, is not limited to the weak coupling regime. Instead, it directly counts the number of neurons in each state, thereby accounting for finite-neuron effects. Second, unlike the blowup behavior of FP equations, dsODE successfully captures different levels of network synchrony (see Sect.~\ref{Sect4.3-1E1I}). This robustness is attributed to dsODEs being based on discrete neuron counts rather than continuous density functions. The difference arises in their derivations: FP methods require $\delta v \to 0$ at first to obtain the probability distribution over a continuous domain of $v$, thus needing to assume low levels of correlation between recurrent inputs in a finite $\delta t$. In contrast, dsODEs first take $\delta t \to 0$, while sacrificing the “spatial resolution” of the membrane potential distribution by introducing discrete $v$-states. This trade-off relieves dsODEs from the low-synchrony assumption, enabling them to handle varying levels of transient synchrony. Additionally, the use of discrete $v$-states accommodates the fire-and-reset physiology (even when $\tau^{\rm{ref}} > 0$) without introducing cumbersome boundary conditions.

On the other hand, the form of dsODEs converges to FP equations when $N \to \infty$ and the bin size $a \to 0$ (while ignoring all noise terms). However, whether the dynamics of dsODEs converge back to those of FP equations is beyond the scope of this study. Nevertheless, we point out that the core assumption of dsODEs is to decorrelate $v$ and $g^E$. Conversely, the kinetic equation (Eq.~4.19 in \cite{cai2006kinetic}) chooses not to decorrelate $v$ and $g^E$, instead employing dimensional reduction techniques such as moment closures. This allows the kinetic theory to better capture low-firing rate regimes (e.g., background activity) where our Assumption 1 may not hold.

Finally, we compare the predictive power of dsODEs to a recent FP-based method by Vinci et al. \cite{vinci2023self}. In this study, Vinci et al. derived a self-consistent stochastic partial differential equation for an FP model of single finite-size E-populations. They first derived the power spectrum of finite-size noise in the firing rate for an uncoupled population and then extended it to coupled populations using a Markov embedding technique. Their results worked well for the parameter regimes they investigated when $N > 1000$. However, we note that Vinci et al. restricted their study to relatively weak coupling regimes (despite their claim to have tested strong coupling). In their setup, the total contribution of a spike to the membrane potentials of all postsynaptic neurons, $KJ$, is restricted to $<0.6(V^{th}-V^{\rm rest})$, where $K$ is the number of postsynaptic contacts per neuron and $J$ is the synaptic coupling weight. This value is significantly lower than the approximately $10(V^{th}-V^{\rm rest})$ observed in the macaque V1 model \cite{chariker2016orientation}, and the tested range in our study is approximately $[4.5, 54]\cdot(V^{th}-V^{\rm rest})$.

\begin{figure}[htbp]
  \begin{center}
    \begin{subfigure}{\textwidth}
      {\bf A}\\
          \includegraphics*[bb=2.5in 0.8in 34in 18.7in,width=.98\textwidth]{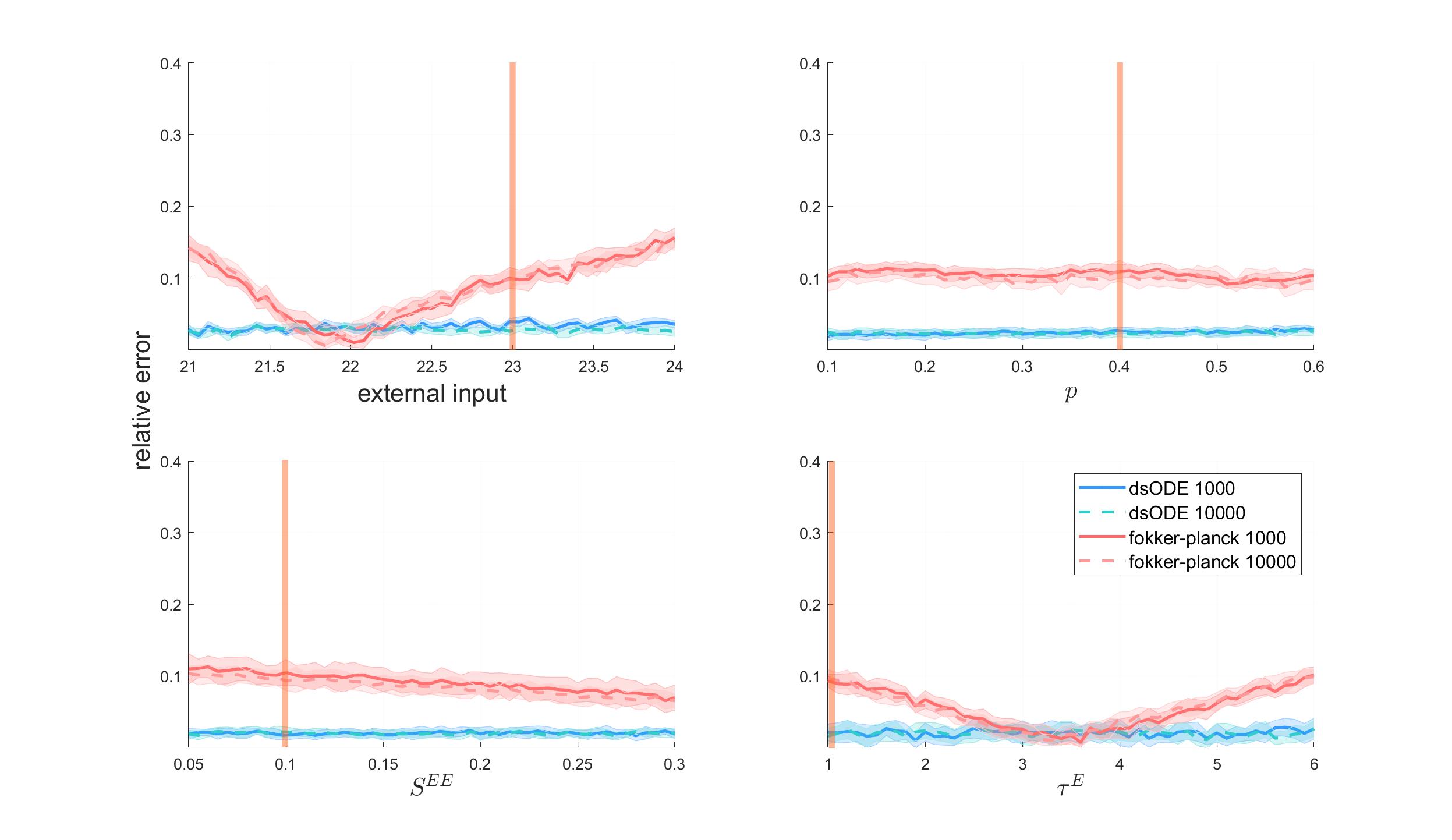} 
    \end{subfigure}\\[1ex]%
    \begin{subfigure}{\textwidth}
        {\bf B}\\
        \includegraphics*[bb=2.5in 0.8in 34in 18.7in,width=.98\textwidth]{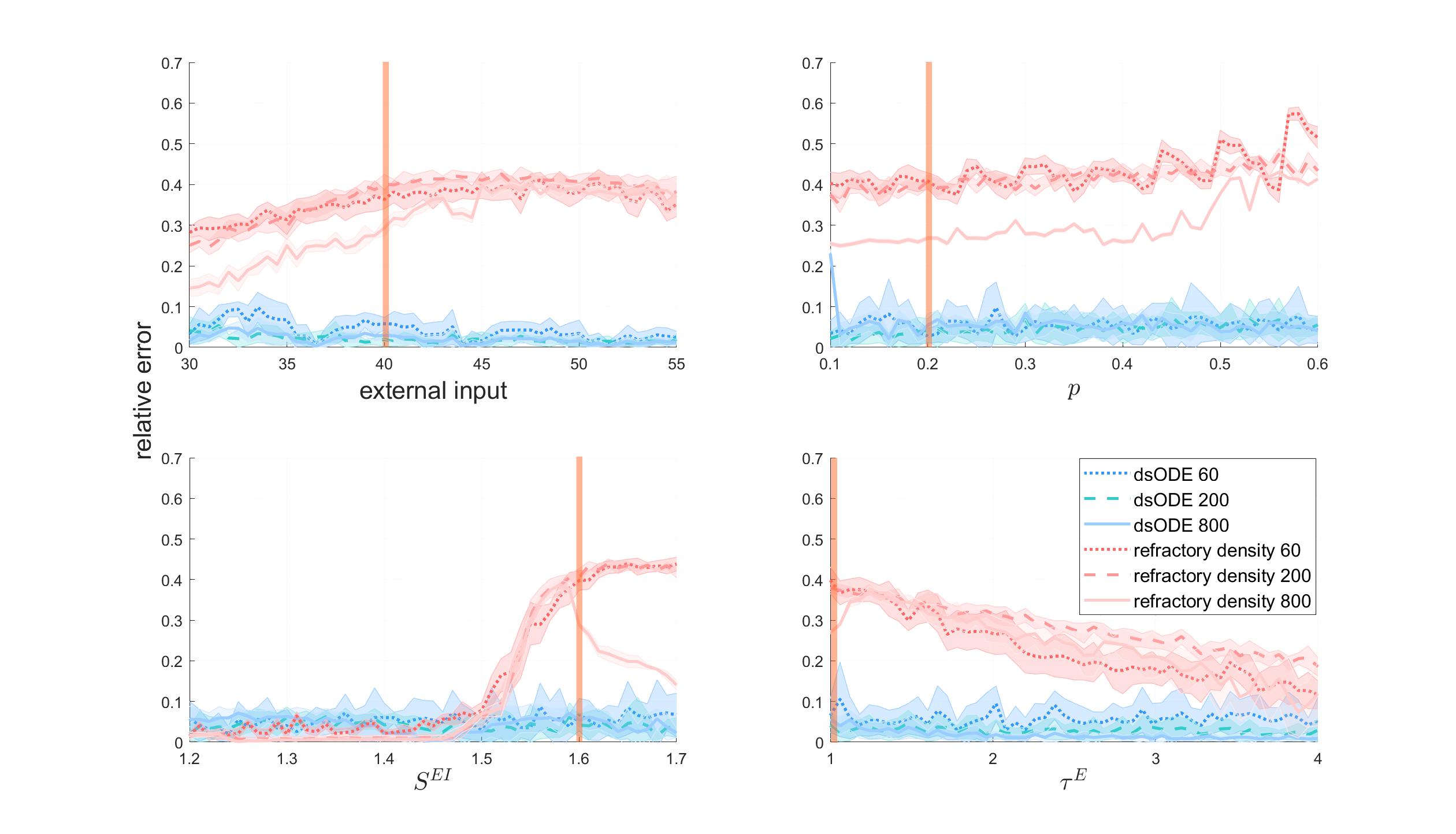}
      \end{subfigure}%
    \caption{The dsODE model yields better predictions to firing rates than Fokker-Planck methods and refractory density methods. Four parameters are varied for the comparison with each method, and the standard parameter choices are indicated by vertical bars. \textbf{A.} Comparison between the Fokker-Planck model and dsODE as parameters vary. \textbf{B.} Comparison between the refractory density model and dsODE across varying parameters.}
    \label{Fig9_Comparison_with_previous_methods}
  \end{center}
\end{figure}

We tested both dsODE and Vinci et al.’s method on the long-term firing rates of single E-populations with $10^3$ and $10^4$ neurons on our parameter sets with stronger couplings between neurons. We choose to vary four parameters, one representing a kind: $\lambda^E$ (rate of external stimuli), $p^{EE}$ (projection probability), $S^{EE}$ (synaptic coupling weight), and $\tau^E$ (synaptic timescale). The ranges of tested parameters are shown in Fig.~\ref{Fig9_Comparison_with_previous_methods}A, with other parameters listed in Table~\ref{Table1_Parameters}. We found that Vinci et al.’s FP method can yield a predictive error exceeding 10\% for firing rates, whereas dsODEs achieve an error less than 5\%.


\subsection{Comparison with refractory density methods}

The Refractory Density Method (RDM) and the dsODE system share a number of conceptual similarities, as both are designed to model homogeneous networks and utilize coarse-graining techniques to simplify the underlying dynamics of spiking neural networks (SNNs). Additionally, both methods discretize neuron states and deal directly with the number of neurons in each state, rather than relying on a continuous density function. This shared feature enables both methods to capture finite-neuron fluctuations, a crucial aspect of accurately representing the dynamics of real neural circuits with relatively small neuron populations.

The key distinction between the two methods lies in how they discretize neuron states. RDM discretizes neuron states based on the timing since the last spike of each neuron, whereas dsODEs discretize neuron states based on physiological variables (for LIF neurons, the membrane potentials). This difference has important implications for the applicability and performance of each method. By focusing on the timing of the last spike, RDM inherently assumes that all neurons within the population receive identical inputs—both external and recurrent. Consequently, any noise or variability in synaptic currents is aggregated and reflected as a “soft threshold,” wherein randomness is introduced in the firing threshold of neurons. While this approach simplifies the analysis and implementation of the method, it introduces limitations when the synaptic noise or recurrent input variabilities are highly correlated across the population.

We validate this statement by directly comparing the numerical simulations of SNNs (one E \& one I-populations, the same as Sect.~\ref{Sect4.3-1E1I}), dsODEs, and RDM with the same choices of parameters. We find that the limitations in RDM become particularly evident in regimes with different levels of synchrony among neurons. When the network exhibits high synchrony (e.g., during $\tau^E<2$), the assumption of independent synaptic inputs breaks down, causing RDM to inaccurately capture the dynamics. This results in prediction errors larger than those of dsODEs, which directly model the interactions between discrete voltage states and account for correlated fluctuations in synaptic currents. On the other hand, both dsODE and RDM converge to similar results under conditions of low synchrony (e.g., during  $\tau^E>2$). We have also compared dsODEs and RDM by varying the expectation of external stimuli and projection probability between neurons. We find that, for different external stimuli, the prediction errors of dsODE are in general one order magnitude smaller than RDM, regardless of the size of SNNs. A similar trend is observed for different $p$. 

In all, our results suggest that while both methods are effective in capturing macroscopic behaviors, dsODEs provide a more nuanced understanding of network dynamics in challenging scenarios such as high synchrony or correlated input variability.

\begin{table*}[htbp]
\begin{center}
    \begin{tabular}{|l|c|l|l|l|}
      \hline
      Parameter Group  & Parameter        & Meaning                       & Standard Value & Tested Range\\ \hline
      Network          & $N^{E}$          & number of E cells             & 1000  & 1000 or 10000\\
      architecture     & $P^{EE}$         & E-to-E coupling probability   & 0.4  & 0.1-0.6 \\ 
                       & $S^{EE}$         & E-to-E synaptic weight        & 0.1 & 0.05-0.30 \\ \hline
      Neuronal         & $\tau^{\rm{ref}}$ & refractory period            & 4 ms & - \\ 
      physiology       & $g^{\rm leak}$    & leak conductance             & (20 ms)$^{-1}$ &- \\
                       & $\tau^E$         & E-synapse timescale           & 1 ms & 1-4 ms\\ \hline
      External         & $S^{E\rm{ext}}$  & external-to-E synaptic weight & 1  & - \\ 
      input            & $\lambda^E$      & external-to-E input rate      & 23 kHz & 21-24 kHz  \\ \hline
    \end{tabular}
 \caption{Standard choices and tested ranges of parameters for the comparison between dsODE and Fokker-Planck model.}
 \label{Table2:Parameters-FP}
\end{center}
\end{table*}

\begin{table*}[htbp]
\begin{center}
    \begin{tabular}{|l|c|l|l|l|}
      \hline
      Parameter Group  & Parameter        & Meaning                       & Standard Value & Tested Range\\ \hline
      Network          & $N^{I}$          & number of I cells             & 100  & 15-200\\
      architecture     & $N^{E}$          & number of E cells             & 300  & $3N^I$ \\
                       & $P^{EE}$         & E-to-E coupling probability   & 0.2  & 0.1-0.6 \\ 
                       & $P^{EI}$         & I-to-E coupling probability   & 0.2  & 0.1-0.6 \\ 
                       & $P^{IE}$         & E-to-I coupling probability   & 0.2  & 0.1-0.6 \\ 
                       & $P^{II}$         & I-to-I coupling probability   & 0.2  & 0.1-0.6 \\
                       & $S^{EE}$         & E-to-E synaptic weight        & 0.37 & - \\ 
                       & $S^{EI}$         & I-to-E synaptic weight        & 1.6 & 1.2-1.7\\
                       & $S^{IE}$         & E-to-I synaptic weight        & 0.3 & - \\
                       & $S^{II}$         & I-to-I synaptic weight        & 1.4 & - \\\hline
      Neuronal         & $\tau^{\rm{ref}}$ & refractory period           & 4 ms & - \\ 
      physiology       & $g^{\rm leak}$    & leak conductance             & (20 ms)$^{-1}$ & - \\
                       & $\tau^E$         & E-synapse timescale           & 1 ms & 1-4 ms\\
                       & $\tau^I$         & E-synapse timescale           & 4.5 ms & -\\\hline
      External         & $S^{E\rm{ext}}$  & external-to-E synaptic weight & 1  & - \\ 
      input            & $S^{I\rm{ext}}$  & external-to-I synaptic weight & 1  & $S^{E\rm{ext}}$ \\
                       & $\lambda^E$      & external-to-E input rate      & 40 kHz & 30-55 kHz  \\
                       & $\lambda^I$      & external-to-I input rate      & 40 kHz & $\lambda^E$\\ \hline
    \end{tabular} 
 \caption{Standard choices and tested ranges of parameters for the comparison between dsODE and refractory density method.}
 \label{Table2:Parameters-RDM}
\end{center}
\end{table*}

\section{Discussion}
\label{Sect6-Diss}
This study presented a novel discretized-state ordinary differential equation (dsODE) framework, offering a robust method to model the intricate dynamics of spiking neural networks (SNNs) with minimal information loss. By integrating principles of Markov models and coarse-graining, our dsODE system bridged the gap between microscopic neuronal activities and mesoscopic network behaviors, providing a powerful tool for studying emergent dynamics in local neural circuits. Our framework not only captured the physiological properties of individual neurons but also illuminated their impact on network-level dynamics, positioning dsODE as a valuable addition to the computational modeling of dynamical systems. In the rest of this section, we first consider the details underlying the effectiveness of the dsODE framework. Second, we discuss some technical issues and limitations of the dsODE. Then, we highlight the mathematical advantages of the dsODE, before we conclude by discussing a few promising future directions.

\heading{Effectiveness.}
The dsODE framework builds upon the Markovian modeling methods of spiking networks, which were first introduced by Li et al. \cite{li2019well} and later employed to study SNN dynamics in a series of works \cite{li2019stochastic, cai2021model, zhang2024learning}. These earlier Markov models were not designed to mimic the physiological details of single neurons and the interactions among them. In contrast, the dsODE system explicitly integrates the leaky integrate-and-fire descriptions, providing a robust mathematical tool for approximating spiking network behavior while maintaining fidelity to the underlying neural processes. 

A key innovation of the dsODE system lies in its derivation, which employs three critical approximations: (1) discretization of voltage states to reduce the complexity of the system, (2) a fast self-decorrelation assumption to derive a coarse-grained (CG) Markov model that captures the transition dynamics between voltage states, and (3) a Gaussian approximation for random variables to derive a set of ODEs. These approximations ensure that the framework remains analytically tractable while preserving the essential features of spiking neural network dynamics. Our results show that dsODE maintains a balance between computational efficiency, analyzability, and biological realism. It makes the dsODE system highly appealing for modeling local neural circuits such as minicolumns and hypercolumns in cortex.

The methodological robustness of the dsODE framework is particularly significant when contrasted with traditional machine learning (ML) models. While ML models are often effective at interpolation within the range of training data, they are typically not designed to extrapolate to unseen parameter regimes or to detect bifurcations—dynamical transitions in the system’s phase space \cite{rudy2017data,lusch2018deep,recht2019tour}. Nevertheless, recent studies reveal that embedding information of related physics or system structure can substantially enhance the ability of ML models to extrapolate and detect bifurcations in differential equations \cite{raissi2019physics,brunton2016discovering,schaeffer2017learning,beregi2023using}.
The dsODE system, on the other hand, offers a principled approach for predicting the evolution of SNNs as network parameters are varied. This predictive capability makes dsODEs a promising candidate for integration with ML methods in hybrid modeling approaches. Such a combination could enhance our understanding of how spiking networks transition between different dynamical regimes, potentially providing insights into how the brain integrates diverse functions within the same neural circuits.

\heading{Limitations.}
While the dsODE system presents a significant advancement in mathematical theories of spiking networks, certain limitations warrant further investigation. First, the discretization of voltage states, though computationally efficient, may introduce artifacts that deviate from the continuous dynamics of leaky integrate-and-fire (LIF) neurons. This challenge becomes more pronounced when applying the dsODE system to more complex neuron models, such as FitzHugh-Nagumo or Hodgkin-Huxley, where the underlying dynamics are more intricate \cite{hodgkin1952quantitative,fitzhugh1961impulses}. In such cases, more refined discretization techniques, such as adaptive state partitioning (see, e.g., \cite{munsky2006finite,deuflhard2008adaptive}), may be necessary to accurately model local circuit phenomena, especially those with neurons with adaptation or bursting behaviors.

Second, the fast self-decorrelation assumption (the key simplification that enabled the derivation of the coarse-grained Markov model) may break down in certain scenarios—particularly in networks with highly biased architectures, long synaptic time constants, or regimes with extremely low firing rates. Addressing this limitation will likely require extending the dsODE system to incorporate non-Markovian dynamics, which would likely better capture long-range correlations and temporal dependencies. Furthermore, extending the model to heterogeneous networks—where neurons exhibit varied biophysical properties or diverse connectivity patterns—would enable a more comprehensive analysis of how network heterogeneity influences overall dynamics and stability.

\heading{Mathematical advantages.} 
Despite these challenges, the Markovian framework introduced in this study offers substantial analytical benefits. For example, the invariant measure of a Markov chain, which can be computed from its transition matrix, provides complete information about the system’s asymptotic behavior. This approach has been successfully applied previously \cite{cai2021model}, where the invariant measure was used to capture distinct types of Gamma oscillations in simple spiking networks. Moreover, based on established proofs of ergodicity \cite{li2019well}, a large deviation analysis could be applied to Markovian spiking networks to compute transition waiting times between multiple attractors. This could offer valuable insights into decision-making processes and planning behaviors in animals by linking cognitive functions to underlying spiking network dynamics. Additionally, the Markovian framework simplifies the computation of adjoint flows \cite{plessix2006review}, as it requires only the transposition of the probability transition matrix. This feature enhances the efficiency of parameter exploration in spiking networks, typically a challenging task for biologically detailed models with many parameters.

The dsODE framework opens new avenues for studying complex systems, particularly in neuroscience. Its capacity to model intricate network dynamics, such as partial synchrony in fluctuation-driven, low-rate regimes, makes it an ideal tool for investigating phenomena where synchrony and timing are critical. One promising direction is the incorporation of synaptic plasticity rules, which would enable exploration of learning and adaptation in SNNs from a dynamical systems perspective.  Moreover, dsODEs could serve as building blocks for multi-scale cortical models, efficiently capturing interactions within and across multiple functional regions of the brain.


\section*{Acknowledgements}
This work was partially supported by the National Science and Technology Innovation STI2030-Major Project No. 2022ZD0204600 and the Natural Science Foundation of China through Grant No. 31771147 (J.C., Z.L., Z.W., L.T.). Z.X. was supported by the Courant Institute of Mathematical Sciences at New York University during most of the time for this study. We thank Adi Rangan for helpful discussions.

\renewcommand\thefigure{A\arabic{figure}}    
\setcounter{figure}{0}
\renewcommand\thesection{A\arabic{section}}    
\setcounter{section}{0}
\vspace{0.3in}
\noindent {\large \bf Appendix}
\section{A piecewise linear scheme of voltage configurations}
In the dsODE framework, we represent the detailed voltage configuration $n^Q(v)$ as a piecewise linear distribution formed by connecting a sequence of turning points. Here, we provide a detailed description of the algorithm based on this configuration.

For each combined voltage bin $I_\mbm =[\mbm\ba,(\mbm+1)\ba)$, we select a turning point $(x^Q_\mbm, y^Q_\mbm)$, where $x^Q_\mbm\in I_\mbm$ and $y^Q_\mbm \geq 0$, and then connect each pair of adjacent turning points linearly. Consequently, the voltage configuration is expressed as:
$$
\bm{n}^Q(v) = k^Q_{\mbm}v + b^Q_{\mbm}, \quad x^Q_{\mbm} < v < x^Q_{\mbm+1}
$$
where $\mbm \in \{-\mbm^{I}, -\mbm^{I} + 1, … , \mbm^{th}\}$ represents the voltage bin partitions. Here, $k^Q_{\mbm}=(y^Q_{\mbm+1}-y^Q_{\mbm})/(x^Q_{\mbm+1}-x^Q_{\mbm})$ is the slope, and $b^Q_{\mbm} = y^Q_{\mbm} - k^Q_{\mbm}x^Q_{\mbm}$ is the intercept. To close the piecewise linear representation at both ends, we set $(x^Q_{-\mbm^{I}-1}, y^Q_{-\mbm^{I}-1}) = (-\varepsilon^{I},0)$ and $(x^Q_{m^{th}+1}, y^Q_{m^{th}+1}) = (V^{th},0)$. Thus, the entire voltage configuration is defined by $2\mathbf{M}$ parameters.

The choices of turning points are constrained by the neuron number $N^Q_\mbm$ and mean voltage $\bar{v}^Q_\mbm$:
\begin{align}
\label{SectA.1-Eq24-numofneuron_Redu}
\nonumber
N^Q_\mbm=&\int_{\mbm\ba}^{(\mbm+1)\ba} \bm{n}^Q(v) \mathrm{d}v\\
=& \frac12\left\{[n^Q(\mbm\ba)+y^Q_{\mbm}][x^Q_{\mbm}-\mbm\ba]+[y^Q_{\mbm}+n^Q((\mbm+1)\ba)][(\mbm+1)\ba-x^Q_{\mbm}]\right\}\\
\nonumber
\label{SectA.1-Eq25-meanv_Redu}
\bar{v}^Q_\mbm=&\frac{1}{N^Q_\mbm}\int_{\mbm\ba}^{(\mbm+1)\ba} v\bm{n}^Q(v) \mathrm{d}v\\
\nonumber
=&\frac{1}{6}\left\{[2x_{\mbm}^2-\mbm\ba x_{\mbm}-(\mbm\ba)^2]y_{\mbm}+[x_{\mbm}^2+\mbm\ba x_{\mbm}-2(\mbm\ba)^2]n(\mbm\ba)]\right.\\
+&\left.[2((\mbm+1)\ba)^2-((\mbm+1)\ba) x_{\mbm}-(x_{\mbm}^2)]n((\mbm+1)\ba)+[((\mbm+1)\ba)^2+((\mbm +1)\ba)x_{\mbm}-2x_{\mbm}^2]y_{\mbm}\right\}
\end{align}
In other words, there exist a mapping $\mathbf{F}$ such that $\left\{N^Q_\mbm, \bar{v}^Q_\mbm\right\} = \mathbf{F}\left(\left\{x^Q_{\mbm}, y^Q_{\mbm}\right\}\right)$, where $\mbm\in\Gamma\setminus\{\mathcal{R}\}$ on both sides.

Finally, the numerical solution of dsODE for one timestep should first evolve $N^Q_\mbm$ and mean voltage $\bar{v}^Q_\mbm$, then update the turning points $(x^Q_{\mbm}(t), y^Q_{\mbm}(t))$ summarized following the steps below:
\begin{enumerate}
    \item Compute $\left\{N^Q_\mbm(t), \bar{v}^Q_\mbm(t)\right\} = \mathbf{F}\left(\left\{x^Q_{\mbm}(t), y^Q_{\mbm}(t)\right\}\right)$ using Eqs.~\ref{SectA.1-Eq24-numofneuron_Redu} and \ref{SectA.1-Eq25-meanv_Redu}.
    \item	Evolve $N_{\mbm}^{Q}$ and $\bar{v}_{\mbm}^{Q}$ for one timestep $\delta t$ with the Euler forward scheme, using Eqs.~\ref{Sect3.2-Eq16-J_Redu} through~\ref{Sect3.2-Eq18-vflux-Redu}.
	\item	Infer $\left\{x^Q_{\mbm}(t+dt), y^Q_{\mbm}(t+dt)\right\} = \mathbf{F}^{-1}\left(\left\{N_{\mbm}^{Q}(t+dt), \bar{v}_{\mbm}^{Q}(t+dt)\right\}\right)$ via the Newton-Raphson method \cite{ypma1995historical}. This step involves solving for $2\mathbf{M}$ unknown variables from $2\mathbf{M}$ equations. On the other hand, $\mathbf{F}^{-1}$ is not guaranteed as well defined for arbitrary $\left\{N_{\mbm}^{Q}(t+dt), \bar{v}_{\mbm}^{Q}(t+dt)\right\}$, and the algorithm will return \texttt{failure} in those cases.
\end{enumerate}
Further details are provided in Algorithm 1.

\SetKwComment{Comment}{/*   }{ */}
\RestyleAlgo{ruled}

\begin{algorithm}
\caption{Evolution of dsODE for one timestep $\delta t$}\label{alg:one}
\KwResult{Update turning points $\left\{x^Q_{\mbm},y^Q_{\mbm}\right\}$}
Set network parameters $S^{QR}, p^{QR}, \tau^{R},\tau^{\rm ref}$, etc.; $\mathbf{M} \gets $ number of bins; $\delta t \gets$ timestep\;
$\left\{x^Q_{\mbm}(t),y^Q_{\mbm}(t)\right\} \gets$ initial values of turning points\;
$N^Q_\mathcal{R} \gets$ initial number of neurons in refractory state\;
$\bm{u}^{QE},\bm{u}^{QI},\bm{D}^{QE},\bm{D}^{QI} \gets$ initial mean and variance of ($H^{QE}$,$H^{QI}$) \Comment*[r]{See definitions in Sect. 3.1.1}
$(e, \ba, A_{\rm max},i_{\rm max})\gets $ termination tolerance, noise of initial guess, maximum attempts, maximum iterations in Newton-Raphson method\;
\For{$\mathbf{m} \in\mathbf{\Gamma}\setminus\{\mathcal{R}\}$}{ 
\Comment{Compute $N_{\mathbf{m}}^{Q}$ and $\bar{v}_{\mathbf{m}}^{Q}$ for each voltage bin from turning points $\left\{x^Q_{\mathbf{m}}(t),y^Q_{\mathbf{m}}(t)\right\}$}
    $N^Q_\mbm \gets$  $r.h.s.$ of Eq.~\ref{SectA.1-Eq24-numofneuron_Redu}\;
    $\bar{v}^Q_\mbm\gets$  $r.h.s.$ of Eq.~\ref{SectA.1-Eq25-meanv_Redu}\;
}
\For{$\mathbf{m} \in\mathbf{\Gamma}$}{
    Evolve $N_{\mathbf{m}}^{Q}$ and $\bar{v}_{\mathbf{m}}^{Q}$ for $\delta t$ following Eqs.~\ref{Sect3.2-Eq15-nDiscrete_Redu} and \ref{Sect3.2-Eq18-vflux-Redu} \;
}

\Comment{Update turning points with Newton-Raphson method}
$A \gets 0$ \Comment*[r]{$A$ is the number of attempts that has been made} 
\While {$A=0$ \bf{or} \texttt{isinvalid}$(\left\{x^Q_{\mbm}(t+\delta t),y^Q_{\mbm}(t+\delta t)\right\})$}{
    \Comment{\texttt{isinvalid} is a function to judge the validity of solution. See \bf{Algorithm 2}}
    \If{$A>A_{\rm max}$}{
    \bf{return} \texttt{failure}\;
    }
    \eIf{$A=0$}{
    $\vec{X}_0\gets\left\{x^Q_{\mbm}(t),y^Q_{\mbm}(t)\right\}$
    }{
    \Comment{Starting from the last guess and make another attempt}
    $\delta \vec{X}_0\gets \frac12\ba\cdot$\texttt{rand}$(2\mathbf{M})$ \Comment*[r]{uniform perturbations to the original guess}
    $\vec{X}_0\gets\left\{x^Q_{\mbm}(t),y^Q_{\mbm}(t)\right\} + \delta \vec{X}_0$
    } 
    \texttt{norm}$\gets e+1$ \Comment*[r]{set initial value of norm, which is the $\mathcal{L}^2$ norm of $\delta\vec{X}$}
    $i \gets 0$ \Comment*[r]{$i$ is the iteration number}
        \While {\texttt{norm}$>e$ \bf{and} $i<i_{\rm max}$}{
            $\delta \vec{X}_{i+1}\gets \mathbf{J}_\mathbf{F}^{-1}(\vec{X}_{i})\cdot\delta\vec{X}_{i}$  \Comment*[r]{$\mathbf{J}_\mathbf{F}(\vec{X}_{i})$ is the Jacobian matrix of mapping $\mathbf{F}$ evaluated at $\vec{X}_{i}$}
            $\vec{X}_{i+1} \gets \vec{X}_{i}-\delta \vec{X}_{i+1}$\;
            \texttt{norm} $\gets ||\delta \vec{X}_{i+1}||_{\mathcal{L}^2}$\;
            $i \gets i+1$
        }
        $\left\{x^Q_{\mbm}(t+\delta t),y^Q_{\mbm}(t+\delta t)\right\} \gets \vec{X}_A$\;
        $A\gets A+1$
        
}

\bf{return} $\left\{x^Q_{\mbm}(t+\delta t),y^Q_{\mbm}(t+\delta t)\right\}$\;
\end{algorithm}

\RestyleAlgo{ruled}
\begin{algorithm}
\caption{\texttt{isinvalid}$(\left\{x^Q_{\mbm},y^Q_{\mbm}\right\})$}\label{alg:two}
\KwResult{Judge if the turning points $\left\{x^Q_{\mbm},y^Q_{\mbm}\right\}$ are invalid}
$\mathbf{M}\gets$ number of bins\;
$\ba \gets $ width of bins \Comment*[r]{See Sect. 3.2.1 for details}

\For{$\mathbf{m} \in \Gamma\setminus\{\mathcal{R}\}$}{
\If{$x_\mathbf{m}^Q>(\mathbf{m}+1)\ba$ \bf{or} $x_\mbm^Q<\mbm\ba$}{
\bf{return} \texttt{true}
}
\If{$y_\mathbf{m}^Q<0$}{
\bf{return} \texttt{true}
}
}
\bf{return} \texttt{false}
\end{algorithm}
\section{Deriving flux terms in dsODE}
In Eq.~\ref{Sect3.2-Eq16-J_Redu}, the expectations of upward fluxes due to leakage and external stimuli are straightforward. We here present the details for the expectation of recurrent E-fluxes, and the expectation for recurrent I-fluxes is similar.
\begin{align*}
    &\mathrm{E}[J^{QE}(v)] = \lim_{a\to0}\left\{-\int_v^{v+a\delta^E(v)} \frac{D^{QE}}{2(\tau^E)^2} \bm{n}^Q(x) \mathrm{d}x\, +\, \int_{v^*}^{v} \left[\frac{u^{QE}}{\tau^E} + \frac{D^{QE}}{2(\tau^E)^2}\right] \bm{n}^Q(x) \mathrm{d}x \right\}\\
& =  \lim_{a\to0}\left\{\int_{v^*}^{v} \frac{u^{QE}}{\tau^E}  \bm{n}^Q(x) \mathrm{d}x\,+\,\frac{D^{QE}}{2(\tau^E)^2}\left[\int_{v^*}^{v} \left(\bm{n}^Q(x) - \bm{n}^Q(x + a\delta v^*)\right) \mathrm{d}x\,+\,\int_{v+a\delta^E{v}}^{v+a\delta^E{v^*}} \bm{n}^Q(x)\mathrm{d}x\right]\right\}\\
& =  \frac{\bm{u}^{QE}}{\tau^E}(\varepsilon^{E}-v)\bm{n}^Q(v) +\lim_{a\to0}\left\{ \frac{\bm{D}^{QE}}{2(\tau^{E})^2}\left[\frac{v-v^*}{a}\cdot\frac{\bm{n}^Q(v)-\bm{n}^Q(v^*)}{a} +\frac{a(\delta^E(v^*) - \delta^E(v))}{a^2}\bm{n}^Q(v)\right]\right\} \\
& =  \left(\frac{\bm{u}^{QE}}{\tau^E} + \frac{\bm{D}^{QE}}{2(\tau^{E})^2}\right)\cdot(\varepsilon^{E}-v)\bm{n}^Q(v) - \frac{\bm{D}^{QE}}{2(\tau^{E})^2}\cdot(\varepsilon^{E}-v)^2\frac{\partial \bm{n}^Q}{\partial v} \,.
\end{align*}

\section{A more efficient numerical implementation of dsODE}
There are infinitely many possible ways to infer the detailed voltage configuration $\{n^Q_m|[m/L]=\mbm\}$. However, our empirical findings suggest that, when the combined voltage bin size $\ba$ is sufficiently small, the neuron distribution within each voltage bin can be approximated as uniform. Although this approximation may affect the flux terms at interval boundaries in the dsODE model (e.g., producing infinitely large flux as timestep $\delta t \to 0$ while fixing $\ba$), dsODE remains accurate with the uniform scheme, provided that the timestep $\delta t$ is aligned with the combined voltage bin size $\ba$ (i.e., smaller $\ba$ necessitates a smaller $\delta t$).

To deduce $\{n^Q_m|[m/L]=\mbm\}$, we select a uniform distribution that satisfies both the mean voltage ($\bar{v}^Q_\mbm$) and the total neuron count ($N^Q_\mbm$) for bin $\mbm$. Specifically:
\begin{align*}
    &\text{if } \bar{v}^Q_\mbm>(\mbm+\frac12)L,\quad\text{then } n^Q_m = \begin{cases}
        \frac{N^Q_\mbm}{2[(\mbm+1)L-\bar{v}^Q_\mbm]} &\text{for } 2\bar{v}^Q_\mbm-(\mbm+1)L<m\leq(\mbm+1)L \\
        0 &\text{otherwise}
    \end{cases}  \\
    &\text{if } \bar{v}^Q_\mbm\leq(\mbm+\frac12)L,\quad\text{then } n^Q_m = \begin{cases}
        \frac{N^Q_\mbm}{2(\bar{v}^Q_\mbm-\mbm L)} &\text{for } \mbm L <m\leq 2\bar{v}^Q_\mbm-\mbm L \\
        0 &\text{otherwise } 
    \end{cases} 
\end{align*}
As $a \to 0$, the density configurations $P(v|\bar{v}^Q_\mbm)$ converge weakly to either $\mathrm{U}(\mbm \ba, 2\bar{v}^Q_\mbm - \mbm \ba)$ or $\mathrm{U}(2\bar{v}^Q_\mbm - (\mbm+1) \ba, (\mbm+1) \ba)$, depending on $\bar{v}^Q_\mbm$.

The next step involves computing the sum of all flux terms within a timestep, $J^{Q}_{\mbm, \delta t} = J^{Q\rm{leak}}_{\mbm, \delta t} + J^{Q\rm{ext}}_{\mbm, \delta t} + J^{QE}_{\mbm, \delta t} + J^{QI}_{\mbm, \delta t}$. This is achieved by (1) omitting the leaky flux term $J^{Q\rm{leak}}_{\mbm, \delta t}$ and (2) substituting the average voltage $\bar{v}_\mbm$ for $v$ in the recurrent input terms (Eqs.~\ref{Sect3.2-Eq16-JE_Redu} and \ref{Sect3.2-Eq16-JI_Redu}). Consequently, all fluxes are realized by applying a Gaussian convolution to $P(v|\bar{v}_\mbm)$, with the net input to each neuron in bin $\mbm$ within $\delta t$ treated as an independent normal variable. Thus, the total input current $I^{Q}_{\mbm, \delta t}$ follows a normal distribution $\mathcal{N}(\bm{u}^{Q}_{\mbm, \delta t}, \bm{D}^{Q}_{\mbm, \delta t})$, where
\begin{align*}
    \bm{u}^{Q}_{\mbm,\delta t} &= \delta t\cdot\left[S^{Q\rm{ext}}\lambda^Q + \frac{\bm{u}^{QE}}{\tau^E}(\varepsilon^{E}-\bar{v}^Q_\mbm) + \frac{\bm{u}^{QI}}{\tau^I}(\varepsilon^{I}-\bar{v}^Q_\mbm)\right]  \, , \\
    \bm{D}^{Q}_{\mbm,\delta t} &= \delta t\cdot\left[(S^{Q\rm{ext}})^2\lambda^Q + \frac{\bm{D}^{QE}}{(\tau^E)^2}(\varepsilon^{E}-\bar{v}^Q_\mbm)^2 + \frac{\bm{D}^{QI}}{(\tau^I)^2}(\varepsilon^{I}-\bar{v}^Q_\mbm)^2\right]  \, .
\end{align*}

Now, consider a neuron sampled from the uniform distribution within voltage bin $\mbm$ (whose voltage is $X_\mbm$) at the beginning of the timestep. After receiving inputs over the timestep $\delta t$, the neuron’s voltage becomes $Z_\mbm = X_\mbm + Y_\mbm $, whose probability distribution is $ f_{Z_\mbm}(v) = \int_{-\infty}^{\infty} f_{X_\mbm}(z) \cdot f_{Y_\mbm}(v - z) , dz$. Here, $Y_\mbm = I^{Q}_{\mbm, \delta t}$. Let $a$ represent the lower bound of the uniform distribution interval ($2\bar{v}-(\mbm+1)L$ or $\mbm L$), and $b$ be the upper bound ($(\mbm+1)L$ or $2\bar{v}-\mbm L$), we have
\begin{align*}
    f_{Z_\mbm}(v) = \frac{1}{2(a-b)}\left[\operatorname{erf}\left(\frac{a+\bm{u}^{Q}_{\mbm,\delta t}-v}{\sqrt{2\bm{D}^{Q}_{\mbm,\delta t}}}\right)-\operatorname{erf}\left(\frac{b+\bm{u}^{Q}_{\mbm,\delta t}-v}{\sqrt{2\bm{D}^{Q}_{\mbm,\delta t}}}\right)\right] \, .
\end{align*}

Returning to evolve dsODEs for $N^Q_{m}$ and $N^Q_{m}\bar{v}^Q_\mbm$ (Eqs.~\ref{Sect3.2-Eq15-nDiscrete_Redu} and \ref{Sect3.2-Eq18-vflux-Redu}), after the timestep $\delta t$, the increment in neuron number in bin $\mbk$ due to neurons moving from bin $\mbm$ is denoted by $\Phi^Q_{\mbm \to \mbk}$, and the increment in $(N^Q_\mbk \bar{v}^Q_\mbk)$ due to bin $\mbm$ neurons by $\Psi^Q_{\mbm \to \mbk}$. Based on $Z_\mbm$’s distribution, we have
\begin{align*}
     \Phi^Q_{\mbm\to \mbk} = &N^Q_{\mbm}\int_{\mbk\ba}^{(\mbk+1)\ba} f_{Z_\mbm}(v) \mathrm{d} v \\
      = &N^Q_{\mbm}\sqrt{\frac{\bm{D}^{Q}}{2\pi(a-b)^2}} \times\\
      &\left[e^{-\frac{(b+\bm{u}^{Q}-v)^2}{2 \bm{D}^{Q}}}-e^{-\frac{(a+\bm{u}^{Q}-v)^2}{2 \bm{D}^{Q}}} -(a+\bm{u}^{Q}-v) \operatorname{erf}\left(\frac{a+\bm{u}^{Q}-v}{\sqrt{2\bm{D}^{Q}}}\right)+(b+\bm{u}^{Q}-v) \operatorname{erf}\left(\frac{b+\bm{u}^{Q}-v}{\sqrt{2\bm{D}^{Q}}}\right)\right] \, . \\
    \Psi^Q_{\mbm\to \mbk} = & N^Q_{\mbm}\int vf_{Z}(v) \mathrm{d} v \\ 
     = & \frac{N^Q_{\mbm}}{4(a-b)}\left[a^2 \operatorname{erf}\left(\frac{-a-\bm{u}^{Q}+v}{\sqrt{2\bm{D}^{Q}}}\right)+(\bm{u}^{Q})^2 \operatorname{erf}\left(\frac{-a-\bm{u}^{Q}+v}{\sqrt{2\bm{D}^{Q}}}\right)+\bm{D}^{Q} \operatorname{erf}\left(\frac{-a-\bm{u}^{Q}+v}{\sqrt{2\bm{D}^{Q}}}\right)\right. \\
    & - \sqrt{\frac{2}{\pi}} \bm{D}^{Q}(a+\bm{u}^{Q}+v) e^{-\frac{(a+\bm{u}^{Q}-v)^2}{2 \bm{D}^{Q}}}+v^2 \operatorname{erf}\left(\frac{a+\bm{u}^{Q}-v}{\sqrt{2\bm{D}^{Q}}}\right) \\
    & +2 a \bm{u}^{Q} \operatorname{erf}\left(\frac{-a-\bm{u}^{Q}+v}{\sqrt{2\bm{D}^{Q}}}\right)-b^2 \operatorname{erf}\left(\frac{-b-\bm{u}^{Q}-v}{\sqrt{2\bm{D}^{Q}}}\right)-(\bm{u}^{Q})^2 \operatorname{erf}\left(\frac{-b-\bm{u}^{Q}+v}{\sqrt{2\bm{D}^{Q}}}\right) \\
    & -\bm{D}^{Q} \operatorname{erf}\left(\frac{-b-\bm{u}^{Q}+v}{\sqrt{2\bm{D}^{Q}}}\right)+\sqrt{\frac{2}{\pi}} \bm{D}^{Q} (b+\bm{u}^{Q}+v) e^{-\frac{(b+\bm{u}^{Q}+v)^2 }{2 \bm{D}^{Q}}} \\
    & \left.-v^2 \operatorname{erf}\left(\frac{b+\bm{u}^{Q}-v}{\sqrt{2\bm{D}^{Q}}}\right)-2 b \bm{u}^{Q} \operatorname{erf}\left(\frac{-b-\bm{u}^{Q}-v}{\sqrt{2\bm{D}^{Q}}}\right)\right] \, .
\end{align*}
Finally, dsODEs are evolved as
\begin{align*}
    N^Q_{\mbm}(t+\delta t) &= \sum_{k} \Phi^Q_{\mbk\to \mbm} - \sum_{k\neq m} \Phi^Q_{\mbm\to \mbk}\, , \\
    (N^Q_\mbm\bar{v}^Q_\mbm)(t+\delta t) &= \sum_{k} \Psi^Q_{\mbk\to \mbm} - \sum_{k\neq m} \Psi^Q_{\mbm\to \mbk}\, .
\end{align*}


\section{Simulations of a rate model}
\begin{figure}[htbp]
  \begin{center}
    \begin{subfigure}{\textwidth}
          \includegraphics*[bb=2.5in 0.5in 34in 19in,width=.98\textwidth]{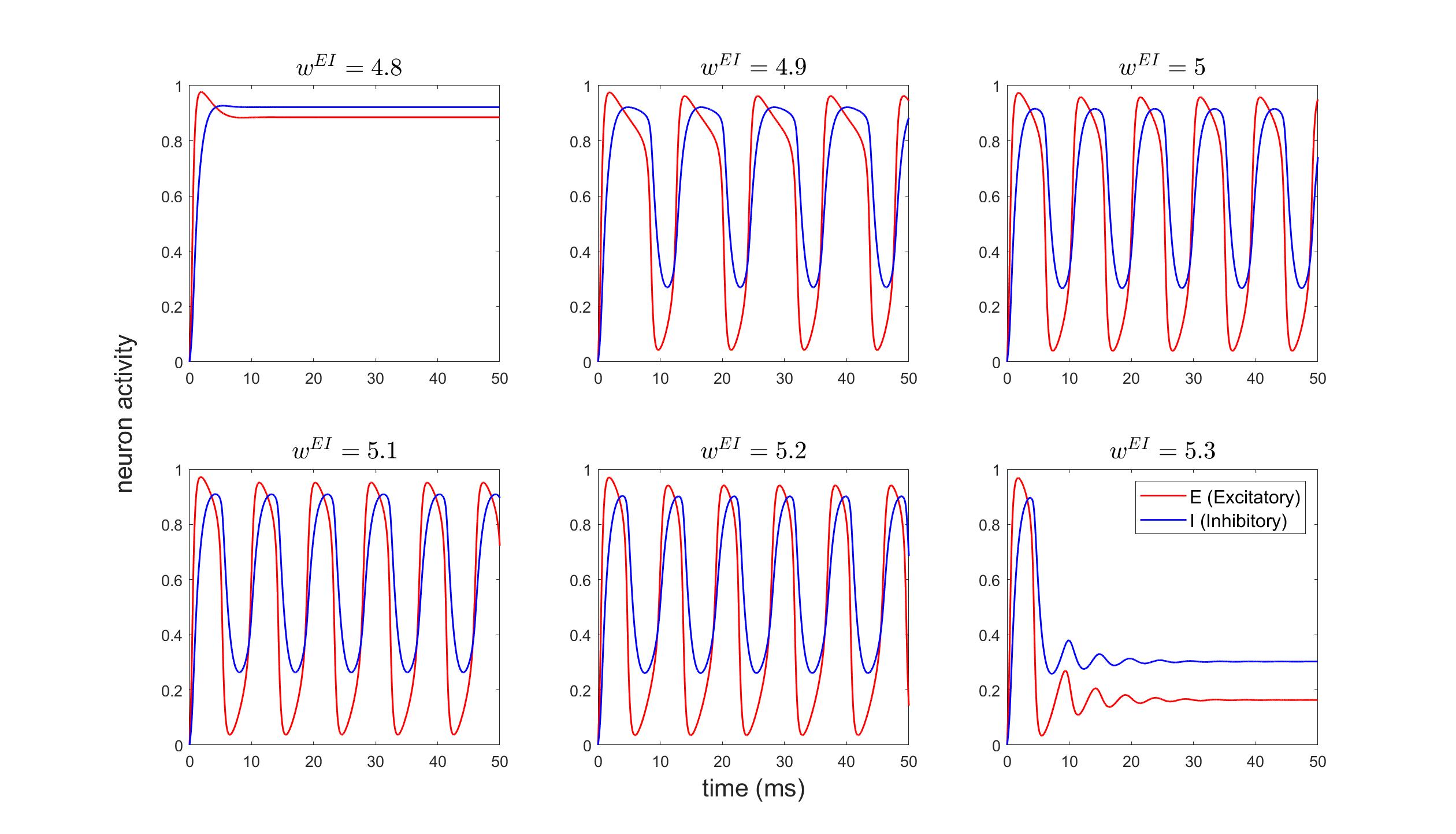} 
    \end{subfigure}
    \caption{Performance of a Wilson-Cowan rate Model for different choices of I-to-E coupling weight. The firing rates of E and I populations are depicted (blue and red curves).}
    \label{FigS2_Rate_Model}
  \end{center}
\end{figure}

A rate model usually directly describes neuronal activity through firing rates. Taking the Wilson–Cowan model as an example, neuronal activity is described by Eq.~\ref{Sect5.1-Eqn22-RateModel}, where $F^{E,I}(\cdot)$ are sigmoid functions \cite{pinto1996quantitative}. To produce the dynamics in Fig.~\ref{FigS2_Rate_Model}, we choose $F^{E,I}(x)=1 / (1 + e^{2.5-x})$ and the rest of the parameters shown in Table 4. The model does not exhibit two-beat dynamics (i.e., the alternation between strong and weak spiking clusters) when varying the $w^{EI}$ parameter.
\begin{table*}[htbp]
\begin{center}
    \begin{tabular}{|c|l|l|} \hline
      \textbf{Parameter} & \textbf{Meaning} & \textbf{Value}\\ \hline
      $\tau^{E}$ & E neuron reaction timescales & 0.3\\
      $\tau^{I}$ & I neuron reaction timescales & 1\\
      $w^{EE}$ & E to E coupling strengths & 9\\
      $w^{EI}$ & E to I coupling strengths & 4.9\\
      $w^{IE}$ & I to E coupling strengths & 5\\
      $w^{II}$ & I to I coupling strengths & 0.5\\
      $I^{E}$ & E external inputs & 1\\
      $I^{I}$ & I external inputs & 1\\ \hline
    \end{tabular}
 \caption{Parameters for the Wilson-Cowan model.}
 \label{Table2:Parameters-Rate}
\end{center}
\end{table*}

\section{Simulations of spiking networks, dsODE, Fokker-Planck, and refractory density methods}
The biological parameters listed in Tables 2 and 3 were used to simulate spiking networks, the dsODE method, and prior population-based methods. For the dsODE simulations, we employed the efficient numerical implementation outlined in Appendix A3, using a voltage interval length of 1 mV. To assess the predictive accuracy of each model, we calculated the relative firing rate errors by comparing each model’s output to that of its corresponding LIF network. The relative error was defined as:
$$
\text{Relative Error} = \frac{|\bar{f}_{\text{test}} - \bar{f}_{\text{LIF}}|}{\bar{f}_{\text{LIF}}} \times 100\%
$$
where $f$'s are defined by Eq.~\ref{Sect5-Eqn20-FRate}. Simulations for each model used a timestep of $\delta t = 0.1$ ms and ran for a total of 10 seconds, with $\bar{f}_{\text{test}}$ calculated every 1 second to obtain the mean firing rate and standard error.

It is worth noting that, in previous publications, refractory density methods (RDM) and Fokker-Planck (FP) equations utilized different physiological setups for spiking networks than those used in this study.  The most significant difference is that \cite{schwalger2017towards} implemented RDM on current-based integrate-fire networks, instead of conductance-based as in this study. Also, \cite{schwalger2017towards} and \cite{vinci2023self} choose set the threshold potential $V^{th} = 15$ mV.
For comparisons with RDM and FP, we adapted the dsODE model to match these setups accordingly. The source code for the Fokker-Planck and refractory density methods are available in \cite{vinci2023self} and \cite{schwalger2017towards}. Code for dsODE, as well as scripts for comparison with RDM and FP, can be accessed at https://github.com/changjie98/dsODE.

\section{Spike synchrony index}
Spike synchrony index (SSI) describes the degree of synchrony of the firing events as follows. In Fig.~\ref{Fig8_finiteN}C, We borrow the definition of SSI from Chariker et al\cite{chariker2018rhythm} and calculate the SSI values for networks of different scales. For each spike occurring at $t$, consider a $w$ ms time window centered by the spike $(t-w/2, t+w/2)$ and count the fraction of neurons in the whole network firing in such window. (In practice, we choose $w$=10ms.) Then, SSI is computed from the fraction averaged over all spikes and all neurons. 

On the other hand, since dsODE does not produce any discrete spike, we use Eq.~\ref{Sect3.2-Eq21-vfluxNoise-Redu} to compute the SSI of dsODE:
\begin{align}
\label{Sect3.2-Eq21-vfluxNoise-Redu}
\text{SSI}=\frac{\int_{w/2}^{t_{end}-w/2}\left(\int_{t-w/2}^{t+w/2}f^Q(\tau)\,\mathrm{d}\tau \right)\cdot f^Q(t)\, \mathrm{d}t}{N^Q\cdot \int_{w/2}^{t_{end}-w/2}f^Q(t)\,\mathrm{d}t}
\end{align}
It is not hard to see that SSI is larger for more synchronous spiking patterns. For the completely synchronized dynamics, every other neuron fires within the time window of each spike hence SSI=1.

\section{Bifurcation analysis of dsODE model with noise}
\begin{figure}[htbp]
\begin{subfigure}{\textwidth}
    \begin{subfigure}{0.5\textwidth}
    {\bf A}\\
        \includegraphics*[bb=2.5in 0in 33in 20in,width=.98\textwidth]{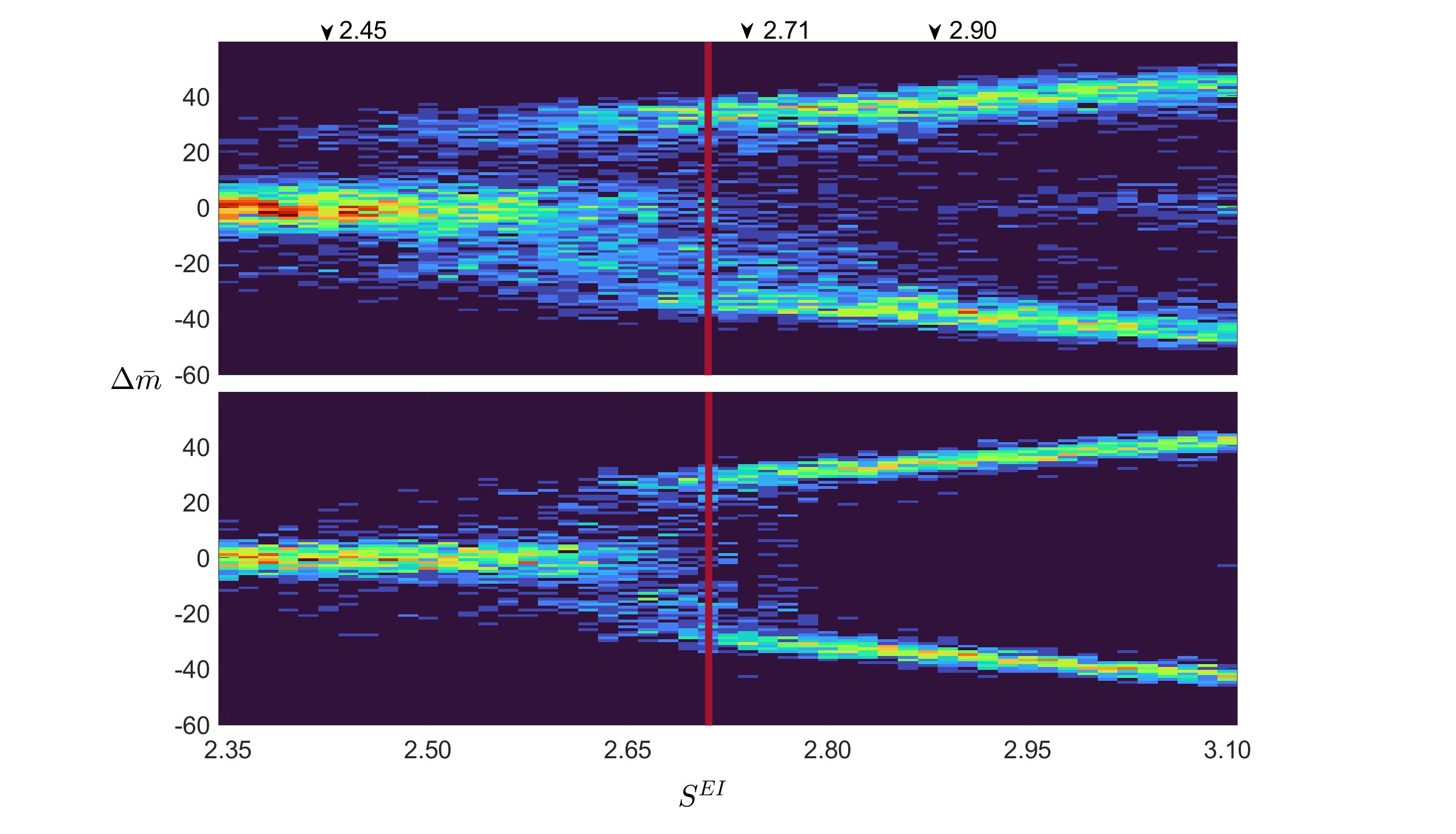}
    \end{subfigure}
    \begin{subfigure}{0.5\textwidth}
    {\bf B}\\
        \includegraphics*[bb=2.5in 0in 33in 20in,width=.98\textwidth]{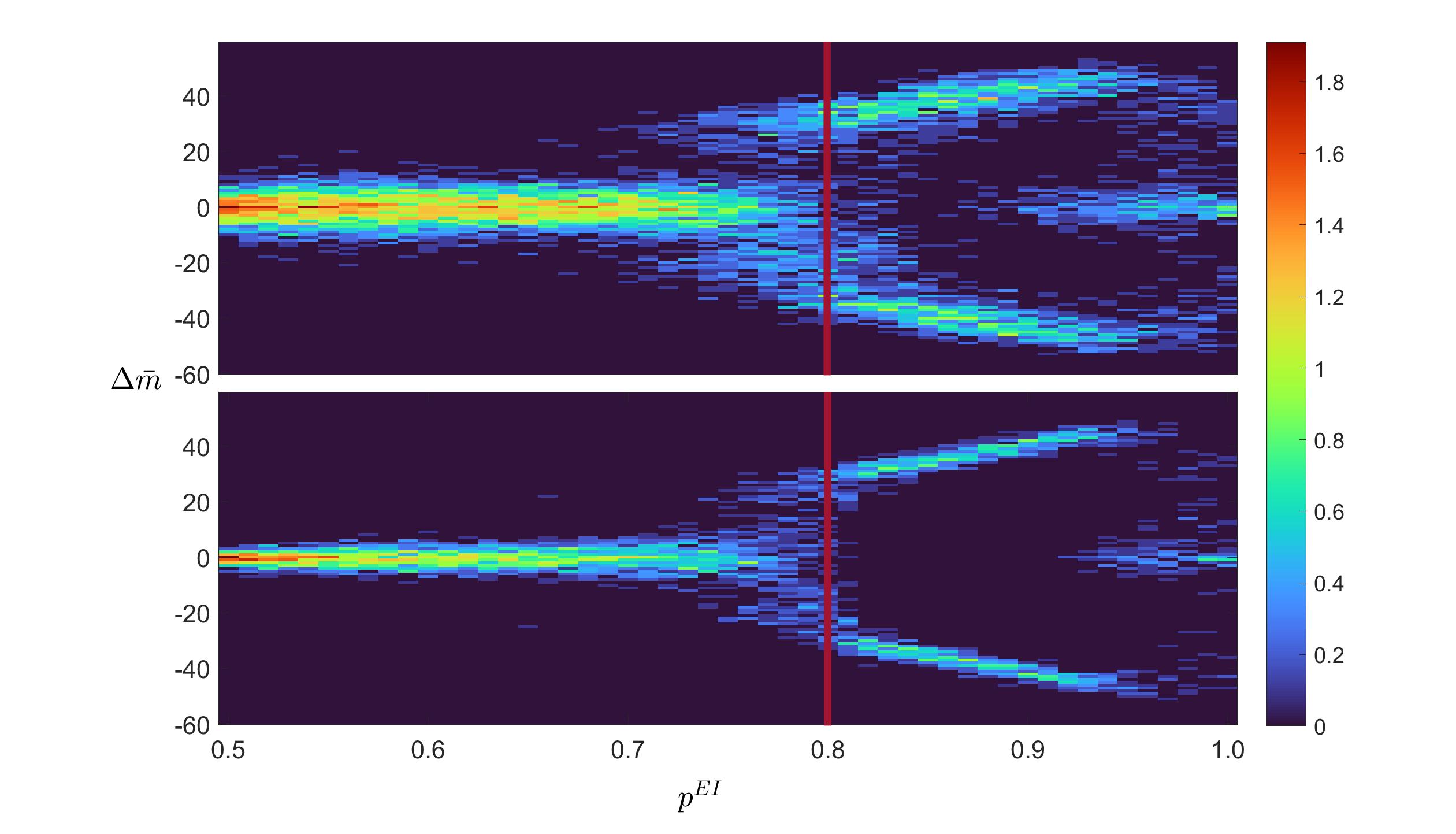}
    \end{subfigure}
\end{subfigure}\\[1ex]
    \begin{subfigure}{0.5\textwidth}
    {\bf C}\\
        \includegraphics*[bb=2.5in 0in 33in 19in,width=.98\textwidth]{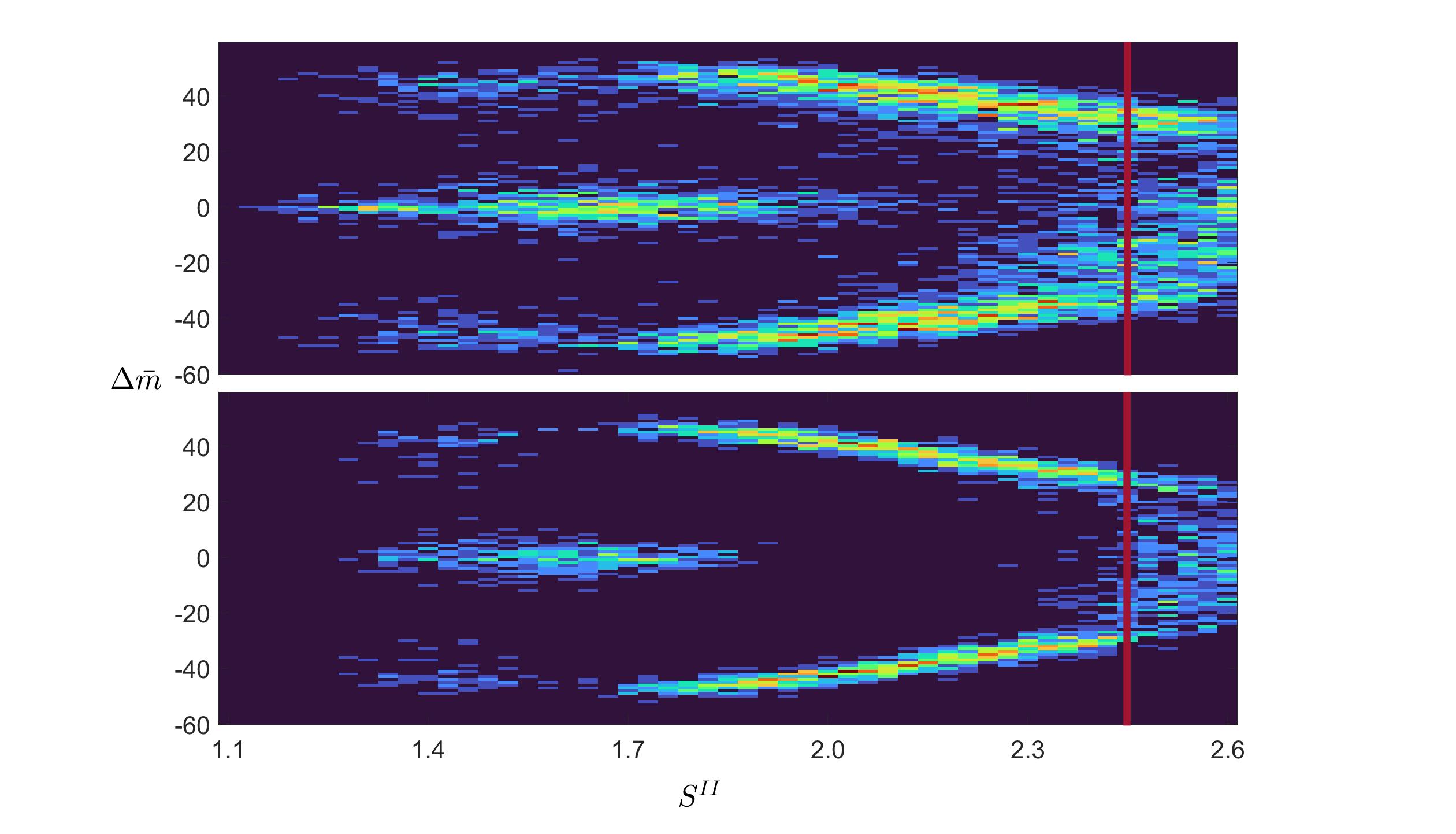}
    \end{subfigure}
    \begin{subfigure}{0.5\textwidth}
    {\bf D}\\
        \includegraphics*[bb=2.5in 0in 33in 19in,width=.98\textwidth]{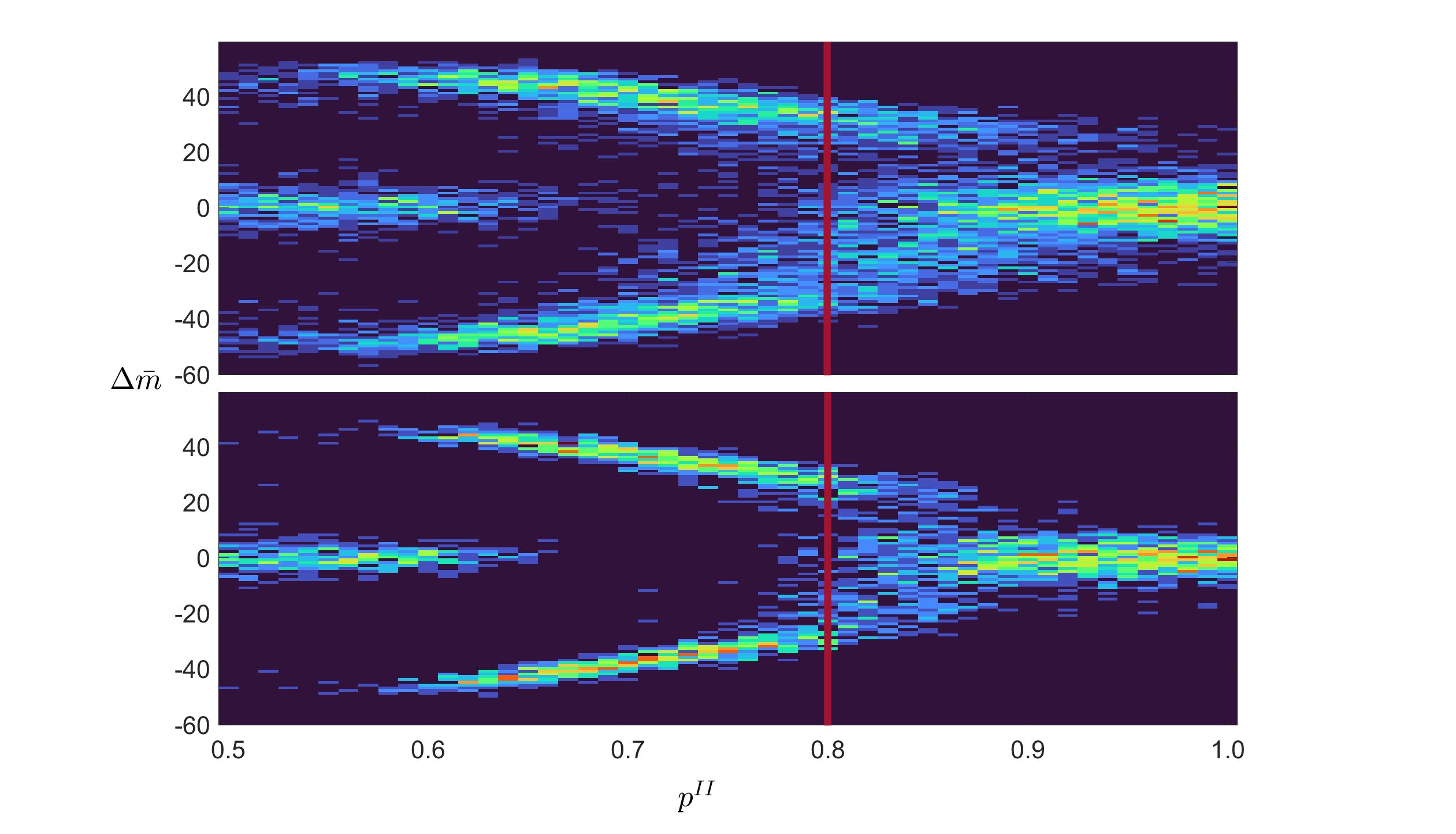}
    \end{subfigure}\\[1ex]
    \begin{subfigure}{0.5\textwidth}
    {\bf E}\\
        \includegraphics*[bb=2.5in 0in 33in 20in,width=.98\textwidth]{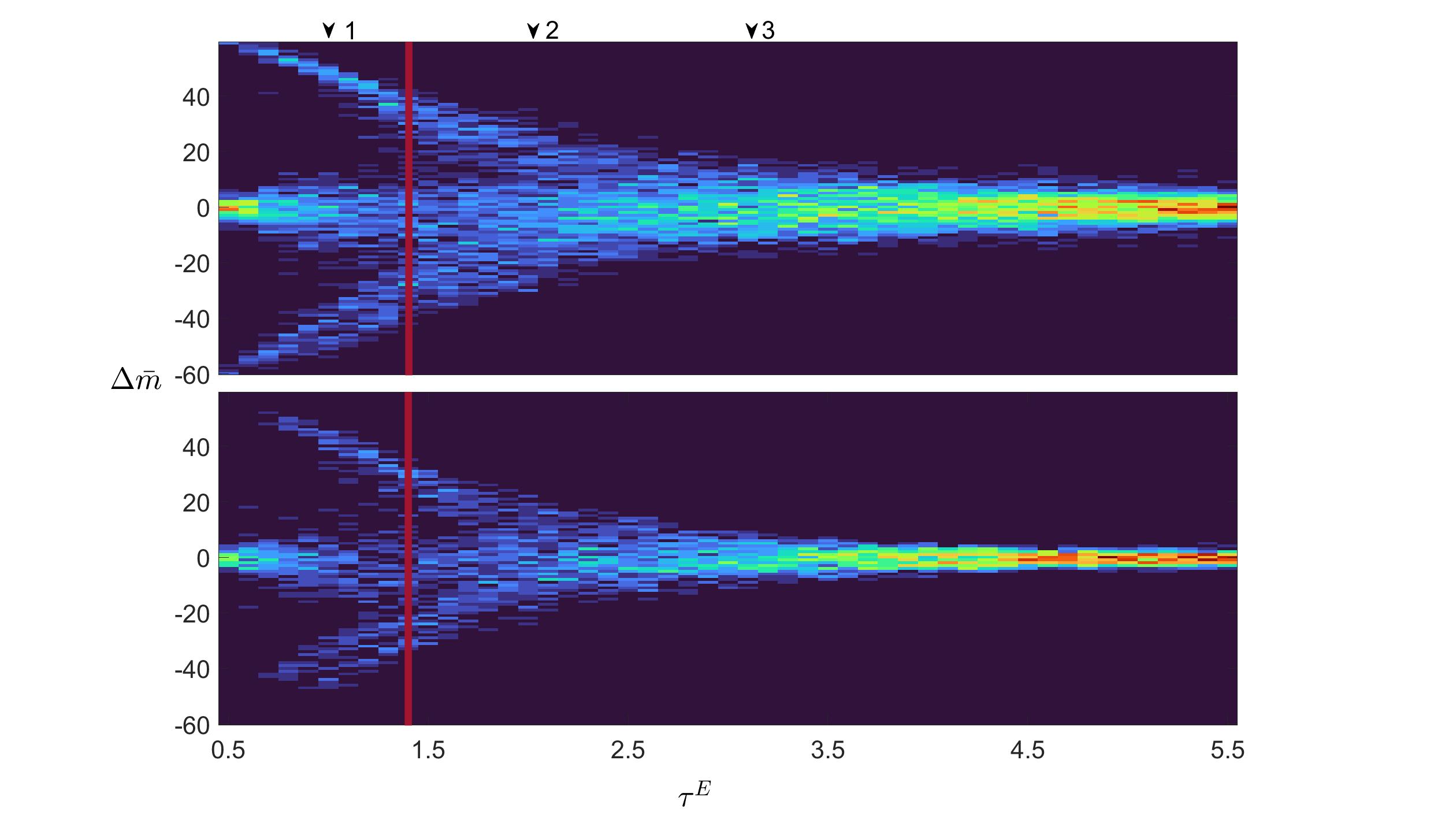}
    \end{subfigure}
    \begin{subfigure}{0.5\textwidth}
    {\bf F}\\
        \includegraphics*[bb=2.5in 0in 33in 20in,width=.98\textwidth]{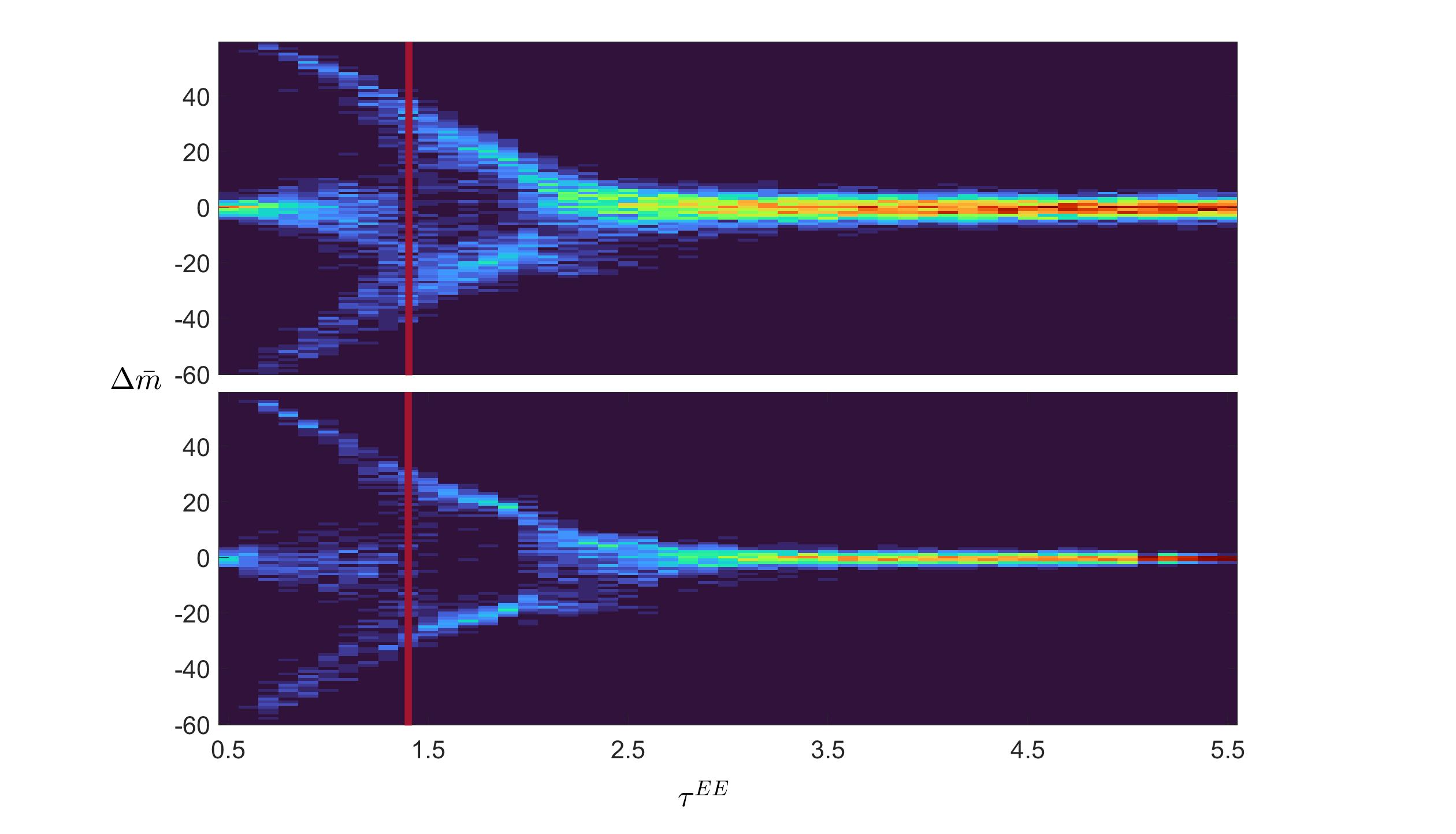}
    \end{subfigure}\\[1ex]
    \begin{subfigure}{0.5\textwidth}
    {\bf G}\\
        \includegraphics*[bb=2.5in 0in 33in 19in,width=.98\textwidth]{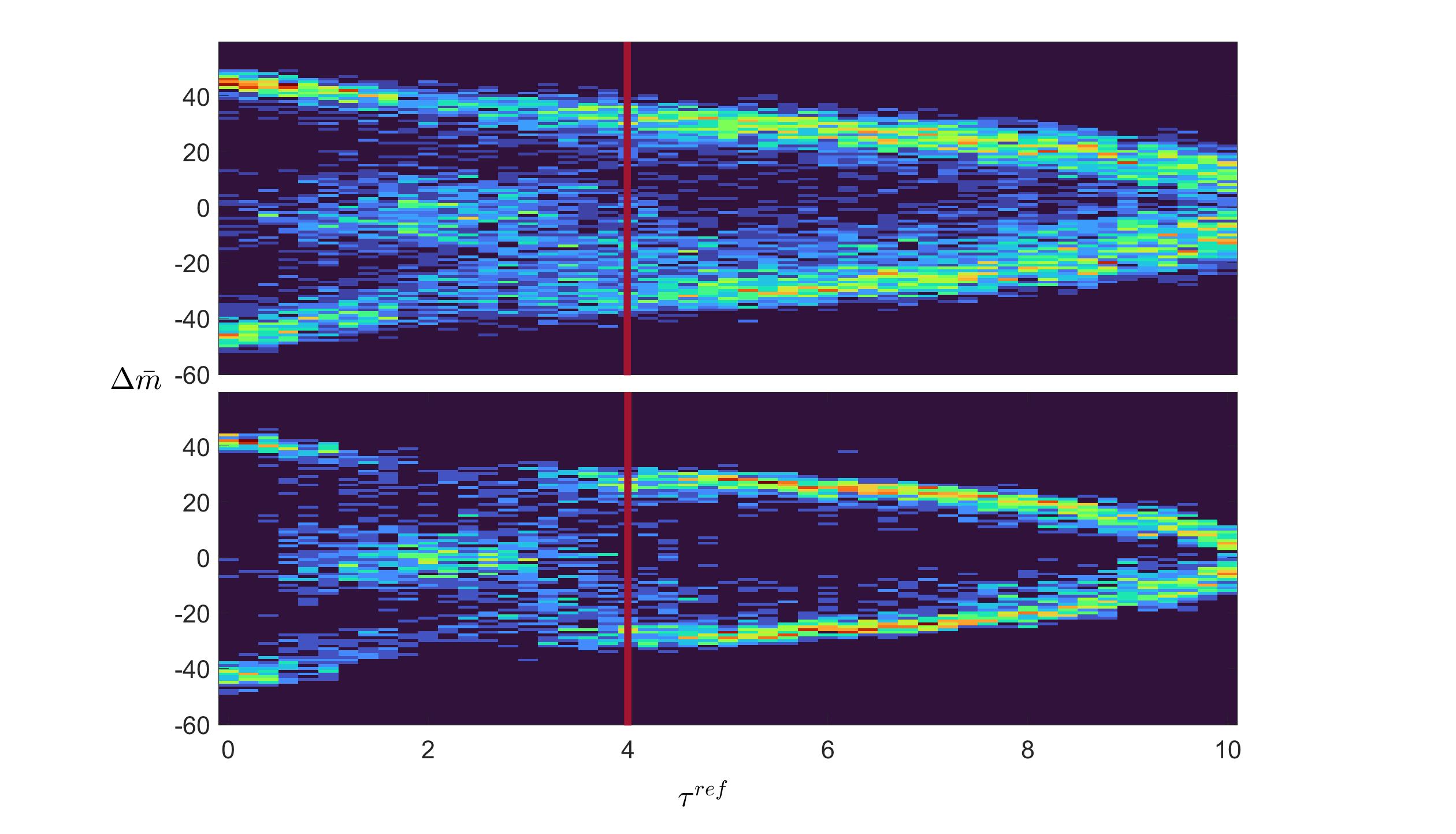}
    \end{subfigure}
    \begin{subfigure}{0.5\textwidth}
        {\bf H}\\
        \includegraphics*[bb=2.5in 0in 33in 19in,width=.98\textwidth]{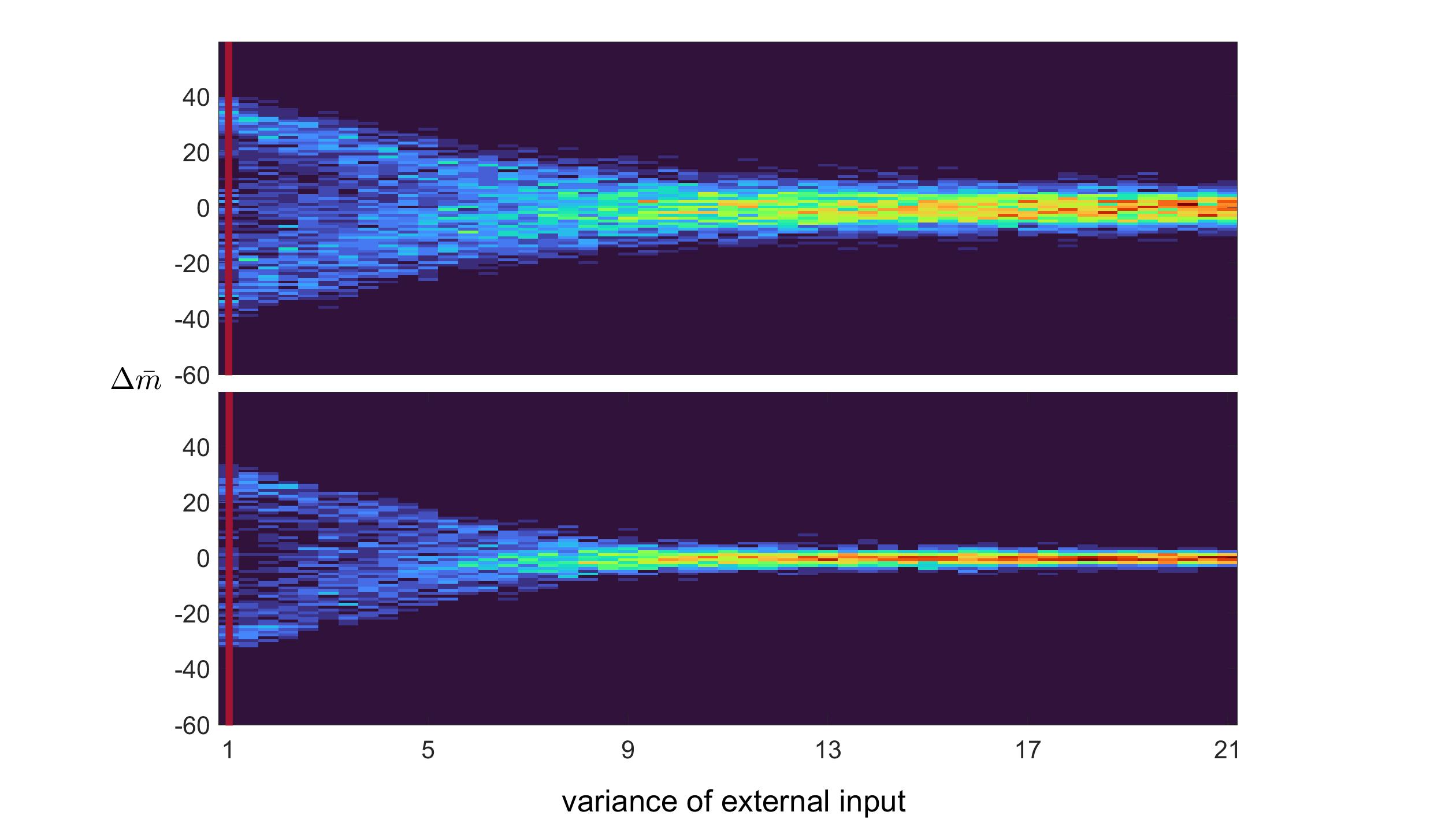}
    \end{subfigure}
    
    \caption{Comparing the bifurcation maps of spiking networks and dsSDEs. Similar to the spiking networks, the dsODE model also displays extra branches and the merging of nearby branches, which are absent in dsODEs without noise. }
    \label{FigS1_bifurcations_of_dsODE_noise}
\end{figure}

Figure~\ref{FigS1_bifurcations_of_dsODE_noise} compares the bifurcation map generated by dsSDE model with that generated by spiking networks. All panels correspond to those in Fig.~\ref{Fig7_bifurcations}, with the dsODE bifurcation maps replaced by dsSDE bifurcation maps. The density maps of spiking clusters are also normalized by 40 Hz. 


\printbibliography

\end{document}